\documentclass[a4paper,12pt]{article}

\usepackage[utf8]{inputenc}
\usepackage[obeyspaces]{url}
\usepackage{physics}
\usepackage{mathtools}
\usepackage{bm}
\usepackage{booktabs,tabularx}
\usepackage{graphicx}
\usepackage{xspace}
\usepackage[usenames]{xcolor}
\usepackage{listings}
\usepackage[absolute]{textpos}
\usepackage[many]{tcolorbox}
\usepackage{xparse}
\usepackage[font=small,labelfont=bf,format=plain]{caption}
\usepackage{bbm}
\usepackage{subcaption}
\usepackage{dsfont}
\usepackage[normalem]{ulem}
\usepackage{dirtree}
\usepackage[text,italic]{esdiff}
\usepackage{multirow}
\usepackage{tabularx}
\usepackage{footnote}
\usepackage{amsmath,amssymb}
\usepackage{fullpage}
\usepackage{lmodern}
\definecolor{darkblue}{rgb}{0.0, 0.0, 0.55}
\usepackage[colorlinks=true,allcolors=darkblue]{hyperref}
\usepackage{cleveref}
\usepackage[numbers,sort&compress]{natbib}

\bibstyle{elsarticle-num}

\makeatletter
\g@addto@macro\bfseries{\boldmath}
\makeatother

\newcommand{\eV}{\ensuremath{\text{e}\mspace{-0.8mu}\text{V}}\xspace}
\newcommand{\MeV}{\text{M\eV}\xspace}
\newcommand{\GeV}{\text{G\eV}\xspace}
\newcommand{\TeV}{\text{T\eV}\xspace}

\newcommand{\MSbar}{$\MSBar$\xspace}
\newcommand{\MSBar}{\overline{MS}}

\newcommand{\gambit}{\textsf{GAMBIT}\xspace}

\newcommand{\fs}{\textsf{FlexibleSUSY}\xspace}
\newcommand{\sarah}{\textsf{SARAH}\xspace}
\newcommand{\softsusy}{\textsf{SOFTSUSY}\xspace}

\newcommand{\email}[1]{\href{mailto:#1}{#1}}

\title{Global fits of SUSY at future Higgs factories}

\author{
Peter Athron$^{b}$\footnote{\email{peter.athron@njnu.edu.cn}}, 
Csaba Balazs$^c$\footnote{\email{csaba.balazs@monash.edu}}, 
Andrew Fowlie$^b$\footnote{\email{andrew.j.fowlie@njnu.edu.cn}}, 
Huifang Lv$^b$\footnote{\email{lvhf@njnu.edu.cn}}, 
\\
Wei Su$^d$\footnote{\email{weisu@kias.re.kr}}, 
Lei Wu$^b$\footnote{\email{leiwu@njnu.edu.cn}}, 
Jin Min Yang$^{e,f}$\footnote{\email{jmyang@itp.ac.cn}}, 
~Yang Zhang$^a$\footnote{\email{zhangyangphy@zzu.edu.cn}}}
\date{}

\begin{document}

\maketitle

% report numbers
\begin{textblock*}{10em}(0.9\textwidth,2cm)
\raggedleft\noindent\footnotesize
KIAS--P22012
\end{textblock*}

\thispagestyle{empty}
\begin{center}
  \it
 $^a$ School of Physics, Zhengzhou University, ZhengZhou 450000, P. R. China  \\[0.5em]
 $^b$ Department of Physics and Institute of Theoretical Physics, Nanjing Normal University, 
       Nanjing 210023, P. R. China \\
 $^c$School of Physics and Astronomy, Monash University, Melbourne 3800 Victoria, Australia\\
 $^d$ Korea Institute for Advanced Study, Seoul 02455, Korea\\
 $^e$ CAS Key Laboratory of Theoretical Physics, Institute of Theoretical Physics, Chinese Academy of Sciences, 
      Beijing 100190, P. R. China  \\
 $^f$ School of Physics Sciences, University of Chinese Academy of Sciences,  Beijing 100049, P. R. China  \\

\end{center}

\begin{abstract}
In this work, we study the impact of electroweak and Higgs precision measurements at future electron-positron colliders on several typical supersymmetric models, including the Constrained Minimal Supersymmetric Standard Model (CMSSM), Non-Universal Higgs Mass generalisations (NUHM1, NUHM2), and the 7-dimensional Minimal Supersymmetric Standard Model (MSSM7). Using publicly-available data from the \textsf{GAMBIT} community, we post-process previous SUSY global fits with additional likelihoods to explore the discovery potential of Higgs factories, such as the Circular Electron Positron Collider (CEPC), the Future Circular Collider (FCC) and the International Linear Collider (ILC). We show that the currently allowed parameter space of these models will be further tested by future precision measurements. In particular, dark matter annihilation mechanisms may be distinguished by precise measurements of Higgs observables.
\end{abstract}

\clearpage
\tableofcontents

\section{Introduction}

After the discovery of the Higgs boson at the Large Hadron Collider (LHC), precisely measuring the properties of the Higgs boson is the next essential task for the high-energy physics community. For this purpose, various next-generation electron-positron colliders have been proposed,  
including the Circular Electron-Position Collider (CEPC)~\cite{CEPCStudyGroup:2018ghi}, the Compact Linear
Collider (CLIC)~\cite{Charles:2018vfv}, the Future Circular Collider (FCC-ee)~\cite{Gomez-Ceballos:2013zzn},
and the International Linear Collider (ILC)~\cite{Baer:2013cma}.  
These state-of-the-art machines can not only scrutinize the nature of the Higgs boson but also shed complementary light 
on new physics, such as low-energy supersymmetry (SUSY; for recent overviews on the status of low-energy SUSY, see e.g.~\cite{Baer:2020kwz,Wang:2022rfd}). Restricted by collision energy, these colliders may not be able to directly 
produce supersymmetric particles,\footnote{For some special case with rather light supersymmetric particles,
see, e.g., \cite{Yuan:2021nsn,Chen:2021omv}.} but they can provide constraints on SUSY models through
high-precision measurements of the Higgs and electroweak (EW) sectors.

As is well known, global fits can provide comprehensive information on new physics models, allowing us to infer 
the maximum amount of information on a given model from the widest range of experimental data \cite{Kvellestad:2019vxm,AbdusSalam:2020rdj}.  
Global fits assess and compare the validity of  models, identify the ranges of model parameters with
the highest likelihood or posterior probability and study the predictions and consequences for future searches. Consequently, global fits are an important part of the Technical Design Report (TDR) for future electron-positron 
colliders.

The GAMBIT Collaboration~\cite{Athron:2017ard} have performed comprehensive global fits on five SUSY models:
the Constrained Minimal Supersymmetric Standard Model (CMSSM) and its Non-Universal Higgs Mass generalisations 
(NUHM1 and NUHM2)~\cite{Athron:2017qdc}, the seven-dimensional weak scale phenomenological MSSM 
(MSSM7) with all parameters defined at the weak scale~\cite{Athron:2017yua}, and a four-dimensional electroweakino sector of the MSSM~\cite{Athron:2018vxy}.  The likelihood functions of the first four global fits include several direct and indirect dark matter~(DM) searches, a large collection of electroweak precision and flavour 
observables, direct searches for SUSY at the Large Electron-Positron collider (LEP), and Runs I and II of the LHC, and constraints from Higgs observables. All GAMBIT input files and generated
likelihood samples for these models are publicly available online through 
Zenodo~\cite{the_gambit_collaboration_2017_843496,the_gambit_collaboration_2017_843535}.

As a comprehensive global fit is computationally expensive, we take full advantage of the massive sets of publicly-available samples that GAMBIT generated by post-processing them with likelihoods of expected precision limits from future electron-positron 
colliders. By comparing the preferred regions and best fits before and after applying such likelihoods, 
we can estimate the prospective reaches of future colliders. This procedure is very different from previous global 
fits of SUSY with future electron-positron colliders (see, e.g., \cite{Ellis:2015sca,Fan:2014txa,Fan:2014axa,Su:2016ghg,Liu:2017msv,Li:2020glc,Arbey:2021jdh}).  

The paper is structured as follows. 
We briefly introduce the theoretical framework of the CMSSM, NUHM1, NUHM2 and MSSM7 along with their parameters 
in Section~\ref{sec:model}.
We then review the precision expectations of the CEPC, ILC and FCC-$ee$ and present our post-processing strategy in Section~\ref{sec:strategy}.
Each subsection in Section~\ref{sec:result} contains implications for future collider searches on the global fit of one model. 
Finally, we summarize the results and draw our conclusions in Section~\ref{sec:conclusion}.

\section{The supersymmetric models}\label{sec:model}

We restrict our consideration to specific scenarios within the CP-conserving, $R$-symmetric, minimal supersymmetric 
standard model (MSSM)~\cite{Fayet:2015sra}.  The MSSM Lagrangian given in Sec.\ 5.4.3 of \cite{Athron:2017ard} defines the soft-breaking 
and SUSY-preserving parameters that describe the MSSM and fix our notation. Here we consider four distinct versions of the model where different constraints are applied to reduce the number of parameters.\footnote{In all four of these models we assume that all soft-breaking parameters are real, that the explicit CP-violating $M_1^\prime$, $M_2^\prime$, $M_3^\prime$ 
are set to zero and that all elements of the matrices ${\bf C}_u$, ${\bf C}_d$, ${\bf C}_e$ vanish.}

\subsection*{CMSSM}
The CMSSM is the most widely studied subspace of the general MSSM~\cite{Nilles:1983ge}. It is inspired by scenarios where SUSY breaking is transmitted through supergravity interactions, fixing the soft mass parameters at very high energy scales close to the Planck scale.  Specifically, influenced by a minimal form of supergravity where the universal couplings are assumed, 
the universal scalar mass $m_0$, the gaugino mass $m_{1/2}$ and the soft-scalar trilinear $A_0$ masses are defined at the Grand Unified Theory (GUT) scale, $M_X$, through the following constraints:
\begin{align}
    ({\bf m}_F^2)_{ij}(M_X) & = m_0^2 \delta_{ij}, \; \textrm{for} \; F\in\{Q,u,d,L,e\}, \label{Eq:CMSSM_m0}\\
    m_{\phi}^2(M_X) & = m_0^2, \;\textrm{for}\; \phi\in\{H_u, H_d\}, \label{Eq:CMSSM_m0_higgs}\\
    %%m_{H_u}^2(M_X) = m_{H_d}^2(M_X) & = m_0^2\label{Eq:CMSSM_m0_higgs}\\
%%    ({\bf m}_Q^2)_{ij} = ({\bf m}_u^2)_{ij} =  ({\bf m}_d^2)_{ij} =  ({\bf m}_L^2)_{ij} =  ({\bf m}_e^2)_{ij} = m_0^2 \delta_{ij},
    M_i(M_X) & = m_{1/2}, \;\textrm{for}\; i\in\{1,2,3\},\label{Eq:CMSSM_m12}\\
    %% M_1 = M_2 = M_2 = m_{1/2} 
    ({\bf A}_f)_{ij}(M_X) &= \delta_{ij} A_0, \;\textrm{for}\; f\in \{ u,d,e \}\label{Eq:CMSSM_A0} 
\end{align}
The Yukawa and gauge couplings of the MSSM can be fixed from data, while the Higgs sector parameters $\mu$ and $b$ (often written as $m_3^2$ or $B\mu$) are partially constrained by the EW VEV, leaving the sign of the superpotential $\mu$ parameter and the ratio of the vacuum expectation values of the two Higgs doublets,
\begin{equation}
\tan\beta = v_u /v_d,
\end{equation}
as free parameters. In our work we input $\tan\beta$ at the scale $m_Z$, following common conventions used in spectrum generators. This gives us four free continuous parameters and one free sign to fully specify scenarios in the CMSSM:
 \begin{equation}
      \textrm{CMSSM:} \; \{m_0, m_{1/2}, A_0, \tan\beta(m_Z), \textrm{sign}(\mu)\}
 \end{equation}

\subsection*{NUHM1}
In this variant, the supergravity inspiration of the CMSSM is maintained but the strong assumption of minimality is relaxed a little. Specifically the GUT-scale constraint on the soft scalar Higgs masses is relaxed, introducing a new free parameter $m_H$~\cite{Matalliotakis:1994ft,Olechowski:1994gm,Berezinsky:1995cj,Drees:1996pk,Nath:1997qm}. The constraint in Eq.\ \ref{Eq:CMSSM_m0_higgs} is replaced by %%\begin{equation}m_{H_u}=m_{H_d}=m_{H}\end{equation} .
\begin{equation}
    m_{\phi}^2(M_X) = m_H^2 \;\textrm{for}\; \phi\in\{H_u, H_d\} \label{Eq:NUHM1_m0_higgs}
\end{equation}
The constraints in Eqs.\ \ref{Eq:CMSSM_m0}, \ref{Eq:CMSSM_m12} and \ref{Eq:CMSSM_A0} are still applied in the NUHM1. 
This means the predictions of the NUHM1 are determined from five free continuous parameters and one sign,
 \begin{equation}
      \textrm{NUHM1:} \; \{m_H, m_0, m_{1/2}, A_0, \tan\beta(m_Z), \textrm{sign}(\mu)\}
 \end{equation}

\subsection*{NUHM2}
Extending this idea slightly further, in the NUHM2 the constraint on the soft Higgs masses is even further relaxed so that $m_{H_u}(M_X)$ and $m_{H_d}(M_X)$ become independent, real, dimension-one parameters at the GUT scale. Therefore, in the NUHM2 only the constraints in Eqs.\ \ref{Eq:CMSSM_m0}, \ref{Eq:CMSSM_m12} and \ref{Eq:CMSSM_A0} are applied. This leaves six free continuous parameters and one sign to specify the physics predictions of the model,
\begin{equation}
     \textrm{NUHM2:} \; \{m_{H_u}(M_X),m_{H_d}(M_X), m_0, m_{1/2}, A_0, \tan\beta(m_Z), \textrm{sign}(\mu)\}
 \end{equation}

\subsection*{MSSM7}
In the MSSM7 the constraints are no longer applied at the GUT scale, allowing input parameters to be specified close to the weak scale and thus no longer requiring renormalization group (RG) running between this scale and the Planck scale to determine 
the models' physical predictions~\cite{MSSMWorkingGroup:1998fiq}. The number of parameters is reduced through the following constraints
\begin{align}
 ({\bf m}_F^2)_{ij}(Q) & = m_{\tilde{f}}^2 \,\delta_{ij}, \; \textrm{for} \; F\in\{Q,u,d,L,e\}, \label{Eq:MSSM7_mf}\\
 M_1(Q) &= \frac{ \sin^2\theta_W}{3/5 \cos^2\theta_W} M_2 \label{Eq:MSSM7_M1} \\
 M_3 (Q) &= \frac{\sin^2\theta_W \alpha_s}{\alpha} M_2 \label{Eq:MSSM7_M3} \\
 %%$3/5 \cos^2\theta_W M_1 & = M_2 = \alpha/\alpha_s M_3$. 
({\bf A}_f)_{ij}(Q) &= 0 \; \forall \;  (i,j) \neq (3,3), \textrm{for} \;  f\in \{u,d\} \label{Eq:MSSM7_Au} \\
{\bf A}_e(Q) &= 0   \label{Eq:MSSM7_Ade}
\end{align}
where for the weak mixing angle we use a fixed value of 
$\sin^2\theta_W = \frac12 - \sqrt{\frac14 - \pi \alpha /(\sqrt{2} m_Z^2 G_F)}$, 
which is independent of precise extractions of the gauge couplings carried out during spectrum generation.
Although in principle the scale $Q$ can be arbitrary, Eqs.\ \ref{Eq:MSSM7_M1} and \ref{Eq:MSSM7_M3} are inspired by the impact of the GUT relation in Eq.\ \ref{Eq:CMSSM_m12} on the gaugino masses at the weak scale. So it does not make much practical sense to choose $Q$ too far away from this scale, and in \cite{Athron:2017fvs} and here we fix $Q= 1$ TeV. 
For the sfermion sector, we assume a common sfermion mass at 1 TeV with no flavour changing and assume only the third generation up- and down-type soft trilinear contribute non-negligibly. In the MSSM7 we also make the trade between VEVs and $|\mu|$, $b$. As in the NUHM2 both soft Higgs masses are free inputs, $m_{H_u}(Q)$ and $m_{H_d}(Q)$, albeit with the parameters input at very different scales which will substantially change the prior distributions they are drawn from. Similarly the sfermion masses are also no longer split by the RG flow,
and the soft trilinears are treated differently with $A_t$ and $A_b$ as free parameters and $A_\tau=0$. 
The final list of free parameters then has seven continuous parameters and one sign:
\begin{equation}\textrm{MSSM7:} \; \{M_2(Q), A_{u_3}(Q), A_{d_3}(Q), m_{\tilde{f}}^2(Q), m_{H_u}^2(Q), m_{H_d}^2(Q), \tan\beta(m_Z), \textrm{sign}(\mu)\}. 
\end{equation}

Global fits of these four models have previously been performed with \gambit in \cite{Athron:2017qdc,Athron:2017yua}. 
Here we will post-process the results from these global fits to explore the impact of future electron-positron colliders.  For the samples we use, the constraints listed above have been applied in \fs \cite{Athron:2014yba,Athron:2017fvs}, which calculates the pole masses and $\overline{\text{DR}}$ couplings used in the global fit. The values of the parameters at different scales are connected by two-loop RGEs which 
\fs obtains\footnote{\fs also uses some code pieces from \softsusy \cite{Allanach:2001kg,Allanach:2013kza}.} 
with the help of \sarah~ \cite{Staub:2009bi,Staub:2010jh,Staub:2012pb,Staub:2013tta}.  
The scanned parameter ranges can be found in \cite{Athron:2017qdc,Athron:2017yua}. All dimensionful parameters are allowed to vary up to 10 \TeV, safely covering scenarios that are motivated by the hierarchy problem and the most phenomenologically interesting regions within reach of colliders. 
% Besides the model parameters, we plan to include nuisance parameters to account for uncertainties associated with the local Dark Matter~(DM) distribution, nuclear matrix elements relevant for direct detection, and SM parameters.

\section{Study Strategy}\label{sec:strategy}

In our analysis, we post-process the publicly available data on  \textsf{Zenodo} for global fits of the GUT scale SUSY models~\cite{the_gambit_collaboration_2017_843496}, and MSSM7~\cite{the_gambit_collaboration_2017_843535}, with additional likelihoods for precision measurements of the SM Higgs observables at the proposed Higgs factories. The total likelihood is thus,
\begin{equation}
    \mathcal{L} = \mathcal{L}_{\gambit} \cdot \mathcal{L}_{\text{Higgs factories}}
\end{equation}
where the likelihood already computed in the publicly available data, $\mathcal{L}_{\gambit}$, includes contributions from a large collection of present constraints on dark matter, electroweak 
precision, flavour observables, sparticles and the SM-like Higgs boson. Some of the constraints have been improved since 
the publication of the data, but they will not qualitatively affect our calculation here. Since we post-process the data, the parameter ranges and priors are the same as in the original studies~\cite{Athron:2017qdc,Athron:2017yua}. 

A Higgs factory with $e^+e^-$ collisions at a center-of-mass energy of 240-250\,GeV exploits the Higgsstrahlung process
\begin{equation}
e^+e^- \to hZ.
\end{equation}
With millions of Higgs bosons produced and the clean experimental conditions at lepton colliders, both the
inclusive cross section $\sigma_{hZ}$ independent of the Higgs decays, and the exclusive channels of individual 
Higgs decays in terms of $\sigma_{hZ}\times {\rm Br}$, can be measured to impressive precision. The cross sections of vector boson fusion processes for the Higgs production are relatively small at low values of center-of-mass energy. Only the main decay modes can be measured. With the center of mass energy increasing, the cross sections of vector boson fusion processes grow logarithmically and can provide crucial complementary information.

In Table~\ref{tab:mu_precision}, we list the anticipated precision of measurements of the Higgs boson rates in the running scenarios of various machines in terms of their center of mass energies and the corresponding integrated luminosities. For the center-of-mass energy of $240-250\,\GeV$, only the measurement of $h\to b\bar{b}$ is provided in the vector boson fusion process, which is listed in the last row.

\begin{table}[tb]
\scriptsize
 \begin{center}
  \begin{tabular}{lrrrrrrrrrrr}
    \toprule
    ~ & \multicolumn{1}{c}{ILC} & \multicolumn{2}{c}{ILC}  & \multicolumn{2}{c}{ILC} & \multicolumn{1}{c}{FCC-$ee$} & \multicolumn{2}{c}{FCC-$ee$} & \multicolumn{1}{c}{CEPC}  & \multicolumn{2}{c}{CEPC}\\
    ~ & \multicolumn{1}{c}{250\,\GeV} & \multicolumn{2}{c}{350\,\GeV}  & \multicolumn{2}{c}{500\,\GeV} & \multicolumn{1}{c}{240\,\GeV}  & \multicolumn{2}{c}{365\,\GeV} & \multicolumn{1}{c}{240\,\GeV}   & \multicolumn{2}{c}{360\,\GeV} \\
    ~ & \multicolumn{1}{c}{2 ab$^{-1}$} & \multicolumn{2}{c}{200 fb$^{-1}$}  & \multicolumn{2}{c}{4 ab$^{-1}$} & \multicolumn{1}{c}{5 ab$^{-1}$}  & \multicolumn{2}{c}{1.5 ab$^{-1}$} & \multicolumn{1}{c}{20 ab$^{-1}$} & \multicolumn{2}{c}{1 ab$^{-1}$} \\
    \midrule
    $\sigma_{Zh}$ & 0.71\% & \multicolumn{2}{c}{2.0\%} & \multicolumn{2}{c}{1.05\%} & 0.5\%  & \multicolumn{2}{c}{0.9\%} & 0.26\%  &  \multicolumn{2}{c}{1.4\%}\\ 
    \midrule
    Decay mode & $\sigma_{Zh}$Br & $\sigma_{Zh}$Br & $\sigma_{\nu\bar{\nu}h}$Br & $\sigma_{Zh}$Br & $\sigma_{\nu\bar{\nu}h}$Br & $\sigma_{Zh}$Br & $\sigma_{Zh}$Br & 
    $\sigma_{\nu\bar{\nu}h}$Br & $\sigma_{Zh}$Br & $\sigma_{Zh}$Br &  $\sigma_{\nu\bar{\nu}h}$Br\\
    \midrule
    $h\to b\bar{b}$ & 0.46\% & 1.7\% & 2.0\% & 0.63\% & 0.23\% & 0.3\% & 0.5\% & 0.9\% & 0.14\% & 0.9\% & 1.1\%\\
    $h\to c\bar{c}$ & 2.9\% & 12.3\% & 21.2\% & 4.5\% & 2.2\% & 2.2\% & 6.5\% & 10\% & 2.02\% & 8.8\% & 16\%\\
    $h\to gg$ & 2.5\% & 9.4\% & 8.6\% & 3.8\% & 1.5\% & 1.9\% & 3.5\% & 4.5\% & 0.81\% & 3.4\% & 4.5\%\\
    $h\to WW^*$ & 1.6\% & 6.3\% & 6.4\% & 1.9\% & 0.85\% & 1.2\% & 2.6\% & 3.0\% & 0.53\% & 2.8\% & 4.4\%\\
    $h\to \tau^+\tau^-$ & 1.1\% & 4.5\% & 17.9\% & 1.5\% & 2.5\% & 0.9\% & 1.8\% & 8.0\% & 0.42\% & 2.1\% & 4.2\% \\
    $h\to ZZ^*$ & 6.4\% & 28.0\% & 22.4\% & 8.8\% & 3.0\% & 4.4\% & 12\% & 10\% & 4.17\% & 20\% & 21\%\\
    $h\to \gamma\gamma$ & 12.0\% & 43.6\% & 50.3\% & 12.0\% & 6.8\% & 9.0\% & 18\% & 22\% & 3.02\% & 11\% & 16\%\\
    $h\to \mu^+\mu^-$ & 25.5\% & 97.3\% & 178.9\% & 30.0\% & 25.0\% & 19\% & 40\% & - & 6.36\% & 41\% & 57\% \\
    $(\nu\bar{\nu})h\to b\bar{b}$ & 3.7\% & - & - & - & - & 3.1\% & - & - & 1.58\%  & - & -\\
    \bottomrule
  \end{tabular}
  \caption{Estimated statistical precision for Higgs measurements at the proposed ILC program with various 
center-of-mass energies~\cite{Bambade:2019fyw}, the FCC-ee program with 5 ab$^{-1}$ integrated luminosity at $\sqrt{s}=240~\GeV$ and 1.5 ab$^{-1}$ integrated luminosity at $\sqrt{s}=365~\GeV$ 
~\cite{Abada:2019lih,Abada:2019zxq}, and the CEPC program with 20 ab$^{-1}$ integrated luminosity at $\sqrt{s}=240~\GeV$ and 1 ab$^{-1}$ integrated luminosity at $\sqrt{s}=360~\GeV$ 
~\cite{CEPC}. 
   }
\label{tab:mu_precision}
  \end{center}
\end{table}

We define the new likelihoods for the proposed Higgs factories simply as 
\begin{equation}
-2\ln\mathcal{L}_{\text{Higgs factories}} = \frac{(m_h-m_h^{\rm obs})^2}{\sigma_{\mu_h}^2} 
+ \frac{(\sigma_{Zh}-\sigma_{Zh}^{\rm obs})^2}{\sigma_{\sigma_{Zh}}^2}
+ \sum\frac{(\mu_i-\mu_i^{\rm obs})^2}{\sigma_{\mu_i}^2},
\end{equation}
where 
\begin{equation}
\mu_i = \frac{\sigma_i\times {\rm Br}_i}{\sigma_i^{\rm SM}\times {\rm Br}_i^{\rm SM}}
\end{equation}
and the index $i$ runs over all the Higgs search channels in Table~\ref{tab:mu_precision}. We take the values of branching ratios for the 125\,\GeV SM Higgs as ${\rm Br}_i^{\rm SM}$, listed in Table~\ref{tab:sm_higgs}. 

The total uncertainties $\sigma_{m_h}$ and $\sigma_{\mu_i}$ consist of experimental uncertainties and theoretical uncertainties, $\sigma_{\rm tot} = \sqrt{\sigma_{\rm the}^2+\sigma_{\rm exp}^2}$. The parametric uncertainties are involved in through nuisance input parameters. For the SM Higgs mass, the present experimental uncertainty, $\sigma_{m_h}^{\rm exp}=0.17\,\GeV$~\cite{Zyla:2020zbs}, is already small and can be neglected compared with the theoretical uncertainty in \gambit, $\sigma_{m_h}^{\rm the}=2\,\GeV$. Thus, we ignore the contribution from the Higgs mass measurement at future Higgs factories to avoid double counting the likelihood for the Higgs mass. 
We use the anticipated precisions displayed in Table~\ref{tab:mu_precision} as the experimental uncertainties on future signal strength measurements, while the theoretical uncertainty is in the same order and therefore cannot be ignored. 
In Table~\ref{tab:sm_higgs}, we show the present theoretical uncertainties of branching ratios for the 125~\GeV SM Higgs boson, which are expected to be improved in the future. 
We consider theoretical uncertainties $\sigma_{\mu_i}^{\rm the} = k \sigma_{\mu_i}^{\rm SM}$ for BSM predictions and by default we use $k=0.2$ in our likelihood, representing a scenario where the theoretical uncertainties are reduced to the same level as the expected experimental uncertainties.
However, the uncertainties on the branching ratios computed for BSM models in \gambit, which use \textsf{SUSY-HIT}~\textsf{1.5}~\cite{Djouadi:1997yw,Muhlleitner:2003vg,Djouadi:2006bz} 
via \textsf{DecayBit}~\cite{GAMBITModelsWorkgroup:2017ilg}, are obviously larger than those in the SM, as carefully shown in \cite{Athron:2021kve} and discussed in \cite{Arbey:2021jdh}. We discuss the impacts of  SUSY contributions to the uncertainties, and using different choices of $k$ in Section~\ref{sec:nuhm}.

\begin{table}[t]
\begin{center}
\begin{tabular}{l c c}
\toprule
Decay mode & Branching ratio & Theoretical error \\
\midrule
$h\to b\bar{b}$     & 57.7\%  & 3.3\% \\
$h\to c\bar{c}$     & 2.91\% & 12\%  \\
$h\to gg$           & 8.57\%  & 10\%  \\
$h\to WW^*$         & 21.5\% & 4.3\% \\
$h\to ZZ^*$         & 2.64\%  & 4.3\% \\
$h\to \gamma\gamma$ & $2.28\times10^{-3}$ & 5\% \\
$h\to \tau^+\tau^-$ & 6.32\% & 5.7\%\\
$h\to \mu^+\mu^-$   & $2.19\times10^{-4}$ & 6.0\%\\
\bottomrule
\end{tabular}
\caption{SM predictions of the decay branching ratios for the 125 GeV Higgs boson~\cite{LHCHiggsCrossSectionWorkingGroup:2013rie}, as well as the relative theoretical uncertainties.}\label{tab:sm_higgs}
\end{center}
\end{table}

As for the central values of future signal strength measurements, $\mu_i^{\rm obs}$,
in our scenario we assume that signal strengths predicted by the best-fit (BF) points in the \gambit datasets are measured. Since the \gambit data are sampled with \textsf{Diver}~\cite{Martinez:2017lzg} and \textsf{MultiNest}~\cite{Feroz:2008xx,Feroz:2013hea}, the regions around the BF points are densely sampled, such that our post-processing procedure should be reasonable. Since the BF point agrees exactly with our assumed measurements, the BF point should not change through post-processing. The extent to which the confidence regions around the BF shrink, however, will reveal the potential impact of precise Higgs factory measurements.  
We will also use other central values to check the validity of conclusions drawn from this choice in Section~\ref{sec:cmssm}.

The proposed electron-positron colliders are also designed to run at the $Z$-pole, and have an excellent capability for precise measurements of EW observables. These measurements are complementary to the Higgs boson coupling measurements. For instance, in the so-called ``blind spot'' the coupling between the lightest stop and the light Higgs boson vanishes, but the stop contribution to the EW precision observables can be visible~\cite{Fan:2014axa}. In the existing GAMBIT data, the strong coupling at the scale $m_Z$ in the \MSbar scheme $\alpha_s^{\overline{\text{MS}}}(m_Z)$ and the top pole mass $m_t$ are nuisance parameters varying within $\pm3\sigma$ of their observed central values, while the $Z$ pole mass is fixed to $91.1876\,\GeV$. Beside, the $W$ mass and $\sin^2\theta_W$ are outputs calculated in \textsf{PrecisionBit}~\cite{GAMBITModelsWorkgroup:2017ilg}. Therefore, we also investigate the impact of the expected EW precision measurements using 
\begin{align}
-2\ln\mathcal{L}_{\text{$Z$ factories}} ={} & \frac{(m_t-m_t^{\rm obs})^2}{\sigma_{m_t}^2} 
+ \frac{(\alpha_s^{\overline{\text{MS}}}(m_Z)-\alpha_s^{\overline{\text{MS}}}(m_Z)^{\rm obs})^2}{\sigma_{\alpha_s^{\overline{\text{MS}}}(m_Z)}^2} \\ \nonumber
& + \frac{(m_W-m_W^{\rm obs})^2}{\sigma_{m_W}^2}
+\frac{(\sin^2\theta_W-\sin^2\theta_W^{\rm obs})^2}{\sigma_{\sin^2\theta_W}^2}.
\end{align}
For the same reason as the $\mathcal{L}_{\text{Higgs factories}}$, we set the central values of these measurements to those predicted by the BF point. They are listed for the CMSSM in Table~\ref{tab:Z_precision}, along with the central values used in the likelihood of previous GAMBIT fits and the anticipated precisions of measurements of the EW observables at future lepton colliders. As for theoretical uncertainties, we adopt $1.5\times10^{-5}$ for $\sin^2\theta_W$ and 1~\MeV for $m_W$. 

\begin{table}[tb]
 \begin{center}
  \begin{tabular}{lcccccc}
    \toprule
    ~ & CMSSM    & Present & \multicolumn{3}{c}{Precision}  \\
    ~ & BF point & central value  & ILC & FCC-$ee$ & CEPC  \\
    \midrule
    $m_Z$ & 91.1876\,\GeV & 91.1876\,\GeV & 2.1~\MeV & 0.1~\MeV & 0.5~\MeV\\
    $m_t$ & 173.267\,\GeV & 173.34\,\GeV & 0.03\,\GeV & 0.6\,\GeV & 0.6\,\GeV \\
    $\alpha_s^{\overline{\text{MS}}}(m_Z)$ & 0.11862 & 0.1185 &  $1.0\times 10^{-4}$ & $1.0\times 10^{-4}$  & $1.0\times 10^{-4}$ \\
    $m_W$ & 80.3786\,\GeV & 80.385\,\GeV & 5~\MeV & 8~\MeV & 3~\MeV \\
    $\sin^2\theta_W$ & 0.231424 & 0.23155 & 1.3$\times10^{-5}$ & 0.3$\times10^{-5}$ & 4.6$\times10^{-5}$\\
    \bottomrule
  \end{tabular}
  \caption{ The theoretical predictions of EW observables for the BF point of CMSSM and the corresponding central values used in the previous GAMBIT fits, along with estimated statistical precisions for EW measurements at the proposed ILC, FCC-$ee$ and CEPC programs.
   The precisions are summarised in~\cite{Chen:2018shg}, which are originally from~\cite{Baak:2014ora,Lepage:2014fla,Fan:2014vta,Baak:2013fwa}.
  }
\label{tab:Z_precision}
  \end{center}
\end{table}

In total, we post-processed $7.1 \times 10^7$ viable samples for the CMSSM, $9.4 \times 10^7$ samples for the NUHM1, 
$1.2 \times 10^8$ samples for the NUHM2,  and $1.8 \times 10^8$ samples for the MSSM7. For each of the models the post-processing took several days on 1280 supercomputer cores. In order to compare our results with those shown in previous \gambit 
papers~\cite{Athron:2017qdc,Athron:2017yua}, we present and plot our results in the same way. The samples are sorted 
into 60 bins across the range of data values in each dimension, and  the resulting profile likelihoods are 
interpolated with a bilinear scheme in \textsf{Pippi}~\textsf{2.0}~\cite{Scott:2012qh}.

The post-processing is performed using \gambit~\textsf{2.1.1}, while the original global fits employed \gambit~\textsf{1.0}.
The spectrum generators in the two versions are slightly different, as \gambit~\textsf{2.1.1} includes bug fixes in calculations of the Higgs
branching ratios. Therefore, the branching ratios of the SM-like Higgs in the two results are not fully consistent. Fortunately, the difference is slight and can be neglected compared to the current precision of Higgs measurements at the LHC. Using the new branching ratios, the best-fit regions in the old studies will not change qualitatively. Thus, it is still reasonable to use the results shown in papers~\cite{Athron:2017qdc,Athron:2017yua} 
as the global fits of the present likelihood. 

%%%%%%%%%%%%%%%%%%%%%%%%%%%%%%%%%%%%%%%%%%%%%%%%%%%%%%%
%%%%%%%%%%%%%%%%%%%%%%%%%%%%%%%%%%%%%%%%%%%%%%%%%%%%%%%
\section{Results} 
\label{sec:result}

In this section, we compare profile likelihoods with and without the additional likelihood for the Higgs measurements
at future electron-positron colliders, by taking CEPC as an example, in the CMSSM, NUHM1, NUHM2 and MSSM7.
Moreover, the dependence of the results on assumptions about central values of Higgs measurements at future facilities 
and theoretical uncertainties are investigated. We also compare the sensitivities of the CEPC, FCC-$ee$ and ILC.

%%%%%%%%%%%
\subsection{CMSSM}
\label{sec:cmssm}

\begin{figure}[!th]
  \centering
  \includegraphics[width=0.49\textwidth]{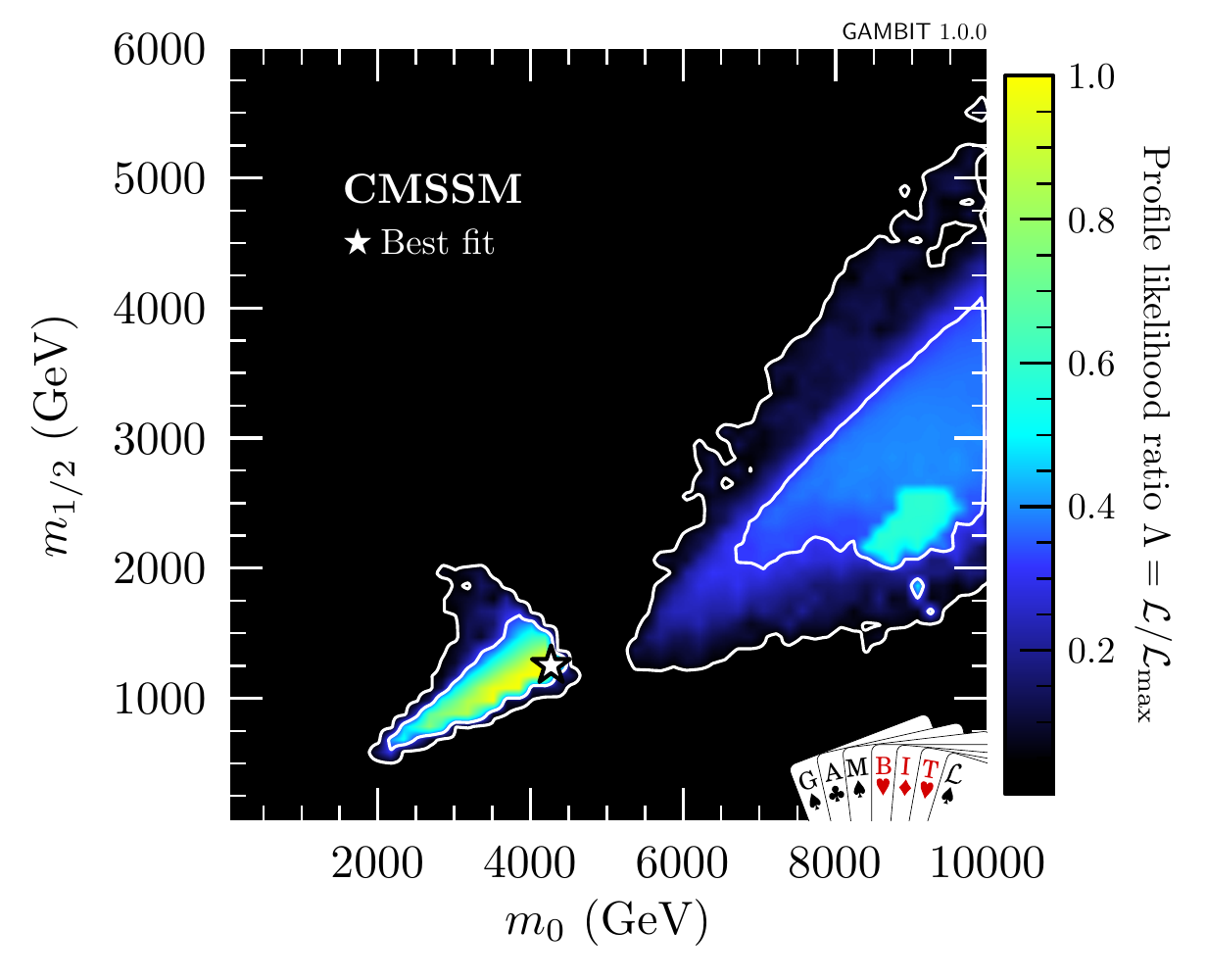}
  \includegraphics[width=0.49\textwidth]{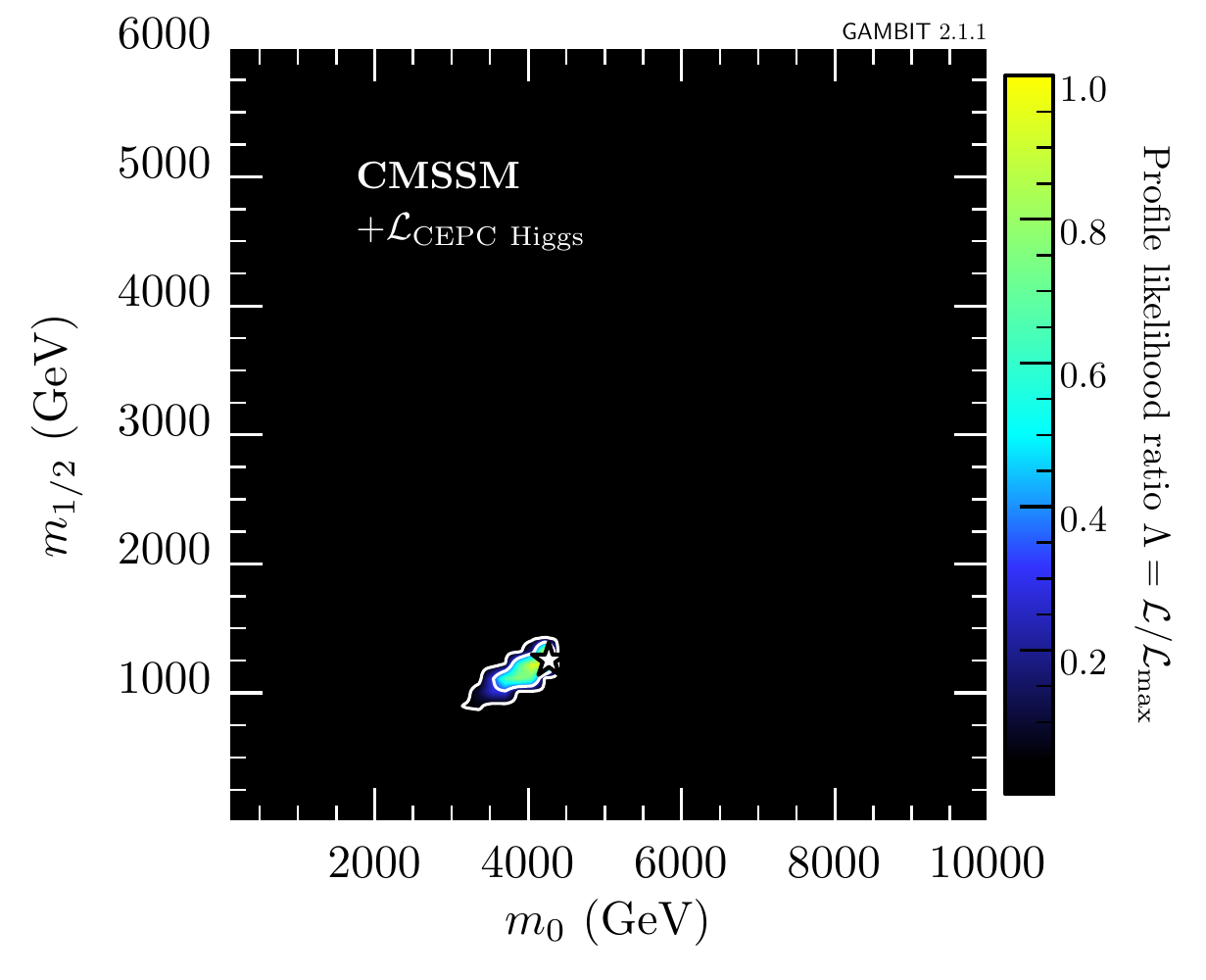}
  \includegraphics[width=0.49\textwidth]{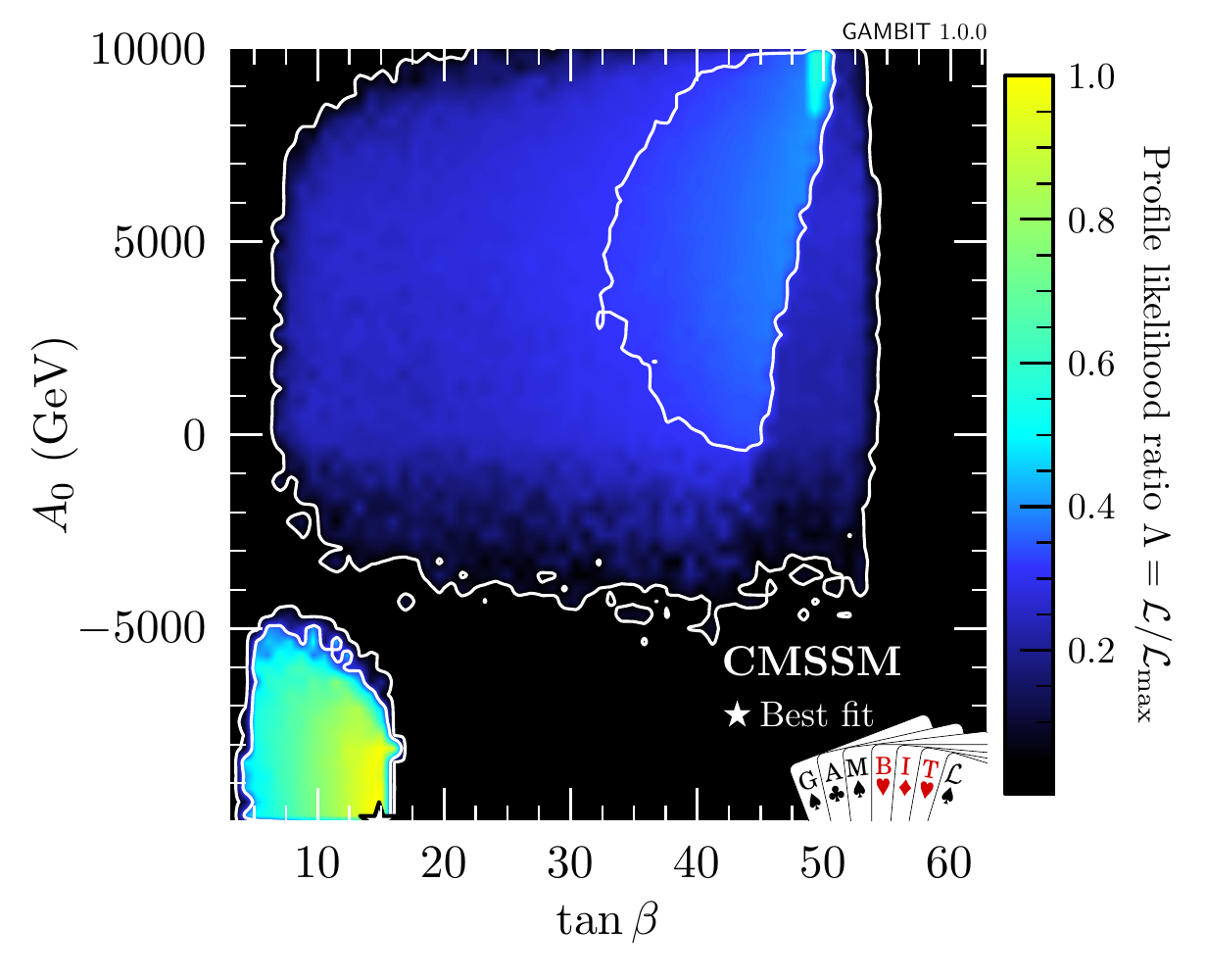}
  \includegraphics[width=0.49\textwidth]{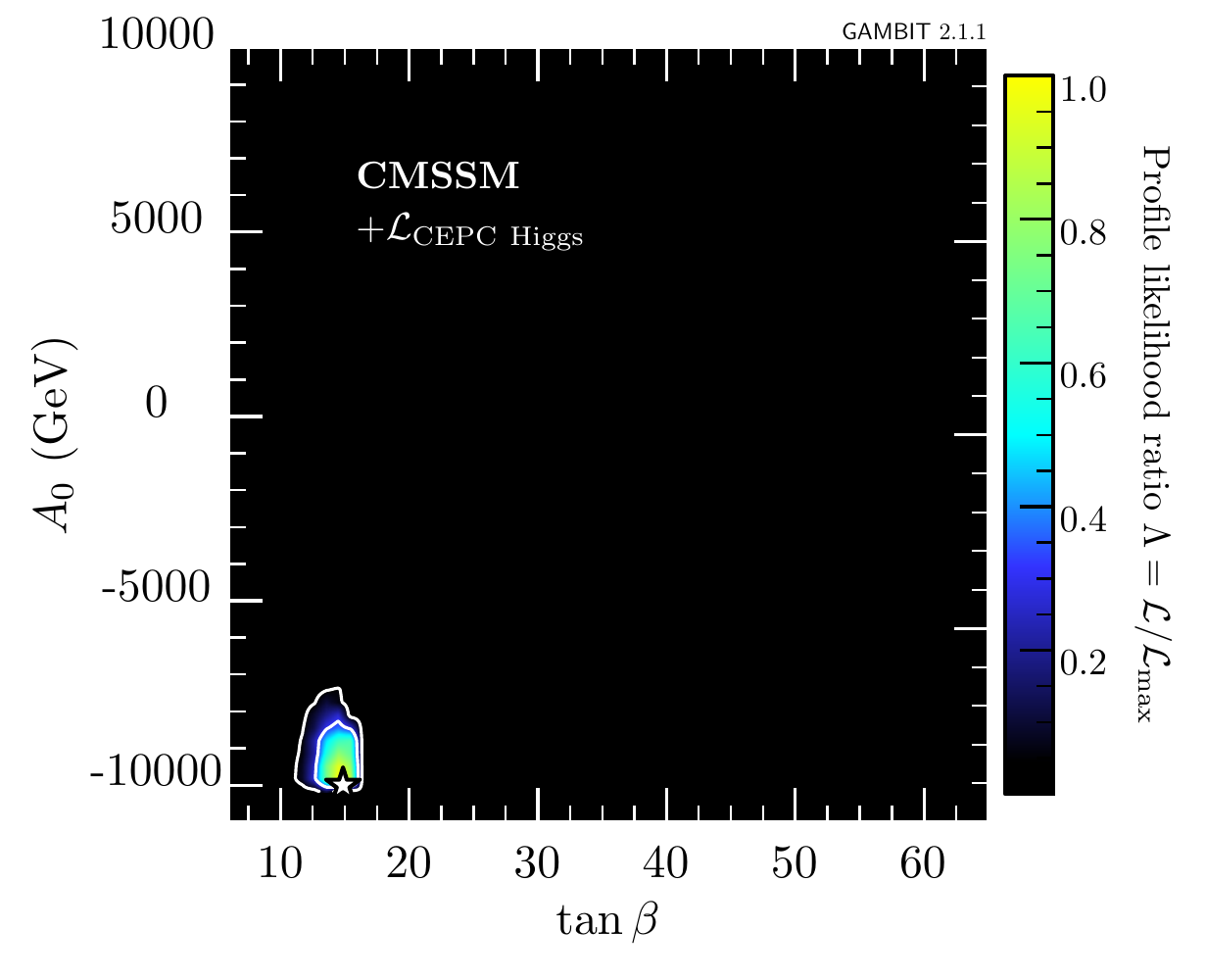}\\
  \caption{ The profile likelihood ratio in the CMSSM for the present constraints  
(left panels, taken from~\cite{Athron:2017qdc}) and including additional CEPC measurements (right panels), 
with 68\% and 95\% CL contours drawn in white, and the best-fit point indicated by a star. 
We use the branching ratios of the best-fit point 
in the CMSSM as the central values of the measurements at the CEPC, and the theoretical uncertainties are $k=0.2$ 
times smaller than the current SM Higgs theoretical uncertainties.}
  \label{fig:2d_parameter_plots_cmssm}
\end{figure}

We show the 2D profile likelihoods in Figure~\ref{fig:2d_parameter_plots_cmssm} for 
the input parameters of the CMSSM without (left panels) and with (right panel) the additional likelihood 
for the Higgs coupling measurements at the CEPC. Here we assume that the central values of measurements 
at the CEPC are same as those predicted by the best-fit point of the CMSSM, and the theoretical uncertainties are $k=0.2$ times smaller than
the current SM Higgs theoretical uncertainties. 

We see that a large part of the region favored by present constraints is excluded by the Higgs precision 
measurements at the CEPC. As we use the best-fit point (the white star in the plots) to set the central 
values of the measurements at the CEPC, it leads to $-2\ln\mathcal{L}_{\rm CEPC}=0$ at the best-fit point, so 
the position of the best-fit point holds still, and the preferred regions shrink significantly towards  
the best-fit point. It is very useful to classify the regions according to the possible DM annihilation 
mechanisms as follows:
\begin{itemize}
\item stop co-annihilation: $m_{\tilde{t}_1} \leq 1.2\,m_{\tilde\chi^0_1}$,
\item $A/H$-funnel: $1.6\,m_{\tilde\chi^0_1} \leq m_\textrm{heavy} \leq 2.4\,m_{\tilde\chi^0_1}$,
\item chargino co-annihilation: $\tilde\chi_1^0$ $\ge50\%$ Higgsino,
\end{itemize}
where `heavy' means a heavy Higgs like $A^0$ or $H^0$. The best-fit point is located in the stop co-annihilation region. 
The stop co-annihilation region prefers large and negative $A_0$, and extends below the lower bound of bottom panels of Figure~\ref{fig:2d_parameter_plots_cmssm}. However, vacuum stability problem need to be examined carefully in this region.
We see that the regions of large $m_0$, $m_{1/2}$ and $\tan\beta$, i.e.\ the $A/H$-funnel and chargino 
co-annihilation regions vanish. Besides, the sign of $\mu$ in the remaining stop co-annihilation regions is always negative. 

The sign of the $\mu$ parameter highly affects several physical observables, such as the recently updated 
anomalous magnetic moment of the muon $a_{\mu}$~\cite{Muong-2:2021ojo}. As mentioned in~\cite{Athron:2017qdc}, 
the fit favors $\mu<0$ versus $\mu>0$ by $\Delta \ln \mathcal{L}=0.4$, mainly because of the LHC Higgs signal likelihood. With the impressive precision of Higgs property measurements at the CEPC, the distinction between 
$\mu>0$ and $\mu<0$ reaches more than $2\sigma$. 

Note that the classification of dark matter annihilation mechanisms is not exclusive. A sample can lie
in more than one region. The classifications provide information about the relationship between sparticle masses. 
Samples in the same region have similar parameter values and mass spectra, as well as  similar SM Higgs branching ratios. 

\begin{table}[t]
\center
\begin{tabular}{lccccc}
\toprule
      & {~~$A/H$-funnel~~} & {~~$\tilde{\chi}_1^{\pm}$ co-ann.~~} & {~~$\tilde{t}$ co-ann.~~} & \\
\midrule
$A_0$       & 9924.435 & 9206.079 & -9965.036 &\\
$m_0$       & 9136.379 & 9000.628 & 4269.402  & \\
$m_{1/2}$   & 2532.163 & 2256.472 & 1266.043  &\\
$\tan\beta$ & 49.048   & 49.879   & 14.857    &\\
sgn($\mu$)  & -        & -         & -         & {Total error} \\
\midrule
$\sigma_{Zh}/\sigma_{Zh}^{\rm SM}$   & 0.975 & 0.977   & 0.975 & 0.5\% \\
$\mu_{(Z)h\to b\bar{b}}$     & 1.018 & 1.031   & 0.995 & 0.71\% \\
$\mu_{(Z)h\to c\bar{c}}$     & 1.135 & 1.125   & 1.154 & 4.08\% \\
$\mu_{(Z)h\to gg}$           & 0.775 & 0.768   & 0.745 & 2.38\%\\
$\mu_{(Z)h\to WW^*}$         & 0.913 & 0.900   & 0.974 & 1.31\% \\
$\mu_{(Z)h\to ZZ^*}$         & 0.934 & 0.919  & 1.003 & 5.17\% \\
$\mu_{(Z)h\to \gamma\gamma}$ & 1.127 & 1.116 & 1.182 & 6.87\% \\
$\mu_{(Z)h\to \tau^+\tau^-}$ & 0.983  & 0.978   & 0.994  & 1.39\% \\
$\mu_{(Z)h\to \mu^+\mu^-}$   & 1.004 & 0.999 &1.015  & 17.04\%\\
$\mu_{(\nu\bar{\nu})h\to \mu^+\mu^-}$   & 0.997 & 1.010 & 0.977  & 2.87\%\\
\bottomrule
\end{tabular}
\caption{SM-like Higgs signal strengths and normalized cross sections of the best-fit points for the present likelihood in the CMSSM,
for each of the regions characterised by a specific mechanism for suppressing the relic density of dark matter. We also give the total uncertainty used in $\mathcal{L}_{\rm CEPC}$ in the last column, assuming $k=0.2$. }\label{tab:cmssm_higgs}
\end{table}

\begin{figure}[!t]
 \centering
 \includegraphics[width=0.49\textwidth]{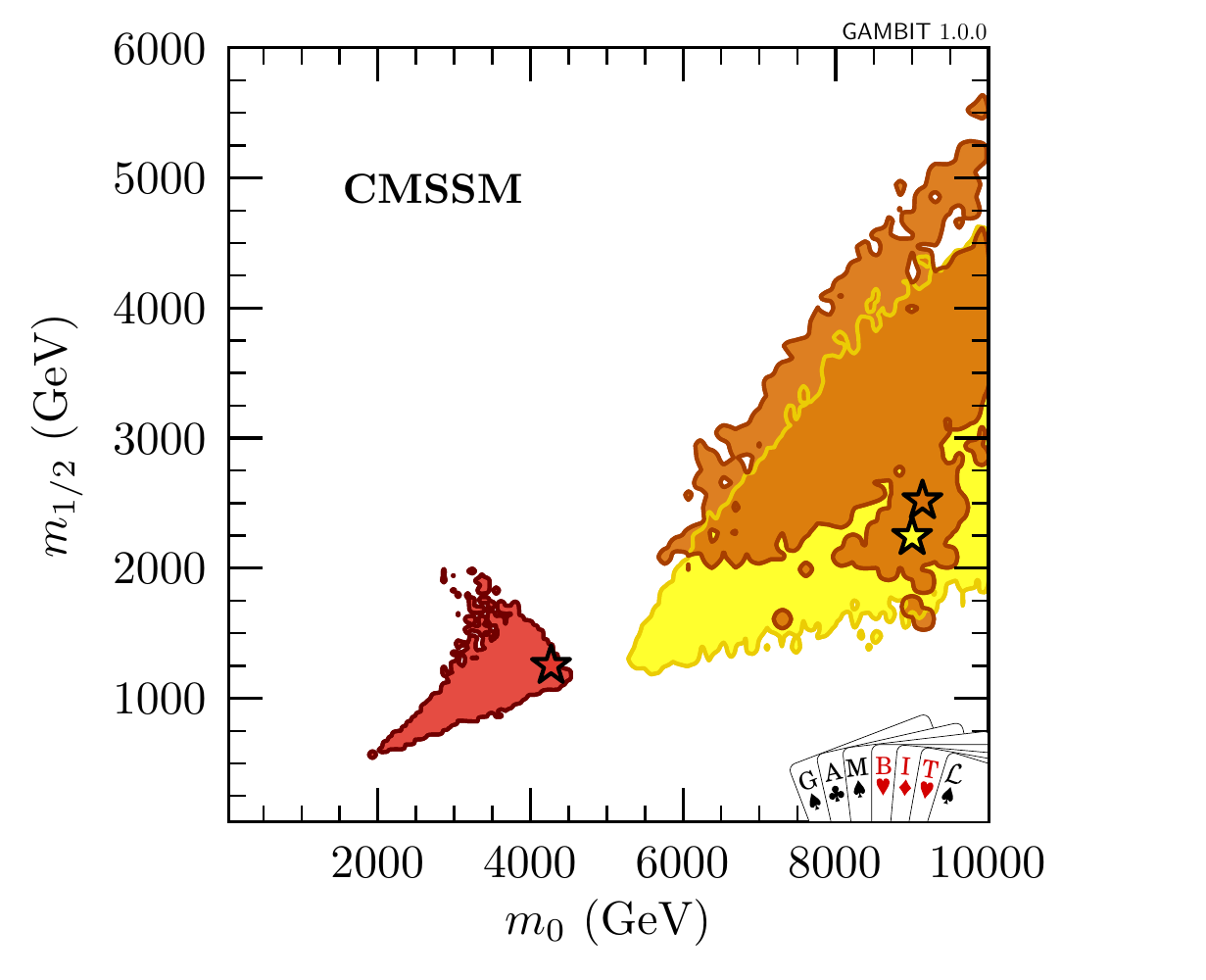}
 \includegraphics[width=0.49\textwidth]{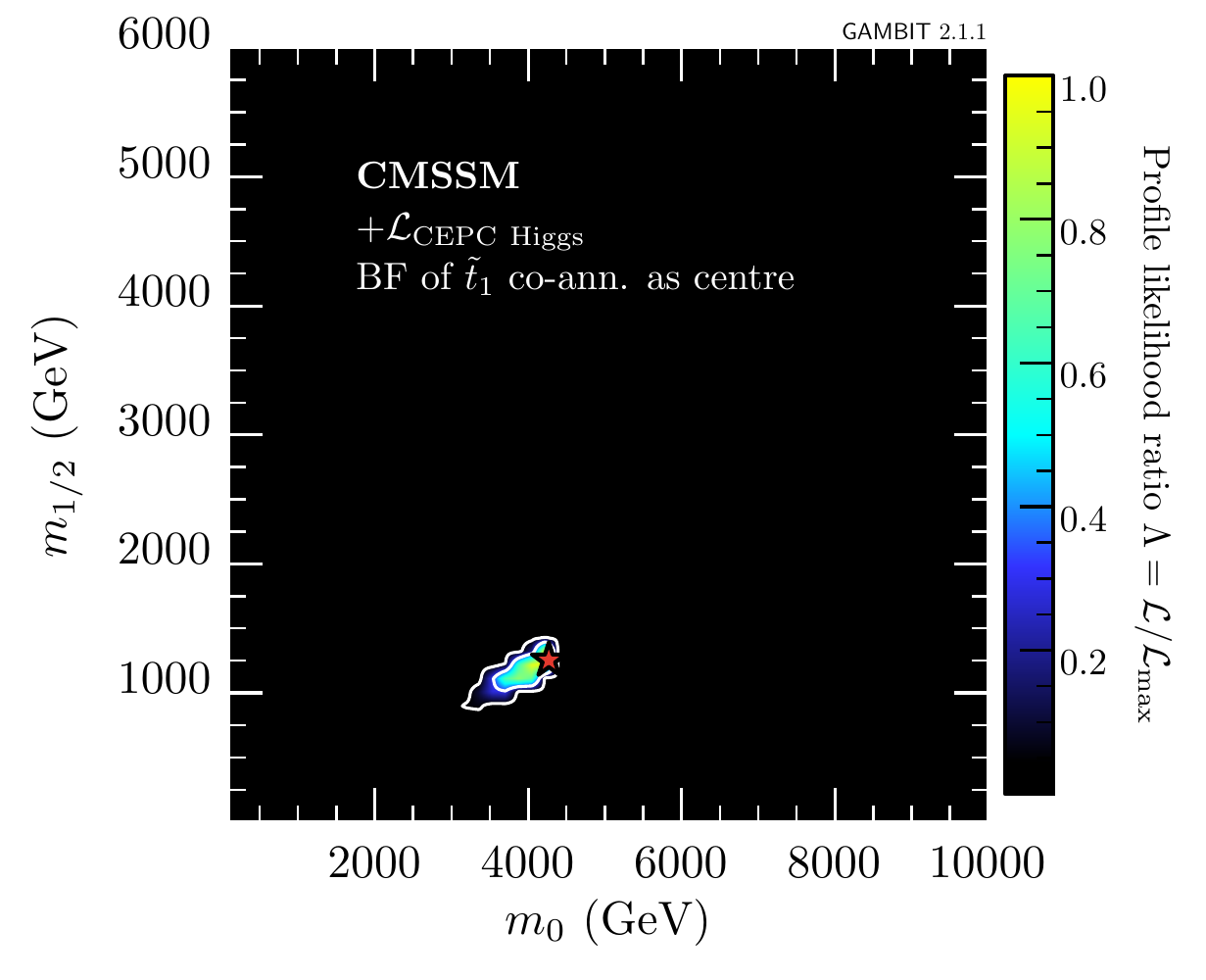}\\
  \includegraphics[width=0.49\textwidth]{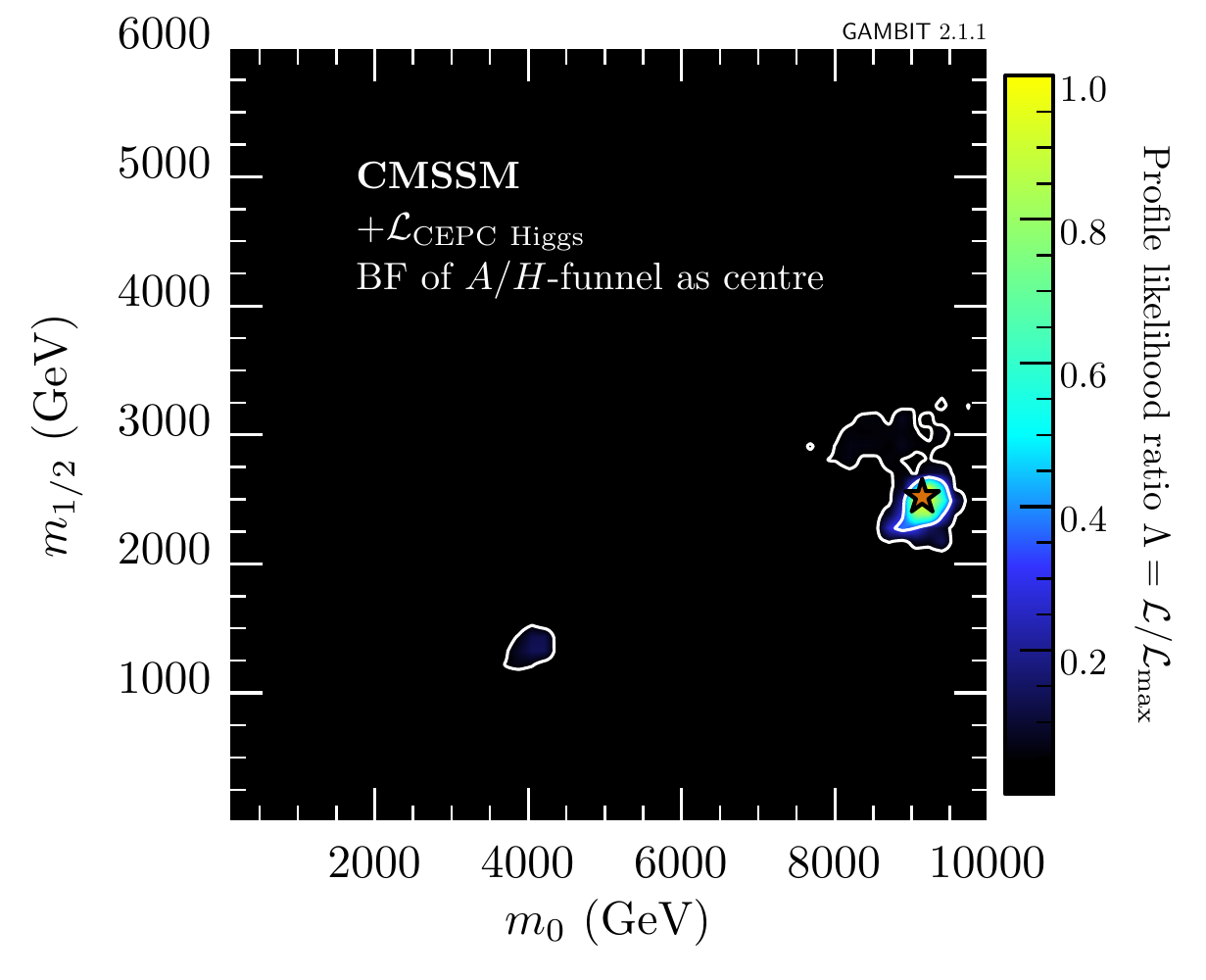}
 \includegraphics[width=0.49\textwidth]{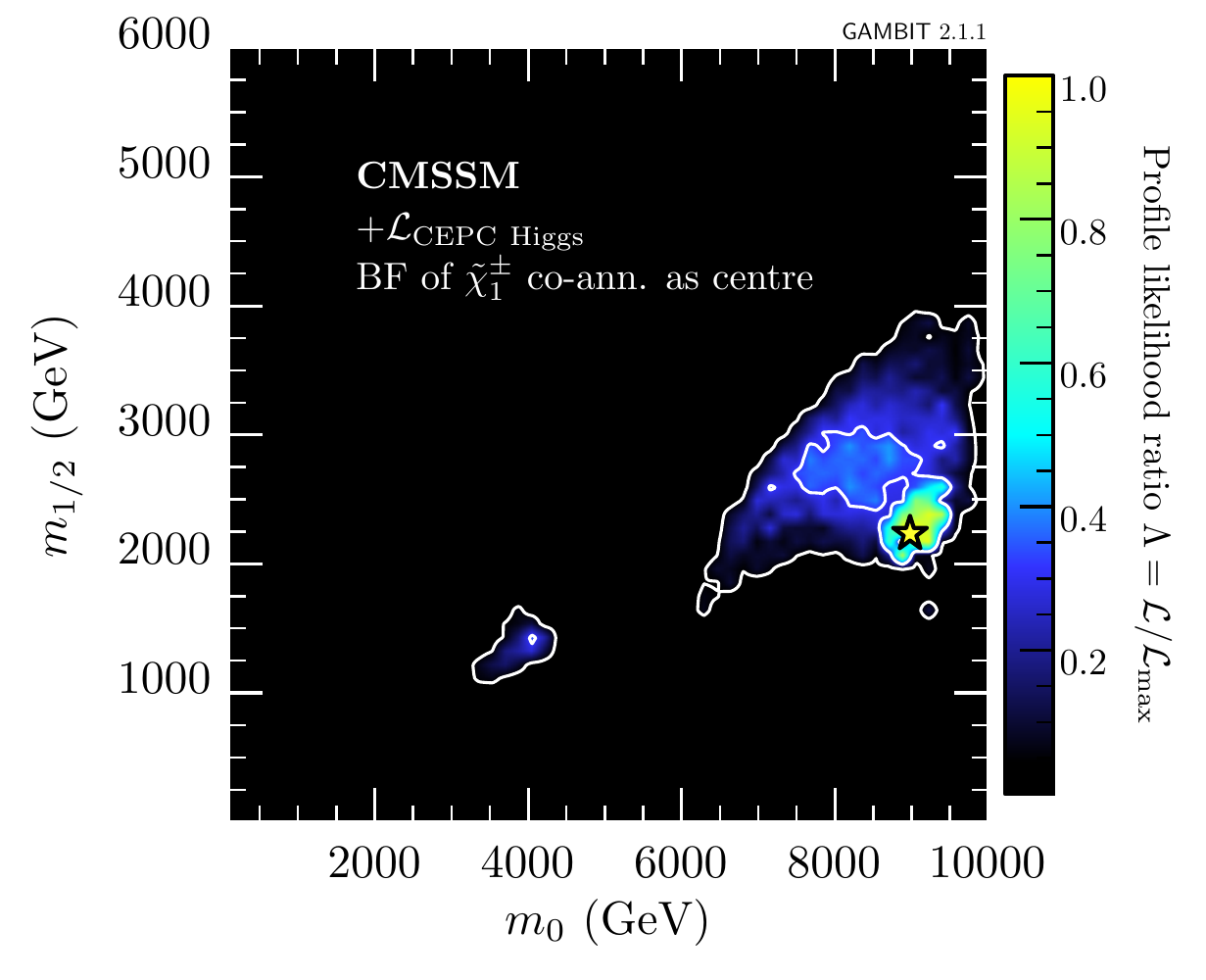}\\
 \includegraphics[height=4mm]{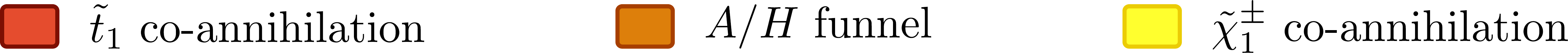}
\caption{Two-dimensional profile likelihoods for the CMSSM with an additional likelihood for CEPC, with different assumptions 
about the central values of Higgs measurements at the CEPC. The top left panel (taken from~\cite{Athron:2017qdc}) 
shows the mechanisms that ensure that the dark matter relic density does not exceed the measured value across the 2$\sigma$ 
contours of the present likelihoods, with the best-fit point in each region indicated by a star. The rest of the panels assume the central values of the Higgs measurements at the CEPC are the values of the best-fit point in the stop co-annihilation 
region (top right), in the $A/H$-funnel region (bottom left) and in the $\tilde{\chi}_1^{\pm}$ co-annihilation 
region (bottom right). }
 \label{fig:impact_of_bf}
\end{figure}

In order to further understand the significant impact of the CEPC, as shown in Figure~\ref{fig:2d_parameter_plots_cmssm},
we display the SM-like Higgs decay branching ratios predicted by the best-fit points in each of the regions in 
Table~\ref{tab:cmssm_higgs}. The combined experimental and theoretical uncertainty is also shown. 
The parameter values, mass spectra, and present likelihood contributions of these best-fit points can be found 
in~\cite{Athron:2017qdc}. We see that the differences in BR($h\to b \bar{b}$), BR($h\to WW^*$) and BR($h\to ZZ^*$) 
between the best-fit points of the stop co-annihilation region and the $A/H$-funnel region or the $\tilde{\chi}_1^{\pm}$ 
co-annihilation regions are significantly larger than the corresponding total uncertainties. This is the reason why the $A/H$-funnel region and the $\tilde{\chi}_1^{\pm}$ co-annihilation region are excluded when we assume that the CEPC measures exactly the central values predicted by the best-fit point in the stop co-annihilation region.

It is obvious that the results shown in Figure~\ref{fig:2d_parameter_plots_cmssm} and the above conclusions depend on
assumptions about central values of the Higgs measurements at the CEPC. Therefore, in Figure~\ref{fig:impact_of_bf} we display the 
2D profile likelihoods assuming the central values of the Higgs measurements at the CEPC to be the values of the best-fit point in each DM annihilation region, not just the overall best-fit point. Whereas the $A/H$-funnel region predicts SM-like Higgs couplings, the $\tilde{\chi}_1^{\pm}$ and stop
co-annihilation predictions for the CEPC are expected to be in about $1\sigma$ and $5\sigma$ tension with the SM, respectively.

It can be seen that the favored regions change dramatically with the central values changing from the values of one best-fit point to another. As expected, the center of the favored regions is on the chosen best-fit point. In the bottom left panel of Figure~\ref{fig:impact_of_bf}, 
where the central values are the same as the best-fit point in the $A/H$-funnel region, the favored regions are not as narrow as before. There is a small stop co-annihilation region inside the 95\% CL region. The samples in the 95\% CL region of $m_0 < 8.5\,\TeV$ and $m_{1/2}>2.6\,\TeV$  mainly satisfy the $\tilde{\chi}_1^{\pm}$ co-annihilation condition, while the 68\% CL region is almost pure $A/H$-funnel region. In the whole region, the $\mu$ parameter is always negative. The confidence regions found when setting the central values of the CEPC measurements to those predicted by the best-fit point in 
the $\tilde{\chi}_1^{\pm}$ co-annihilation region, shown in the bottom right panel of 
Figure~\ref{fig:impact_of_bf}, are even wider. All three DM annihilation regions exist in 
the 68\% CL region. However, there are no samples with $\mu>0$ in the 95\% CL region.

The sfermions in the $A/H$-funnel and the $\tilde{\chi}_1^{\pm}$ co-annihilation regions are heavier than 
about 5\,\TeV. For instance, the best-fit point in the $\tilde{\chi}_1^{\pm}$ co-annihilation region 
has $m_{\tilde{t}_1} \simeq m_{\tilde{\tau}_1} \simeq 6.5\,\TeV$ and $m_{\tilde{q}}\simeq 10\,\TeV$ for the first two generation sfermions. The supersymmetric contributions to the SM-like Higgs branching ratios tend to decouple
in the high mass region. Nevertheless, the SM-like Higgs branching ratios in these regions remain quite different from the SM values listed in Table~\ref{tab:sm_higgs}. Various uncertainties and factors are involved in this discrepancy, such as values of SM parameters used in the calculations. Therefore, it further inspires us not to set the central 
values of the Higgs measurements at future Higgs factories to the SM values listed in Table~\ref{tab:sm_higgs}. 
To some extent, the SM-like Higgs branching ratios of a point in the high mass region can be treated as the SM values, such as the best-fit point of the $A/H$-funnel region or the $\tilde{\chi}_1^{\pm}$ co-annihilation region.

\begin{figure}
 \centering
 \includegraphics[width=0.49\textwidth]{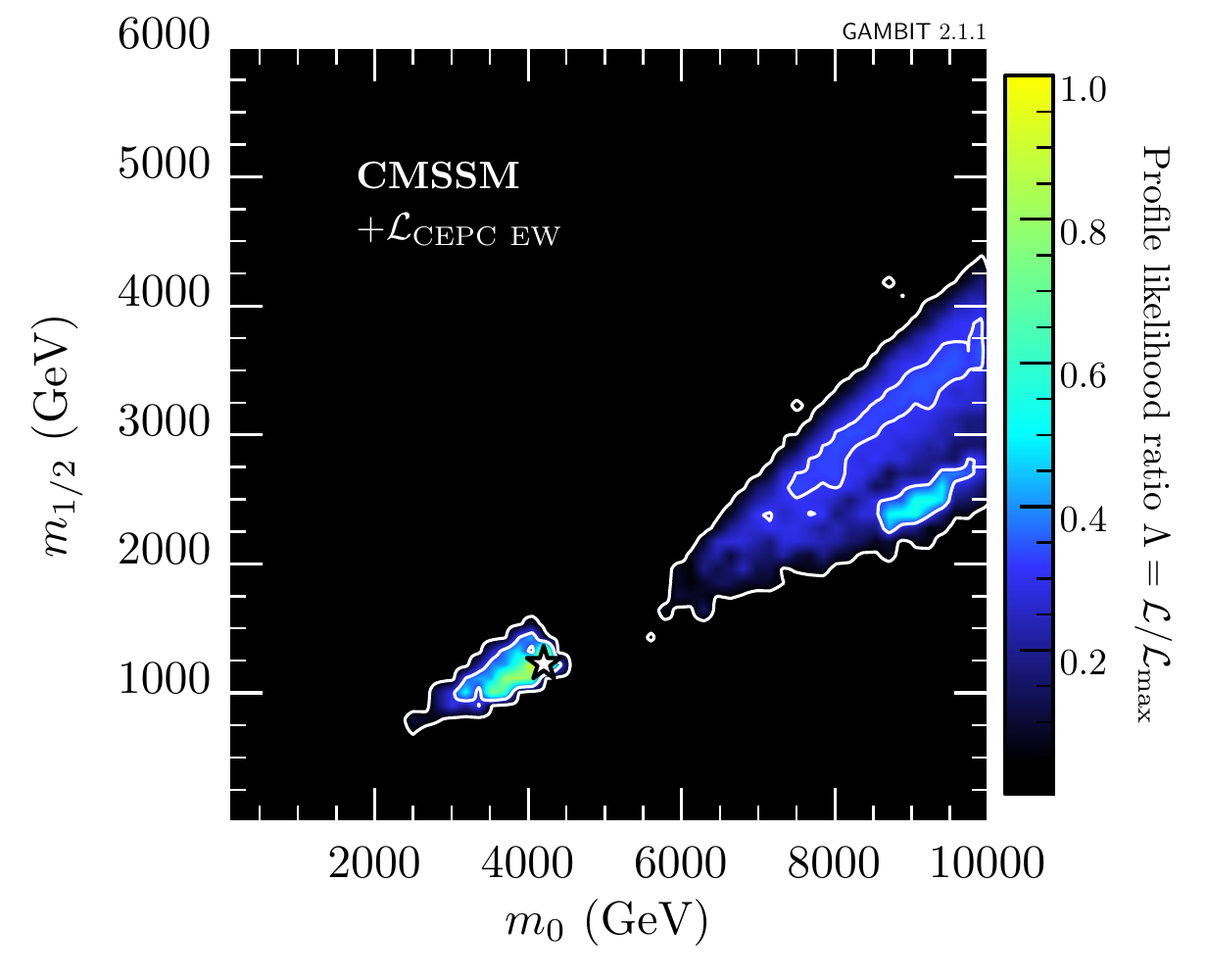}
 \includegraphics[width=0.49\textwidth]{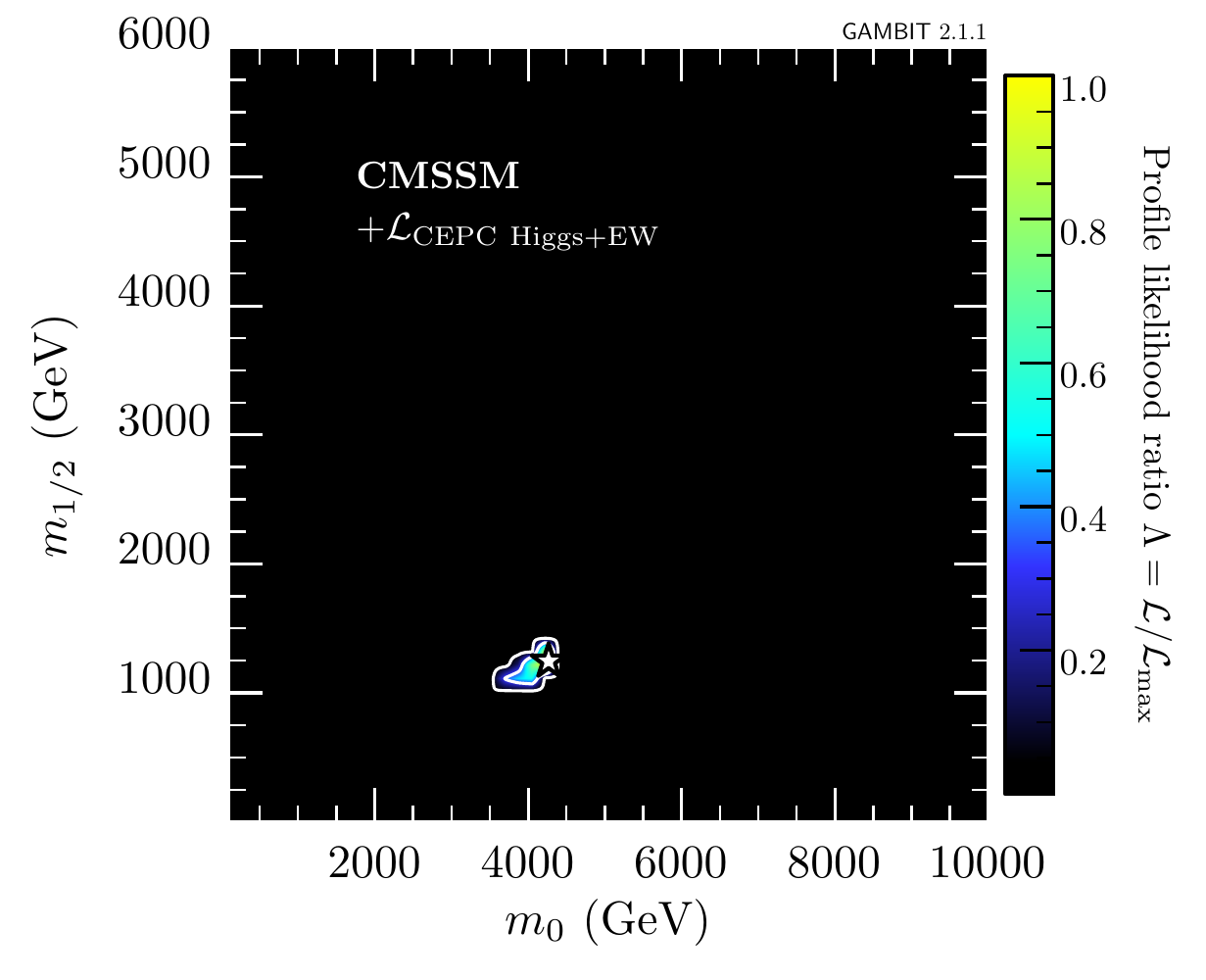}\\
 \caption{2D profile likelihoods for the CMSSM, plotted in the  $m_0-m_{1/2}$ plane, adding only likelihood of EW measurements at CEPC (left) and adding likelihoods of EW measurements and Higgs measurements at CEPC (right). The assumptions about CEPC likelihood for Higgs measurements are as Figure~\ref{fig:2d_parameter_plots_cmssm}. }
 \label{fig:ew}
\end{figure}

We also show the 2D profile likelihood for EW measurements at the CEPC in Figure~\ref{fig:ew}, where the left panel implements the EW measurements on their own and the right panel combines likelihoods for the EW and Higgs measurements. The EW measurements visibly narrow the favored regions, though not to the same degree as the Higgs measurements. The two kinds of measurements complement each other well, as combining them gives extremely strong constraints on the parameter space of the CMSSM, shown in the right panel of Figure~\ref{fig:ew}. The BF point's predictions for EW observables are expected to be in about $3\sigma$ tension with the SM.

%%%%%%%%%%%
\subsection{NUHM1 and NUHM2}
\label{sec:nuhm}

\begin{figure}[!t]
  \centering
  \includegraphics[width=0.32\textwidth]{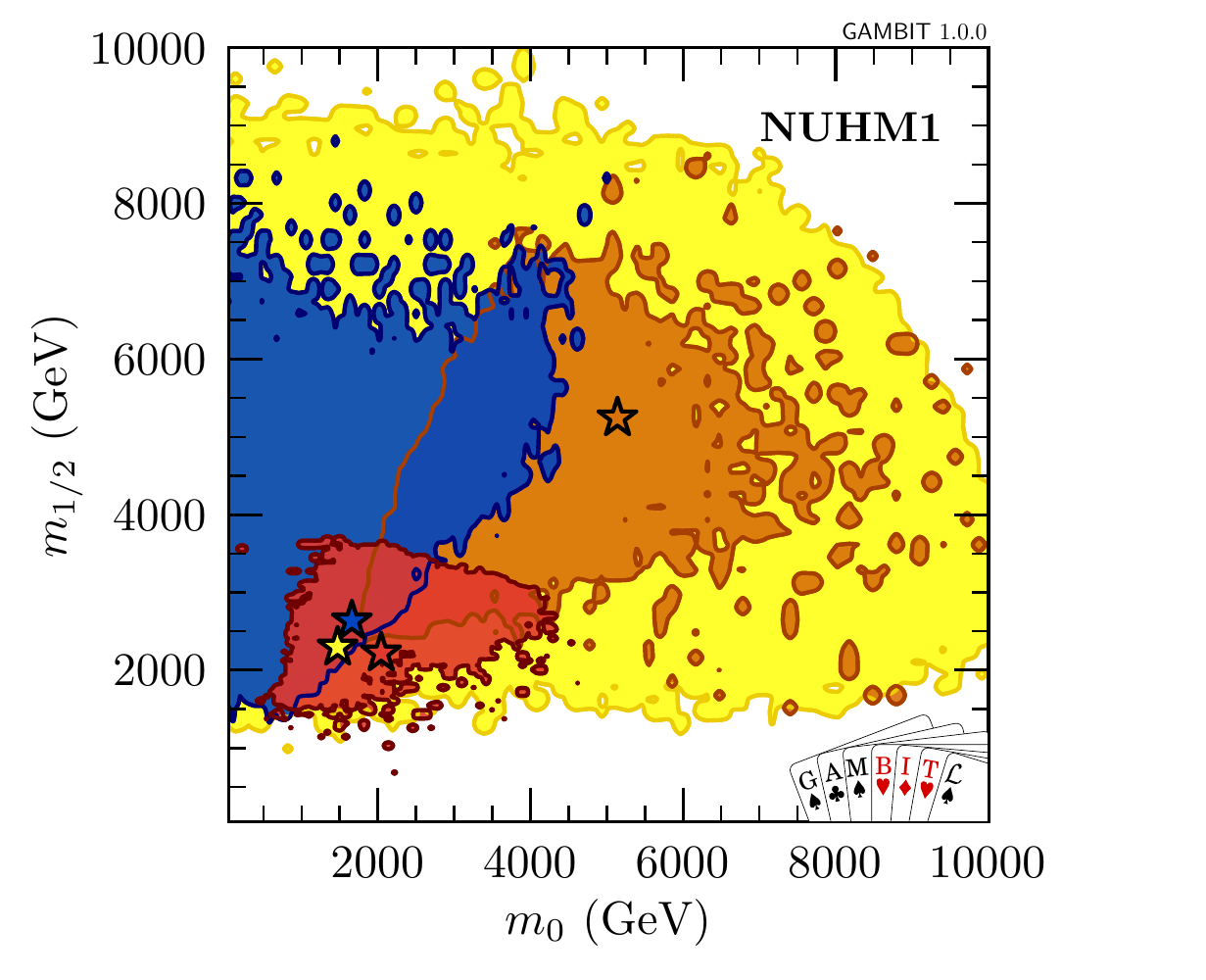}
  \includegraphics[width=0.32\textwidth]{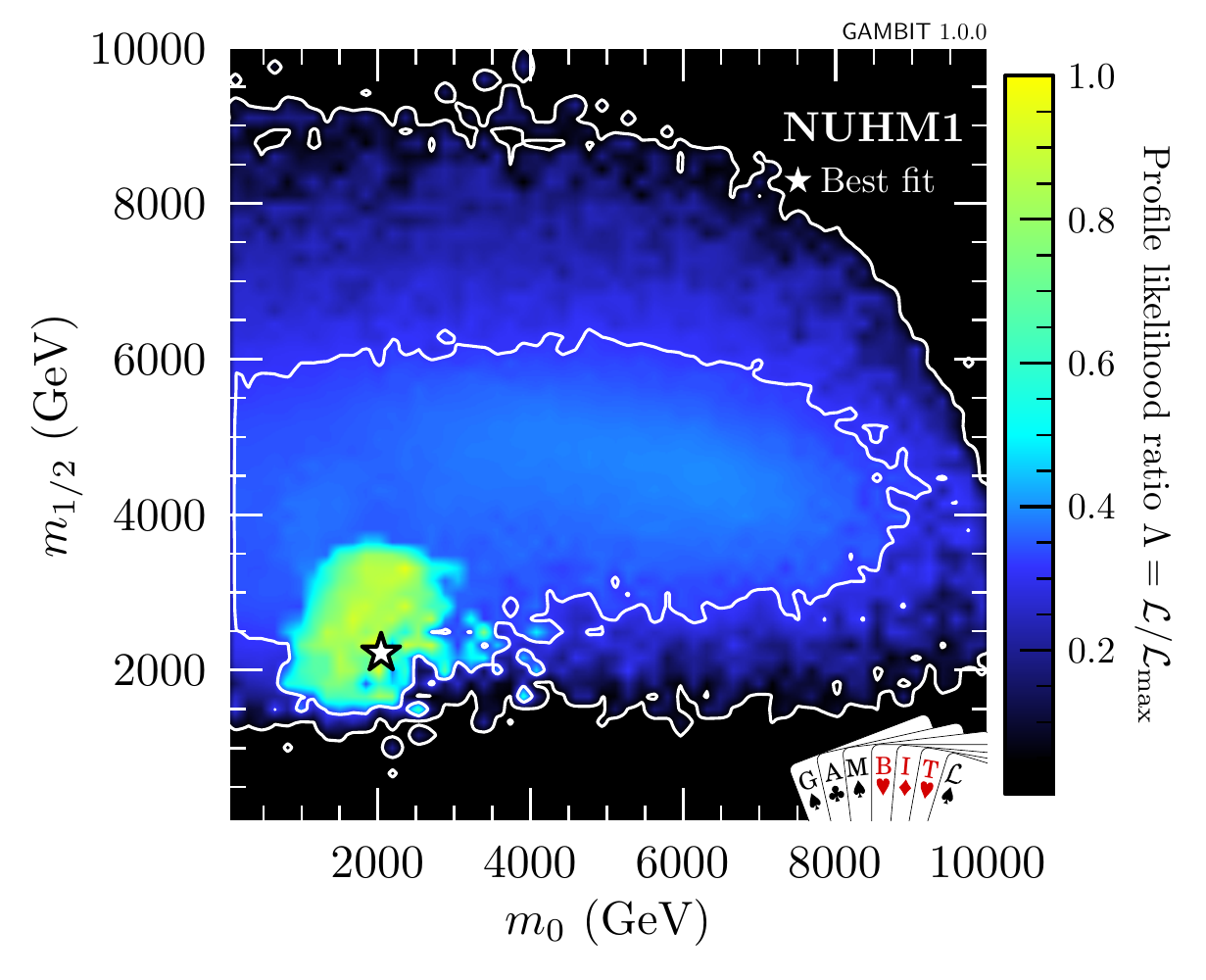}
  \includegraphics[width=0.32\textwidth]{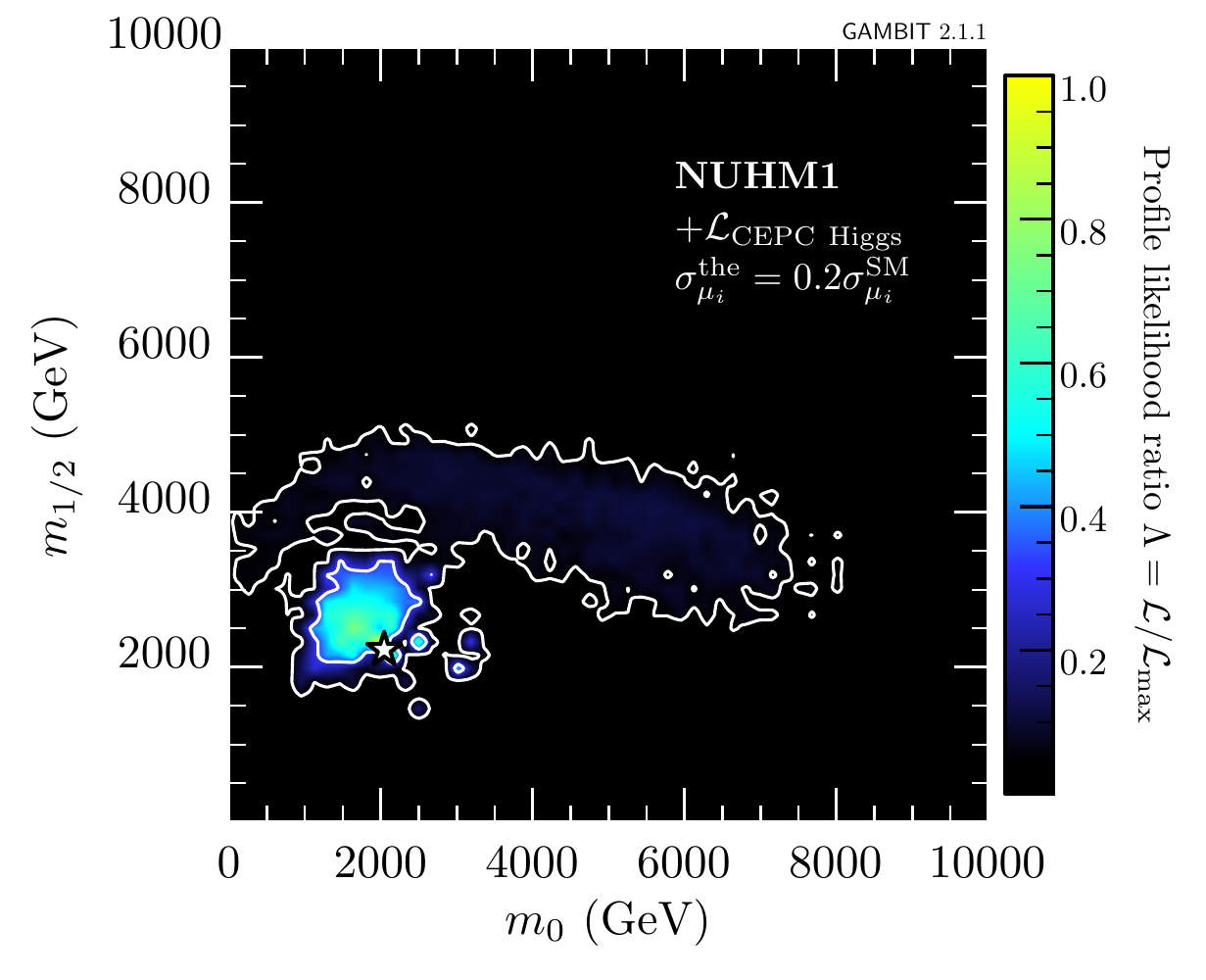}
  \\
  \includegraphics[width=0.32\textwidth]{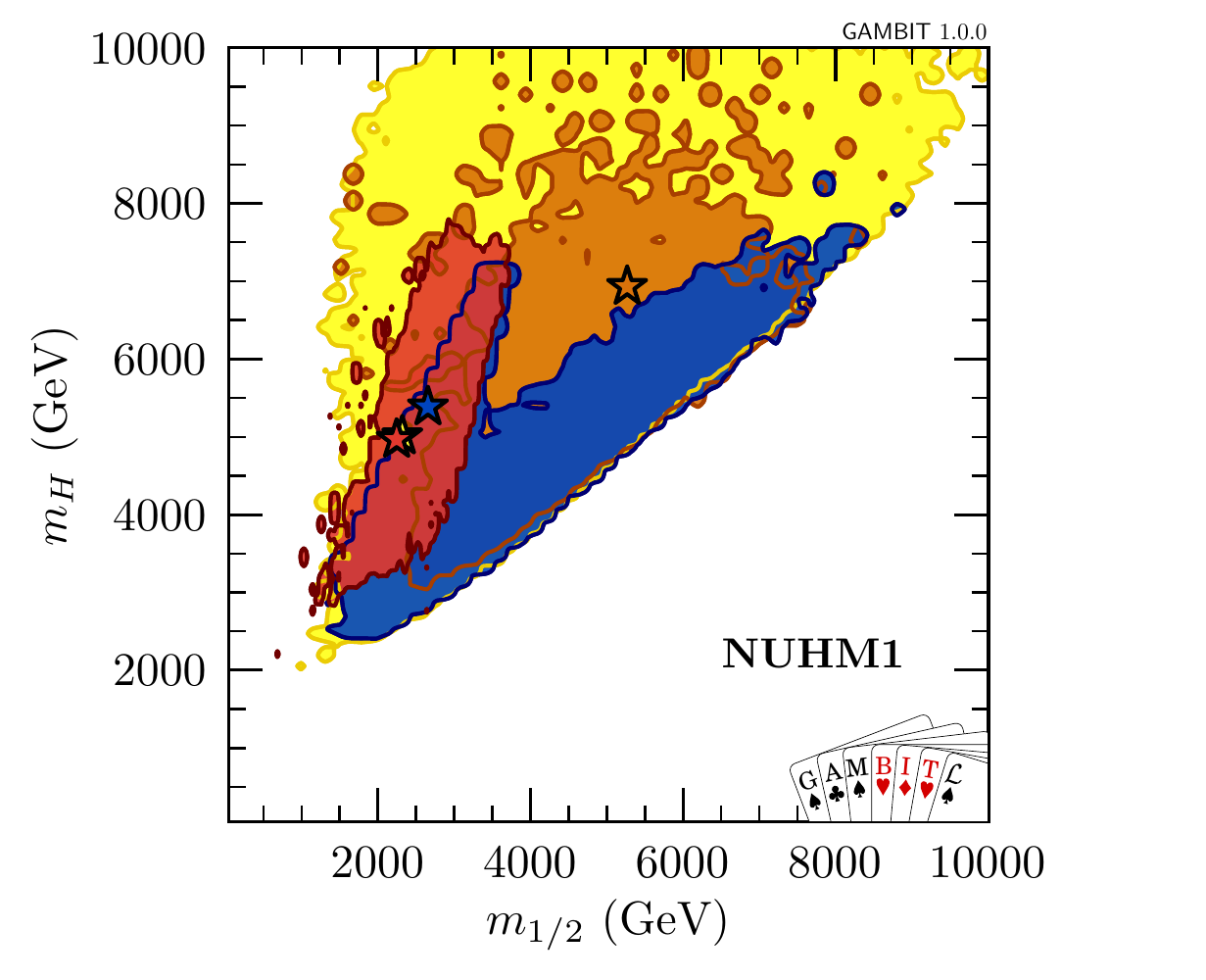}
  \includegraphics[width=0.32\textwidth]{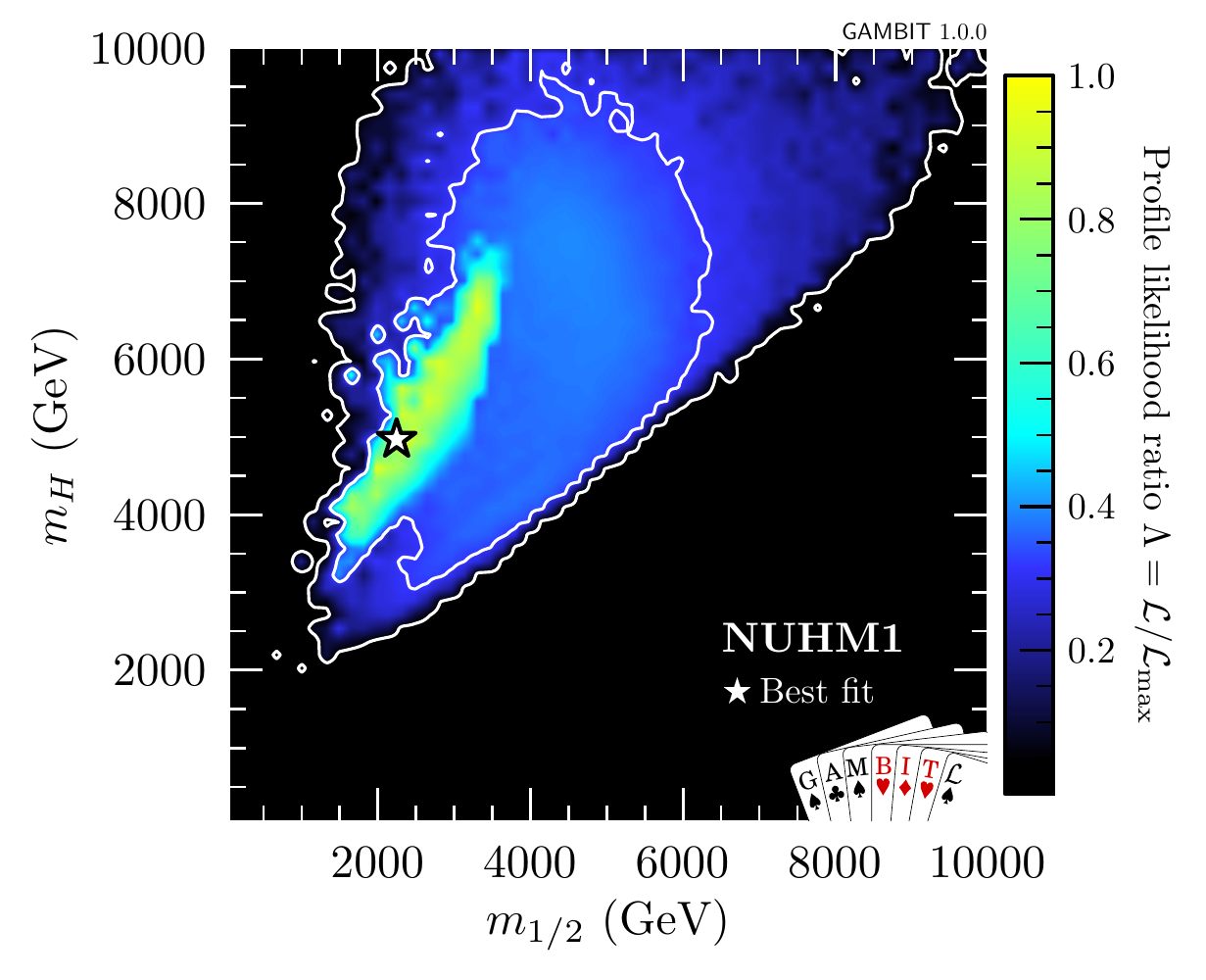}
  \includegraphics[width=0.32\textwidth]{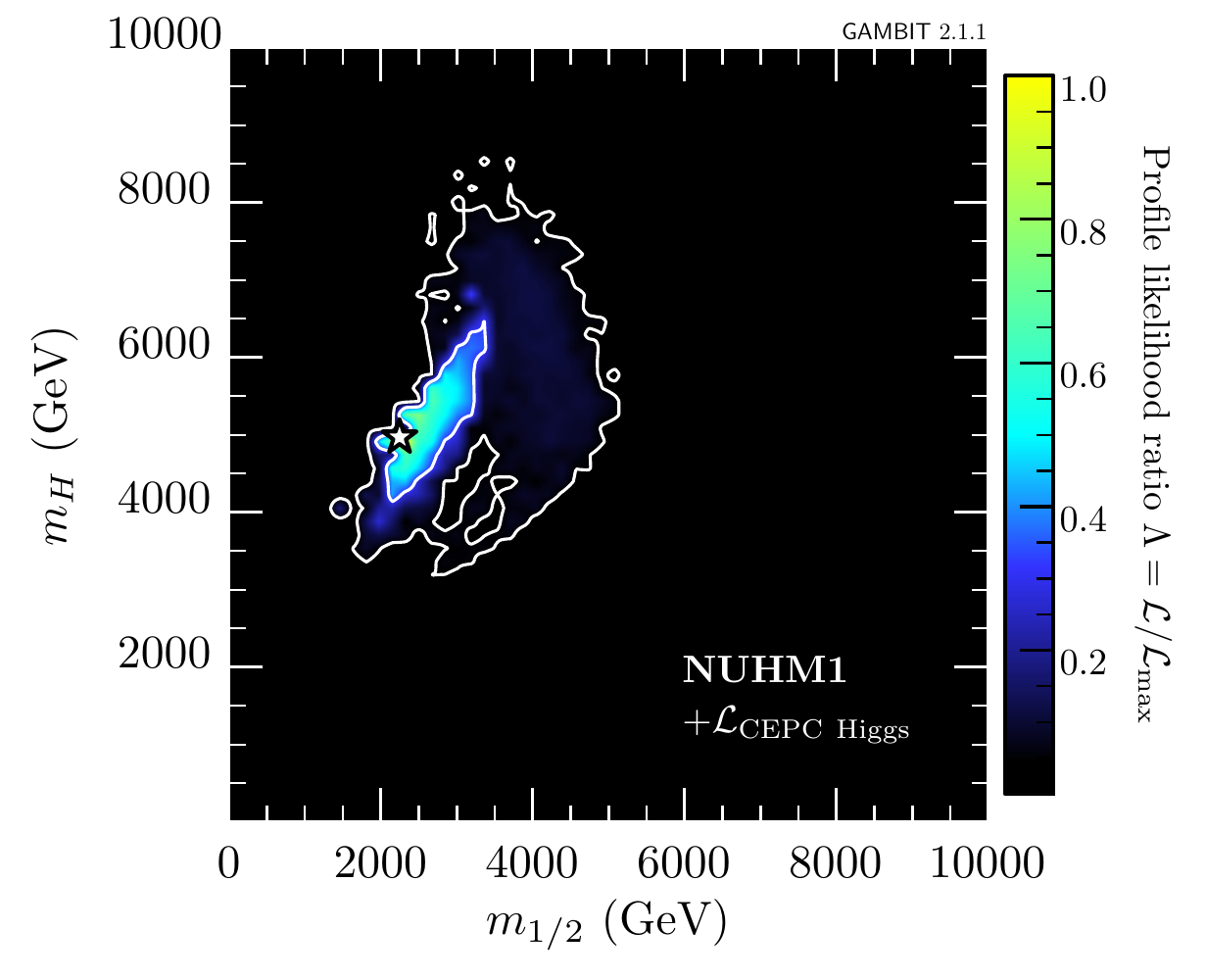}
  \\
  \includegraphics[width=0.32\textwidth]{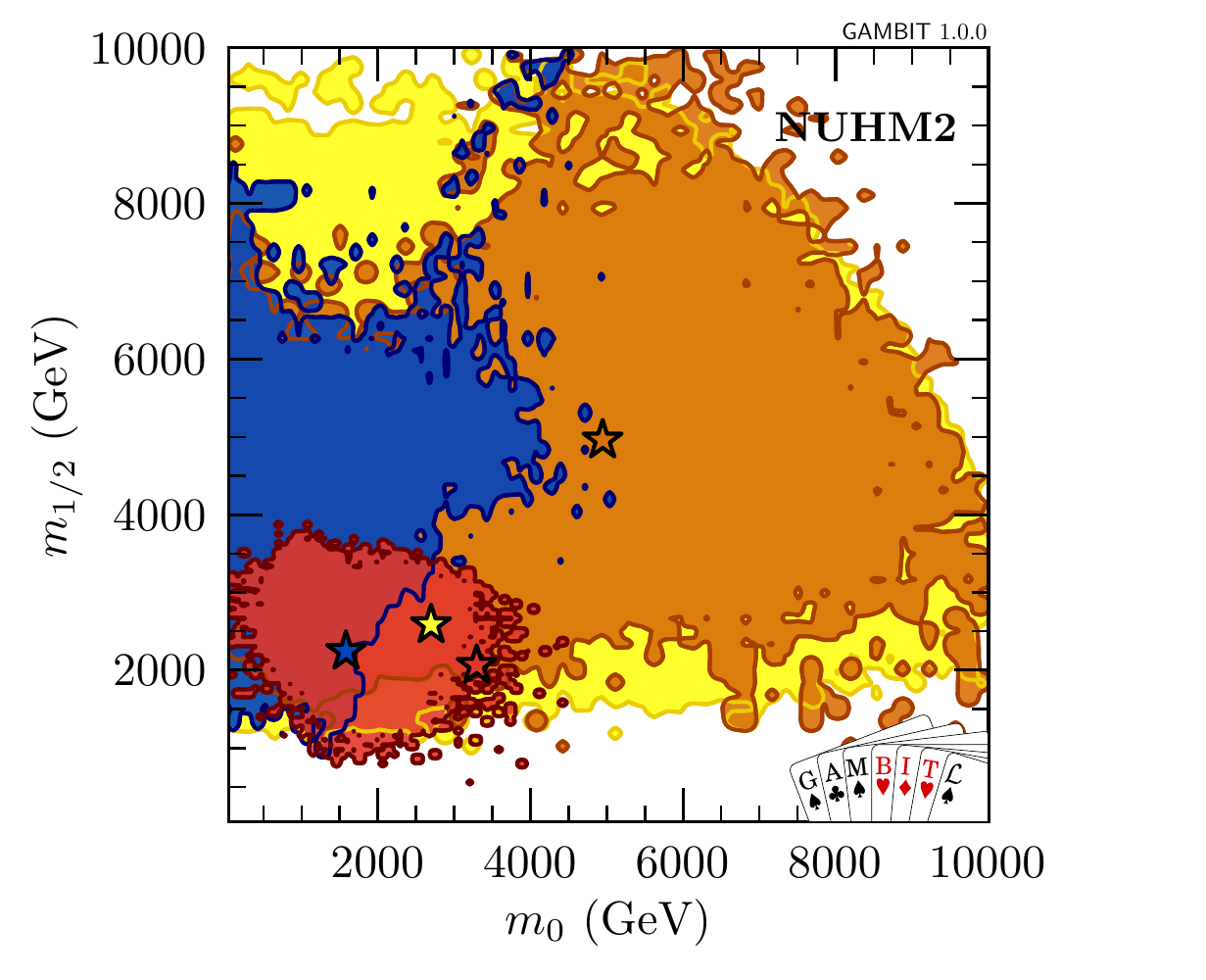}
  \includegraphics[width=0.32\textwidth]{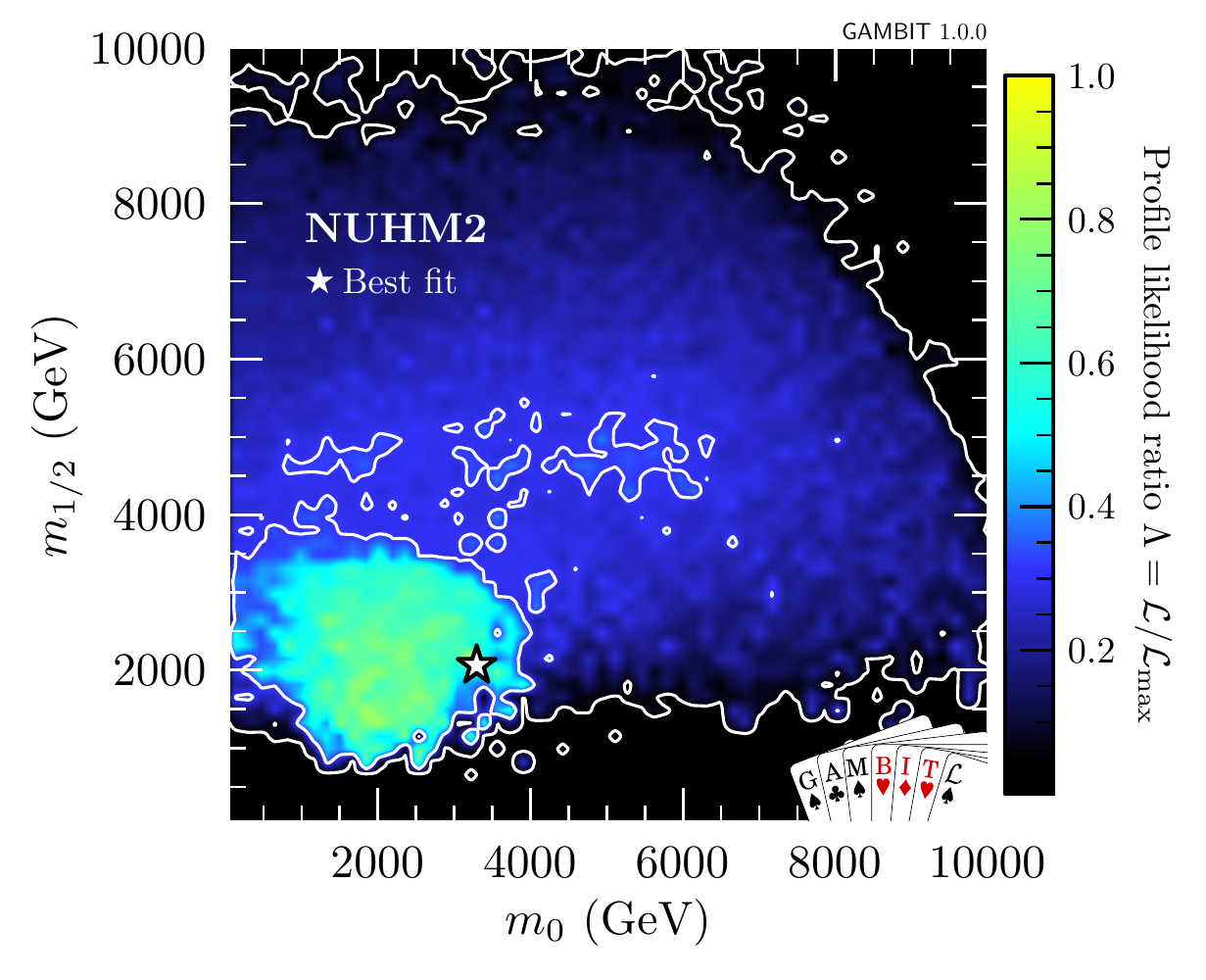}
  \includegraphics[width=0.32\textwidth]{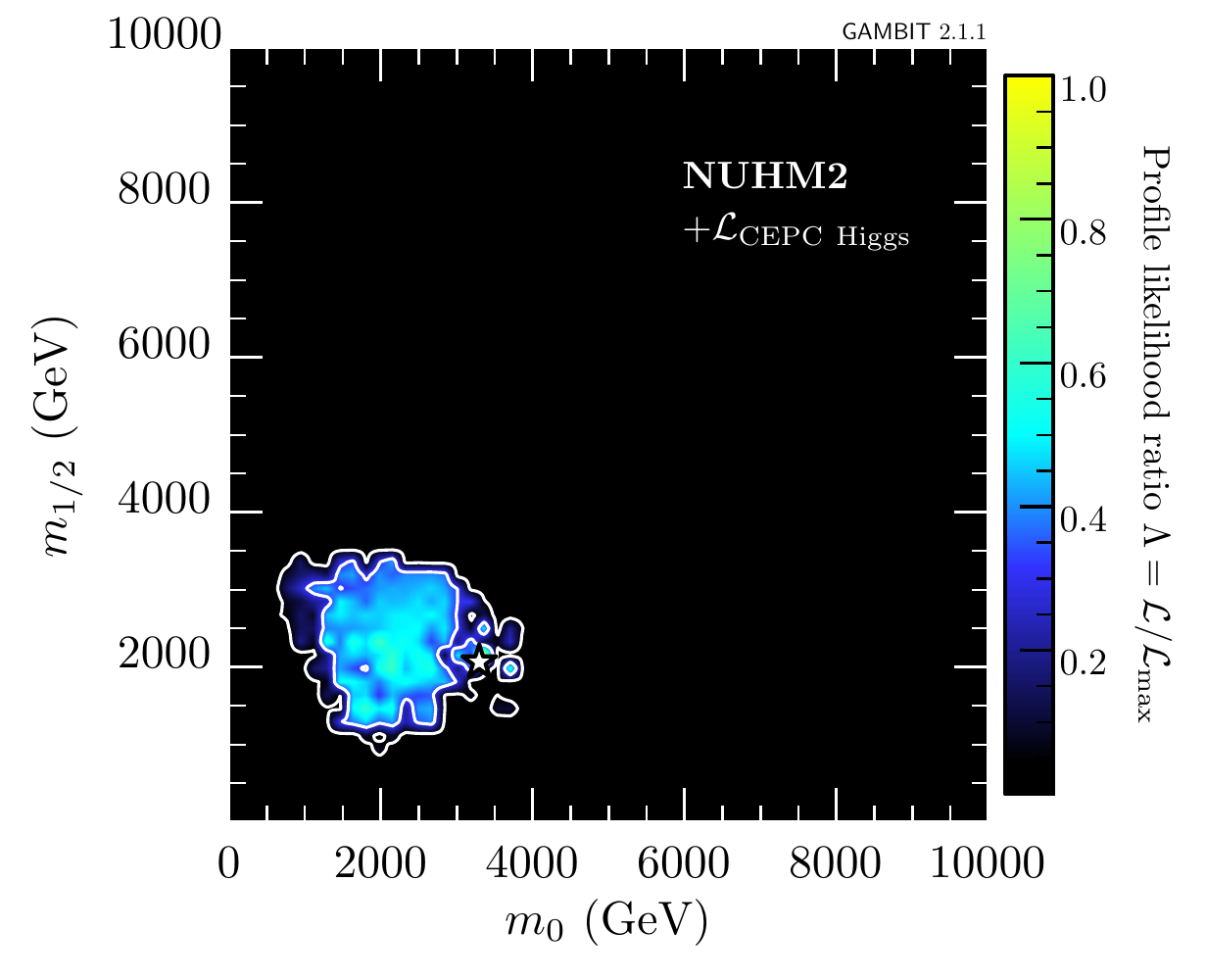}
  \\
  \includegraphics[width=0.32\textwidth]{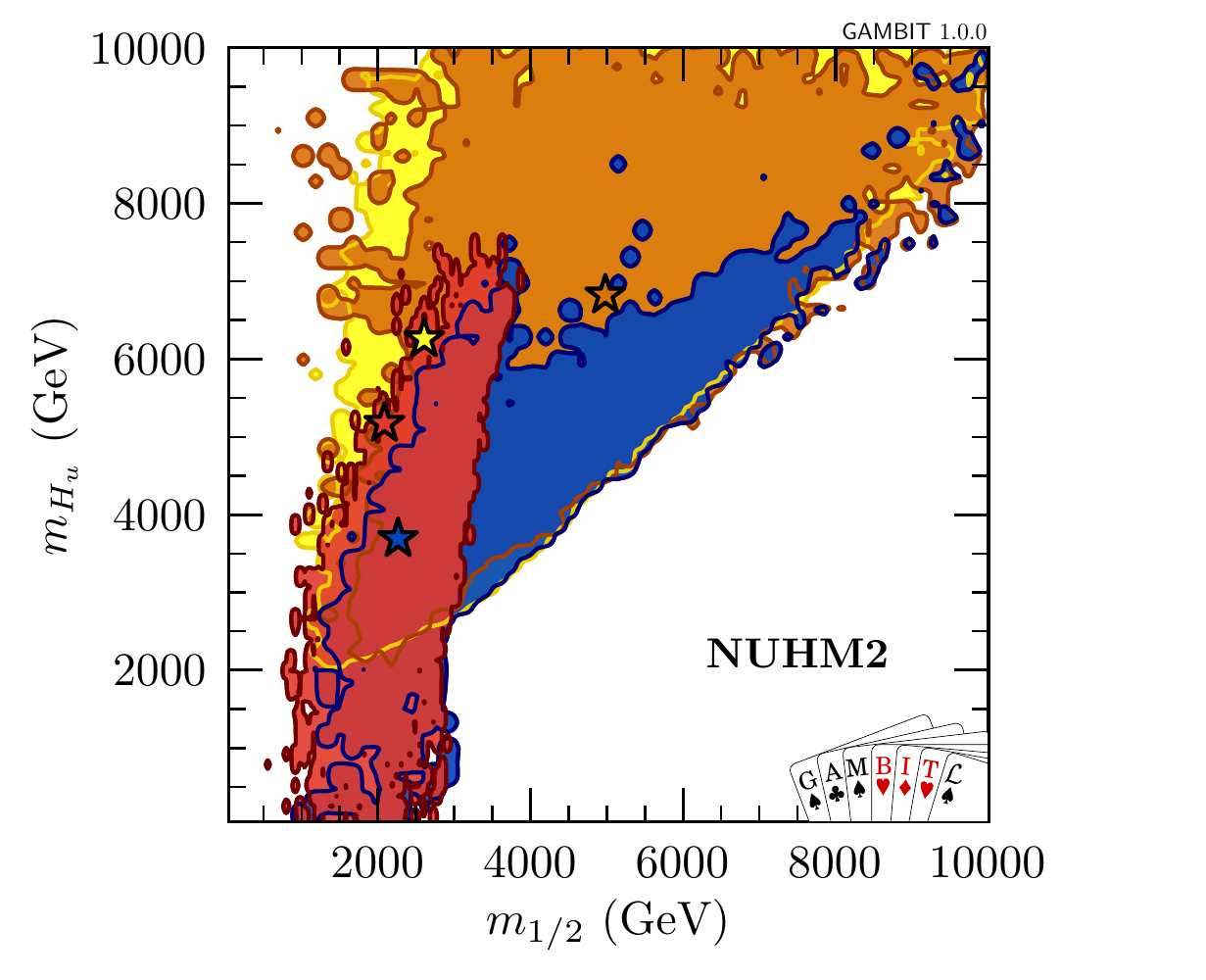}
  \includegraphics[width=0.32\textwidth]{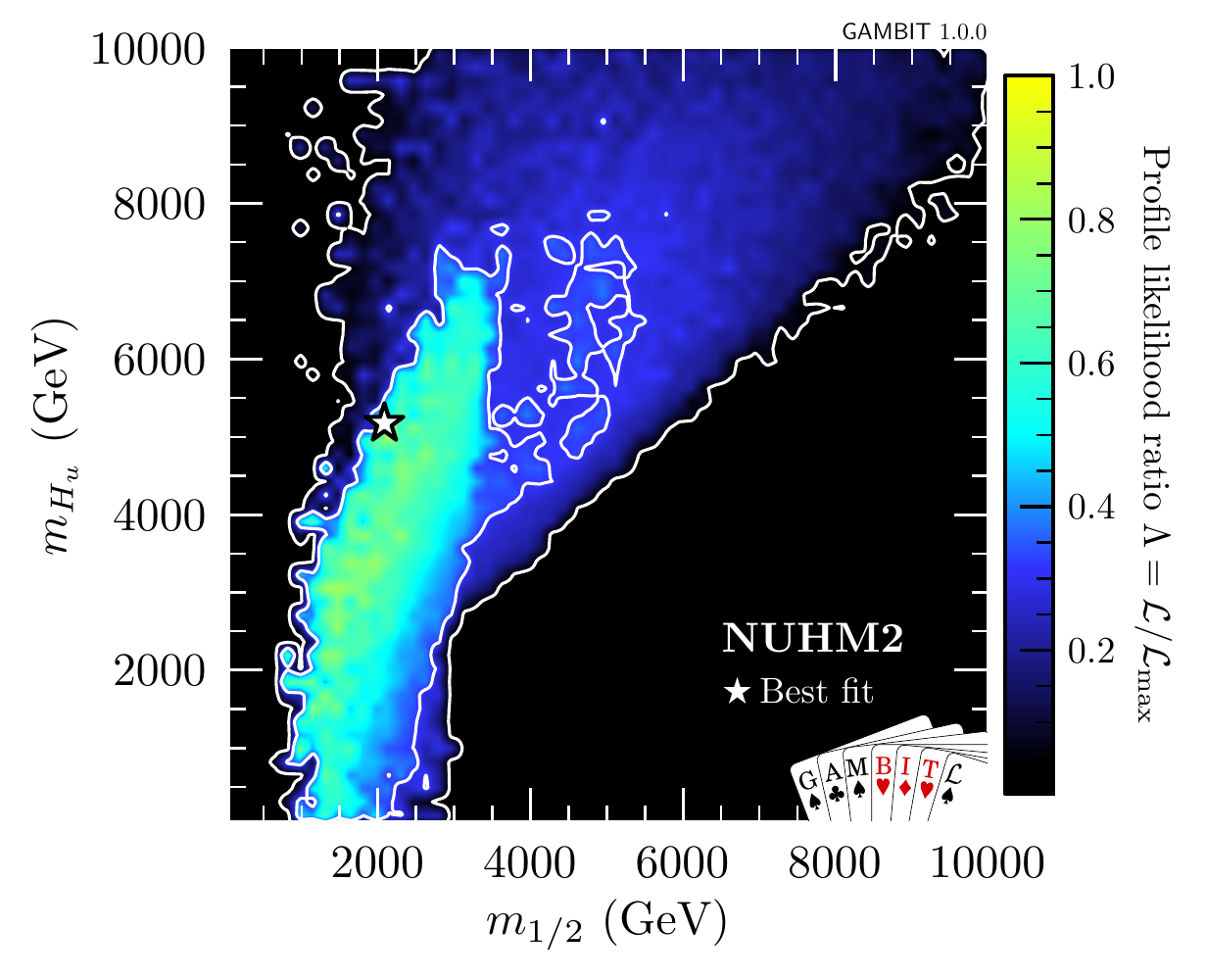}
  \includegraphics[width=0.32\textwidth]{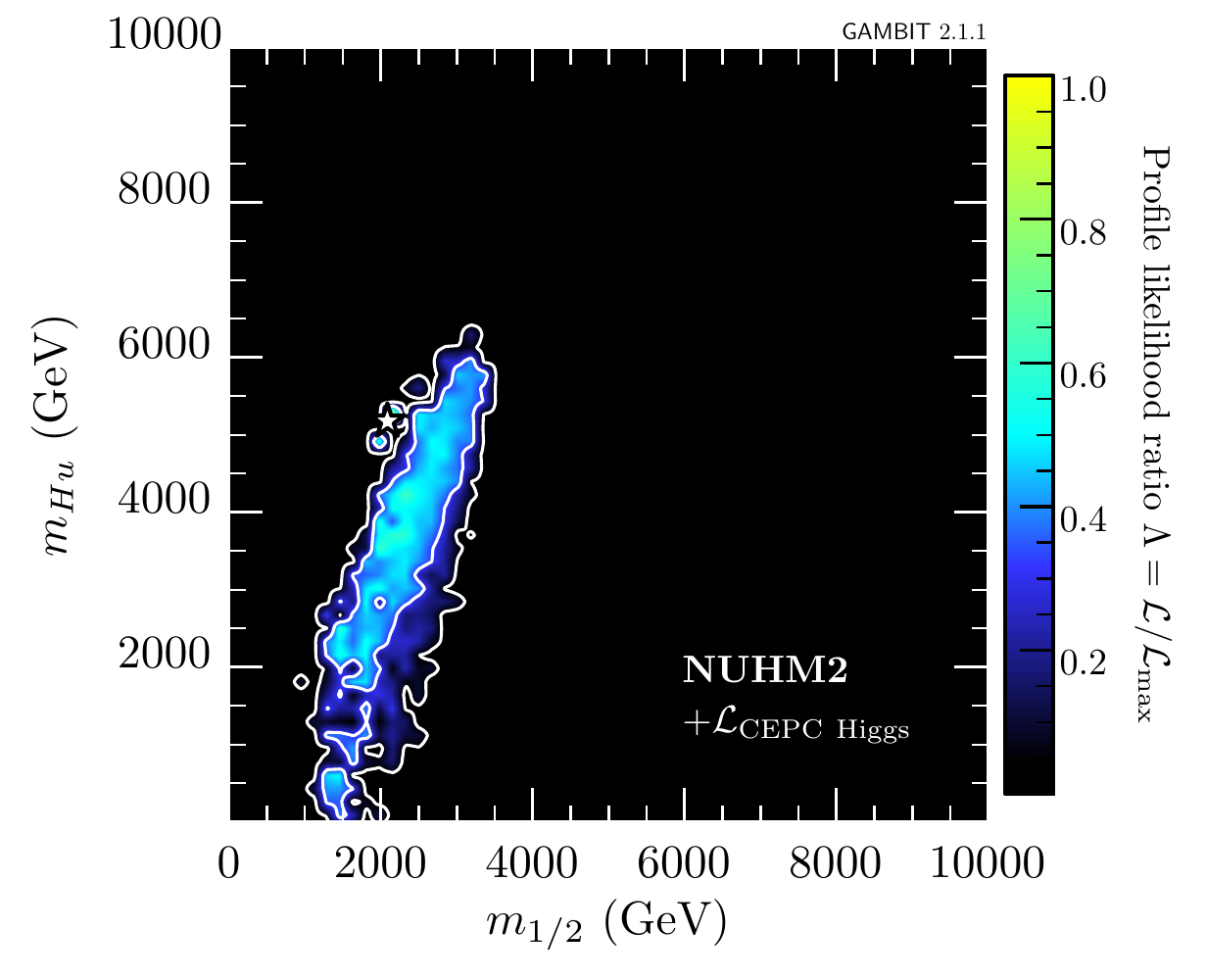}
  \\
  \includegraphics[height=4mm]{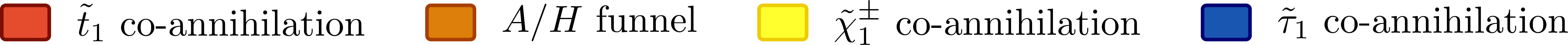}
  \caption{Profile likelihood ratio, without the CEPC likelihood (middle panels, taken from~\cite{Athron:2017qdc}) 
and with the CEPC likelihood (right panels), for the NUHM1 (top two rows) and the NUHM2 (bottom two rows). 
Colour-coding in the left panels (taken from~\cite{Athron:2017qdc}) shows the mechanisms active in models 
within the 95\% CL contour for avoiding thermal overproduction of neutralino dark matter. 
The overall best-fit point is indicated by a white star, while the best-fit points in each region are indicated 
by colored stars. The assumptions about the CEPC likelihood are the same as those in Figure~\ref{fig:2d_parameter_plots_cmssm}.}
  \label{fig:2d_parameter_plots_nuhm}
\end{figure}

In  the right panels of Figure~\ref{fig:2d_parameter_plots_nuhm} we show the joint profile likelihood ratio 
including the proposed CEPC constraints for the input model parameters of the NUHM1 and NUHM2, accompanied by the profile 
likelihood ratio without $\mathcal{L}_{\rm CEPC}$ in the middle panels and the dark matter annihilation mechanisms 
in the left panels. The definitions of the dark matter annihilation regions are same as those in the above subsection, except for an additional category
\begin{itemize}
\item stau co-annihilation: $m_{\tilde{\tau}_1} \leq 1.2\,m_{\tilde\chi^0_1}$,
\end{itemize}
which is absent in the CMSSM results. With the extra freedom presented in the Higgs sector of the Non-Universal 
Higgs Mass models, the $\mu$ parameters decouple from $m_0$, leading to arbitrarily light Higgsinos. 
Thus, the chargino co-annihilation region expands significantly. Meanwhile, the best-fit points in both the 
NUHM1 and NUHM2 results are also located in the stop co-annihilation region and have slightly larger 
likelihoods than the best-fit point in the CMSSM. 

As in the CMSSM, with the addition of $\mathcal{L}_{\rm CEPC}$, the preferred regions shrink significantly towards 
the best-fit points in the NUHM1. However, the stop co-annihilation region here overlaps with all other three 
regions in all of the parameter planes. As a result, the remaining $2\sigma$ regions also contain some chargino 
co-annihilation region. Note that the $1\sigma$ region is a pure stop co-annihilation region. Samples of $\mu>0$ 
are all excluded. In comparison to the NUHM1 results, the NUHM2 results show larger $1\sigma$ regions but smaller $2\sigma$ 
regions, with no chargino co-annihilation region inside the $2\sigma$ region. 

\begin{figure}[t!]
  \centering
  \includegraphics[width=0.49\textwidth]{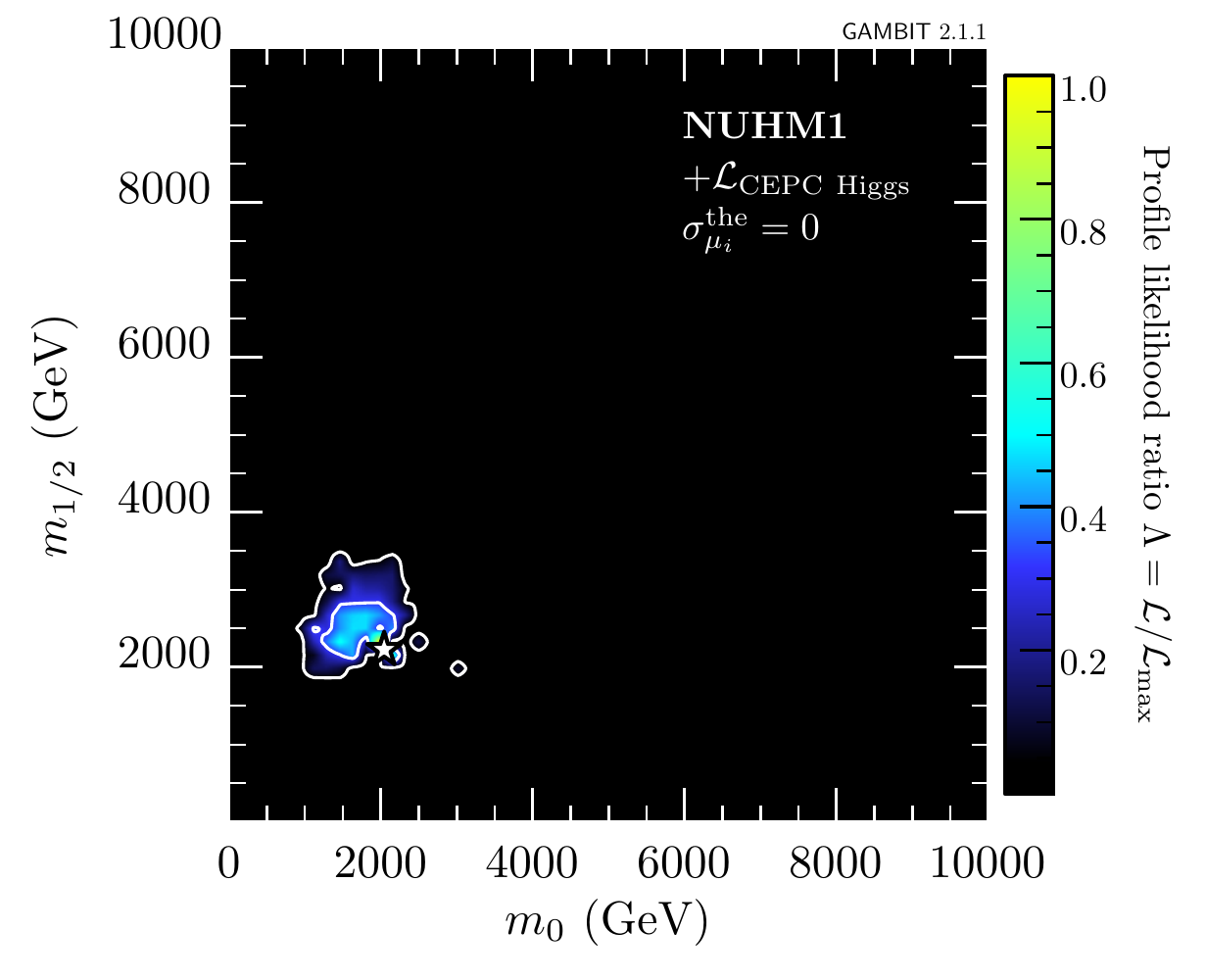}
  \includegraphics[width=0.49\textwidth]{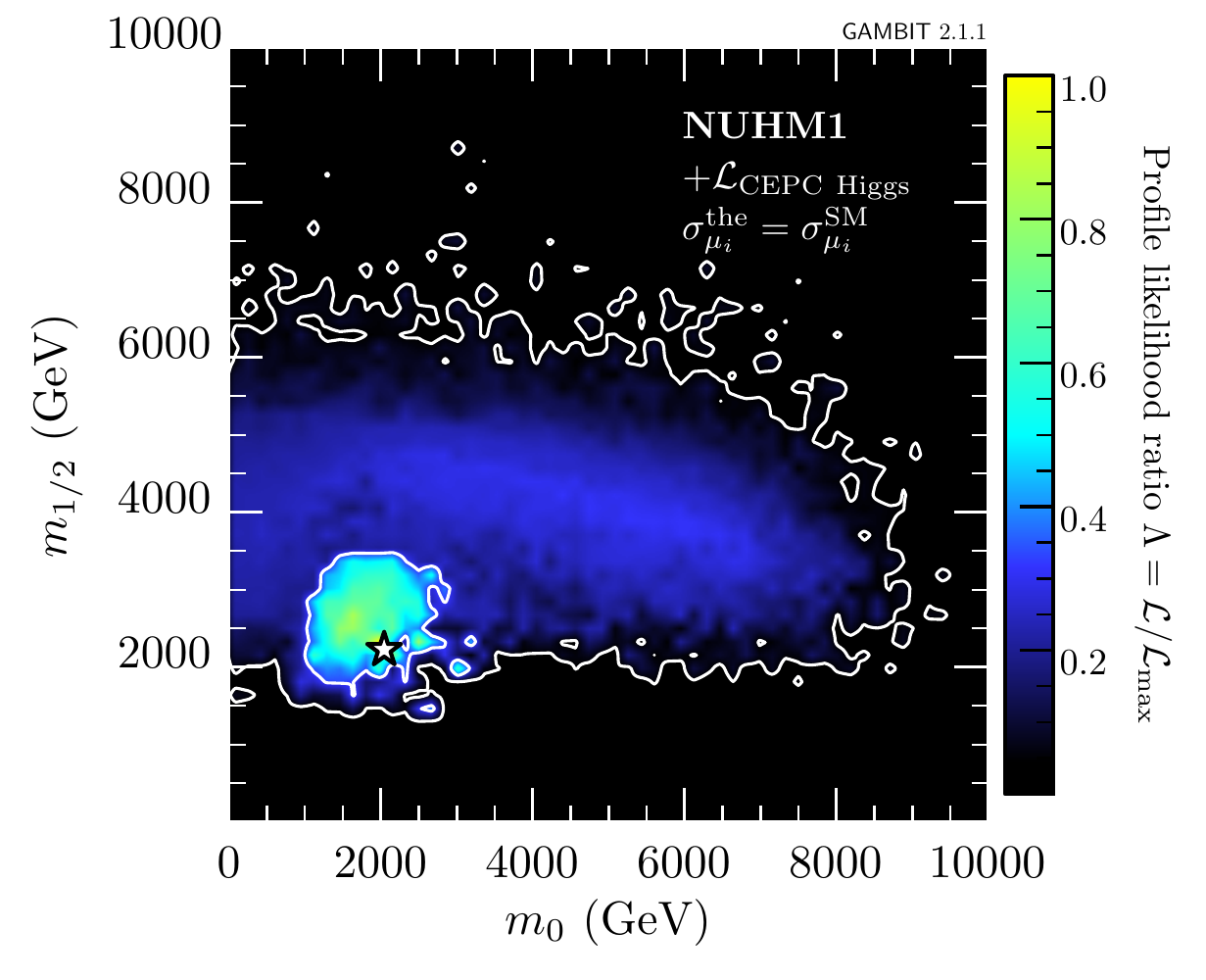}\\
  \includegraphics[width=0.49\textwidth]{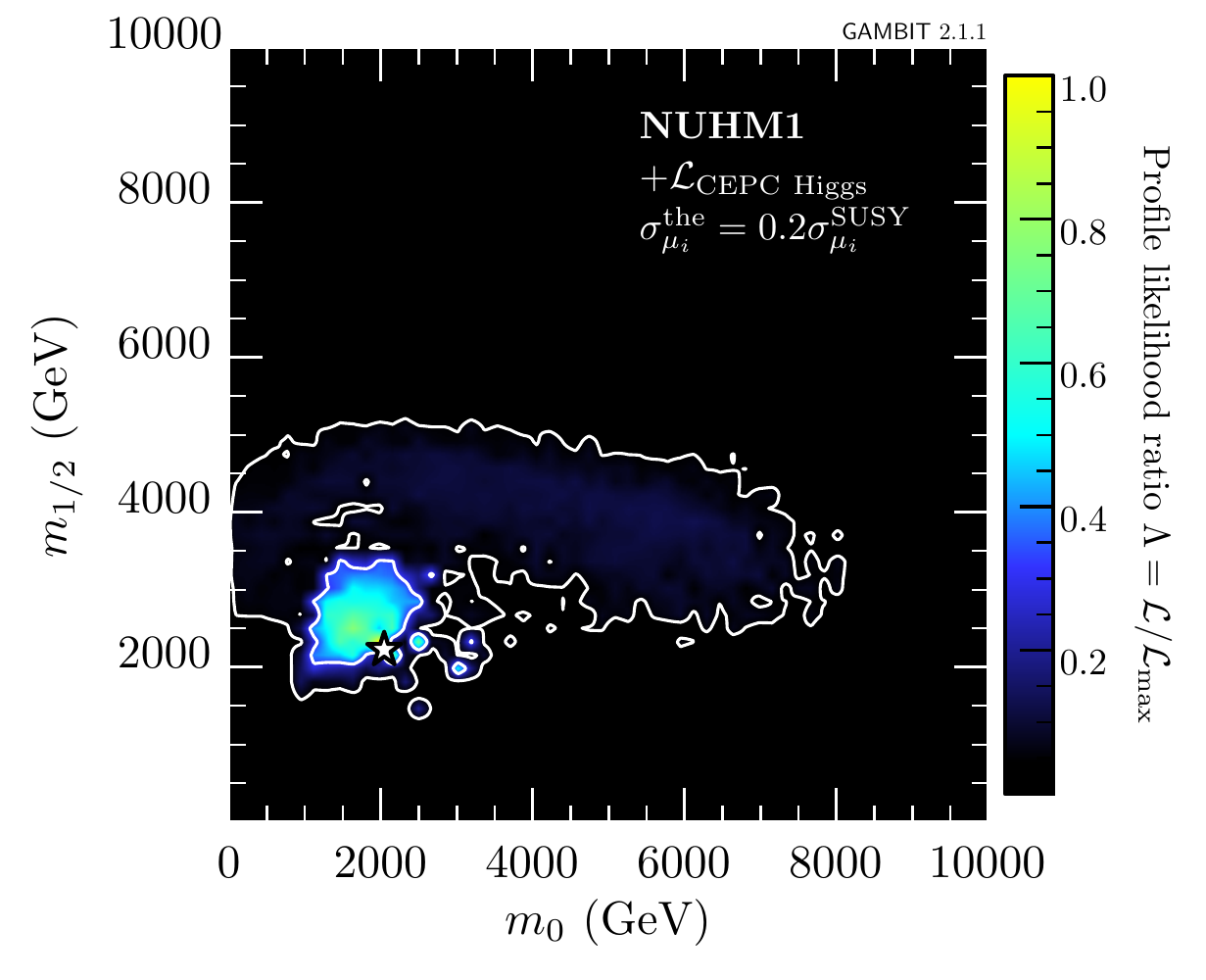}
  \includegraphics[width=0.49\textwidth]{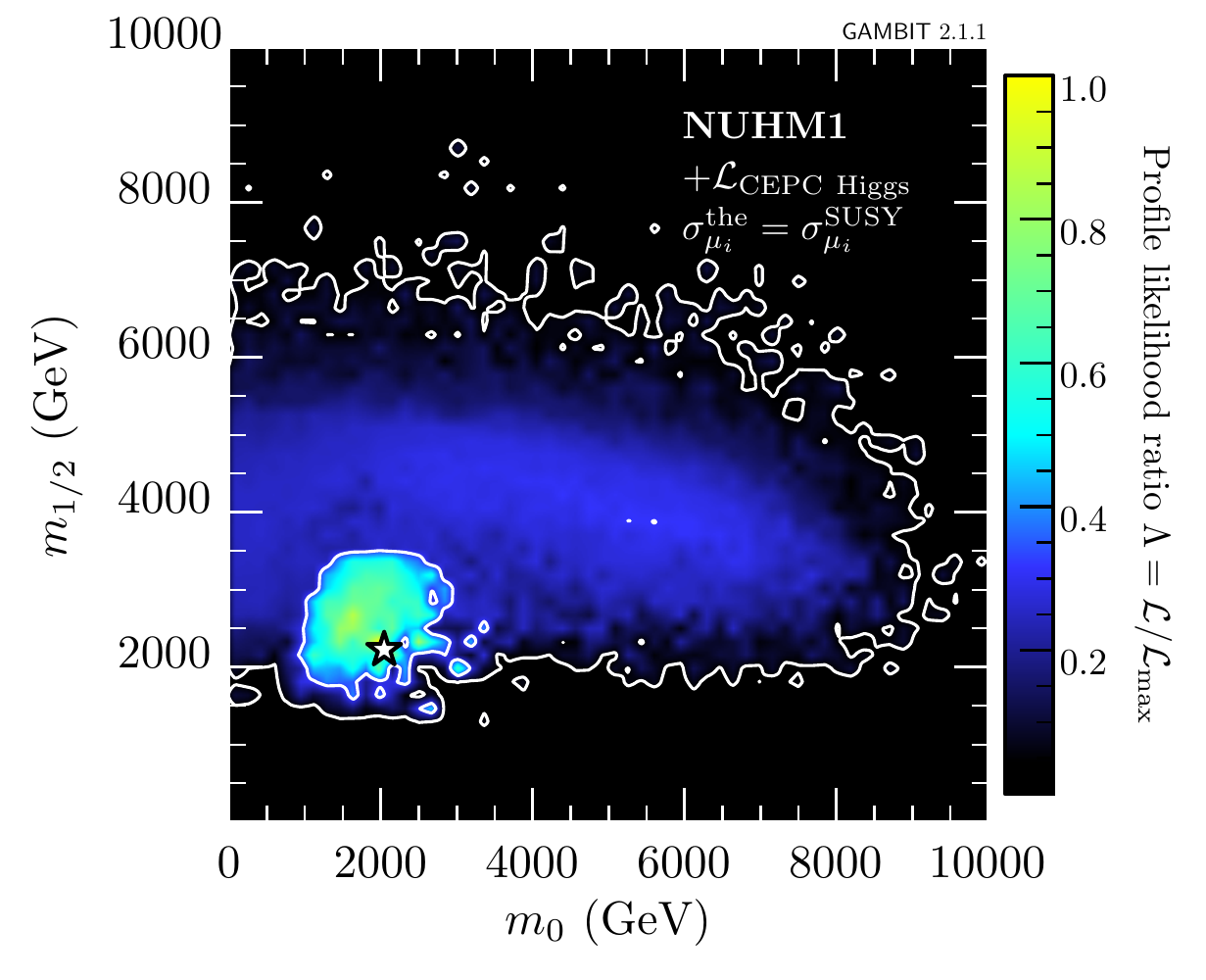}
  \caption{
  Profile likelihood ratio assuming no theoretical uncertainties on the signal strength at CEPC (top left panel), assuming the future theoretical uncertainties equal to the current theoretical uncertainties of the SM Higgs (top right panel), and taking SUSY contributions to the theoretical uncertainties into consideration (bottom panels), plotted in the $m_0$--$m_{1/2}$ plane of the NUHM1. }
  \label{fig:compare_k}
\end{figure}

The extent of the confidence regions surrounding the best-fit point depends to some degree on the assumptions made about the theoretical uncertainties on the Higgs branching ratios.  Comparing with the top right panel of 
Figure~\ref{fig:2d_parameter_plots_nuhm} which uses $0.2$ of the current theoretical uncertainties 
in $\mathcal{L}_{\rm CEPC}$, we show the result without theoretical uncertainties and with full current SM theoretical
uncertainties in the top left and top right panel of Figure~\ref{fig:compare_k}, respectively.  
In general the confidence region shrinks a lot with smaller theoretical uncertainties, though the dark matter 
annihilation mechanisms and the sign of $\mu$ in the $1\sigma$ regions remain the same.
Without theoretical uncertainties, only small regions around the best-fit point remain.
Assuming no improvement on the theoretical uncertainties, the favored regions still shrink, but this would substantially negate the advantage of Higgs factories. 
In addition, we add SUSY contributions to the theoretical uncertainties according to Ref.~\cite{Arbey:2021jdh}, and display the results in bottom panels of Figure~\ref{fig:compare_k}. Here $\sigma_{\mu_i}^{\rm SUSY}$ are the theoretical uncertainties including SUSY contributions based on $\sigma_{\mu_i}^{\rm SM}$, namely 5\% for Higgs decaying to quark pair and lepton pair, 5\% for $h\to WW^*$ and $ZZ^*$, and 3\% for $h\to \gamma\gamma$.  In the bottom left panel we show the impact if the $\sigma_{\mu_i}^{\rm SUSY}$ are reduced by a factor 5, while in the bottom right, in contrast, we use the current uncertainties.  The bottom left panel of Figure~\ref{fig:compare_k} can be  compared with the top right panel of Figure~\ref{fig:2d_parameter_plots_nuhm}. They mildly enlarged the $1\sigma$ and $2\sigma$ regions, as expected.
We see that assumptions about the theoretical uncertainties for the Higgs branching ratios clearly influence the global 
fit results when including the Higgs measurements at the CEPC, though not as much as the choice of which central values are used in the likelihood.

\begin{figure}[tb!]
 \centering
  \includegraphics[width=0.49\textwidth]{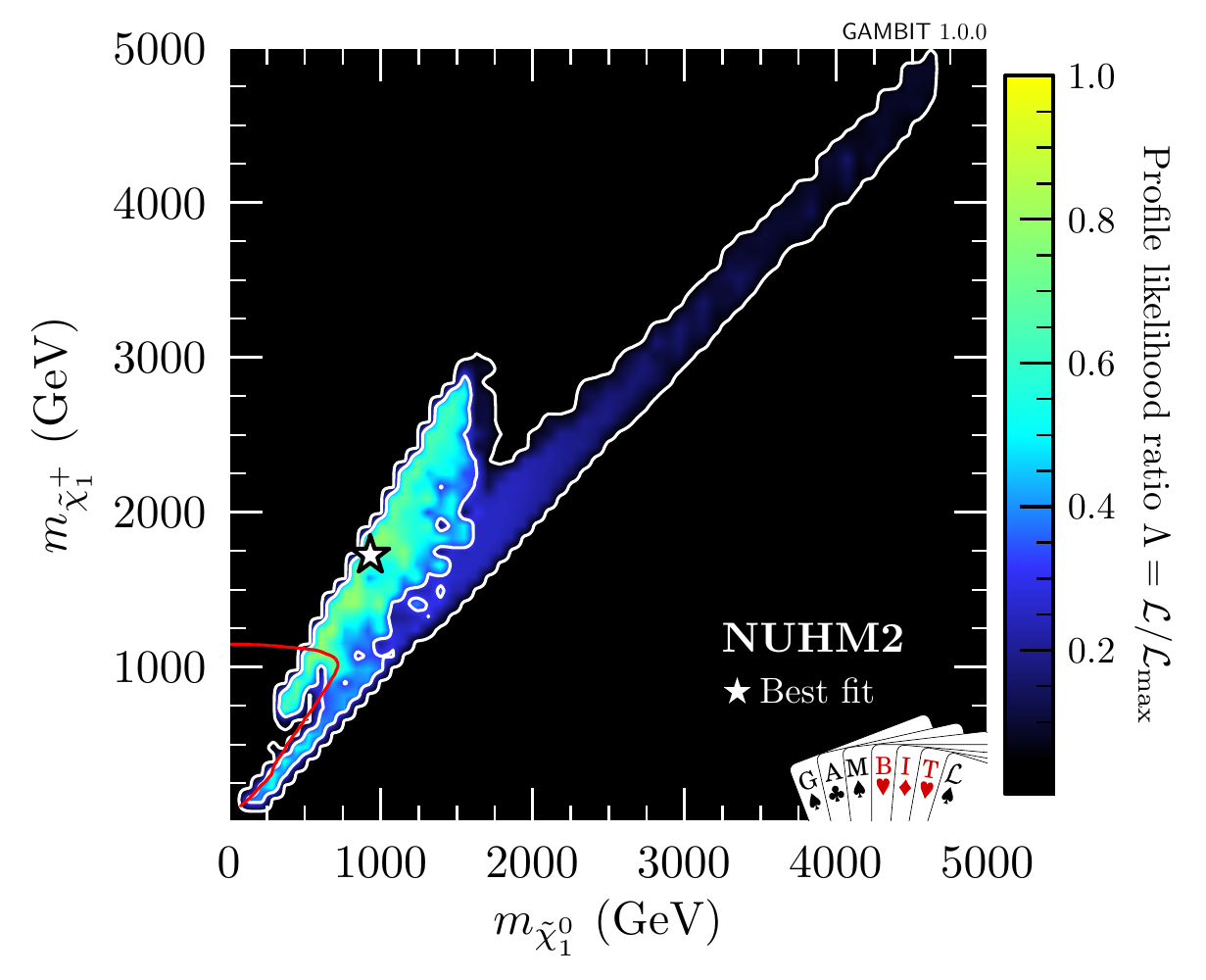}
 \includegraphics[width=0.49\textwidth]{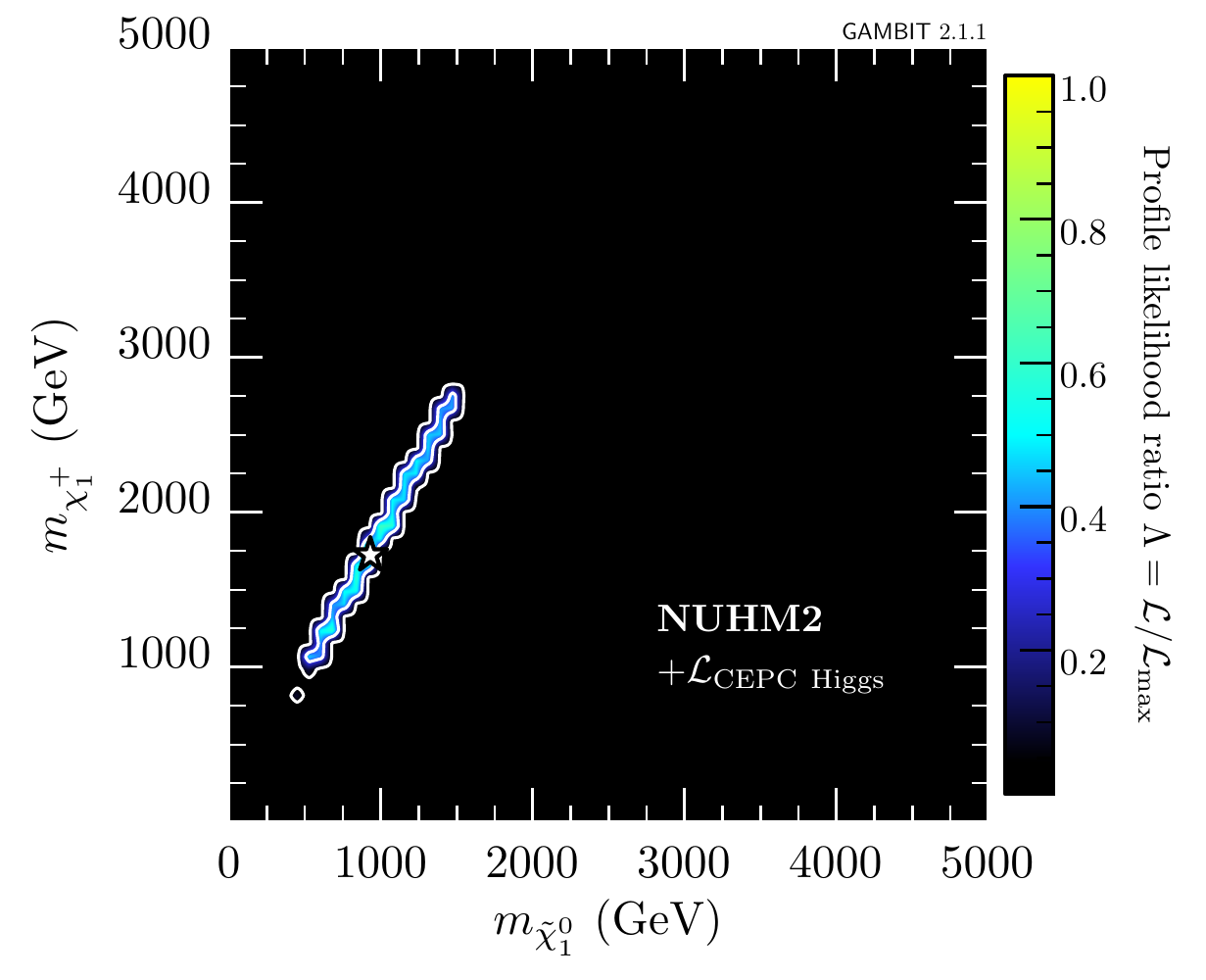}\\
 \caption{Profile likelihoods for the NUHM2 without CEPC likelihood (left panels) and with CEPC likelihood (right panels), 
plotted in the $\tilde{\chi}_1^\pm-\tilde{\chi}^0_1$ mass plane. }
 \label{fig:br}
\end{figure}

In Figure~\ref{fig:br}, we show the results on the  $m_{\tilde{\chi}_1^\pm}$--$m_{\tilde{\chi}^0_1}$ plane. We see that as 
the dark matter annihilation mechanisms constrain the dark matter mass and the relationships between sparticle masses, and that the masses of relevant sparticles are further restricted into limited ranges by $\mathcal{L}_{\rm CEPC}$. 
In the stop co-annihilation region, the wino-dominated chargino mass is about twice as large as the bino-dominated 
dark matter mass, because $M_1:M_2\simeq 1:2$ at the low scale produced by $M_1=M_2$ at the GUT scale. 
As the chargino co-annihilation regions are fully excluded, it sets upper limits of 1.7\,\TeV on $m_{\tilde{\chi}_1^0}$ 
and 3\,\TeV on $m_{\tilde{\chi}_1^\pm}$.

\subsection{MSSM7}
\label{sec:mssm7}

\begin{figure}[t!]
  \centering
  \includegraphics[width=0.32\textwidth]{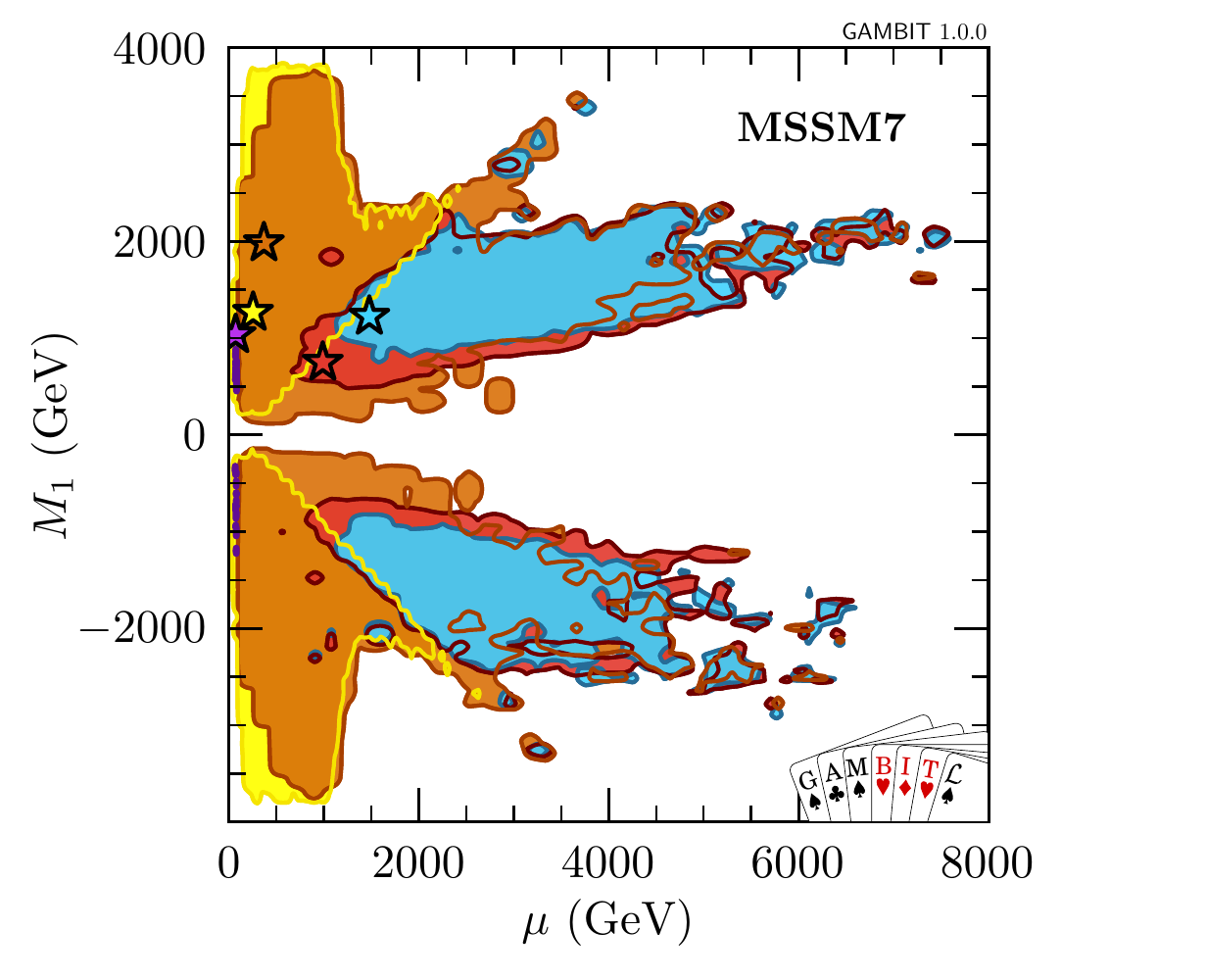}
  \includegraphics[width=0.32\textwidth]{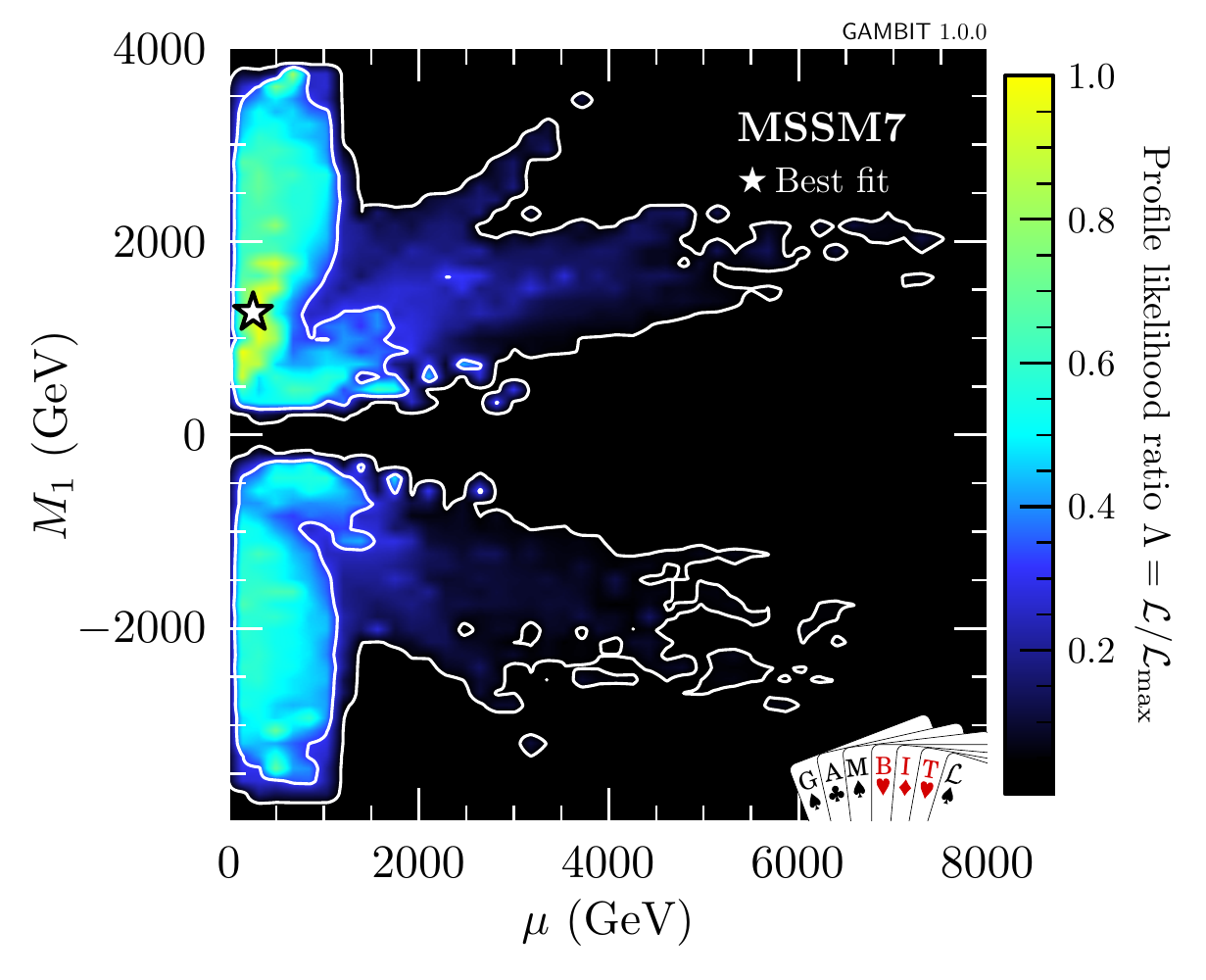}
  \includegraphics[width=0.32\textwidth]{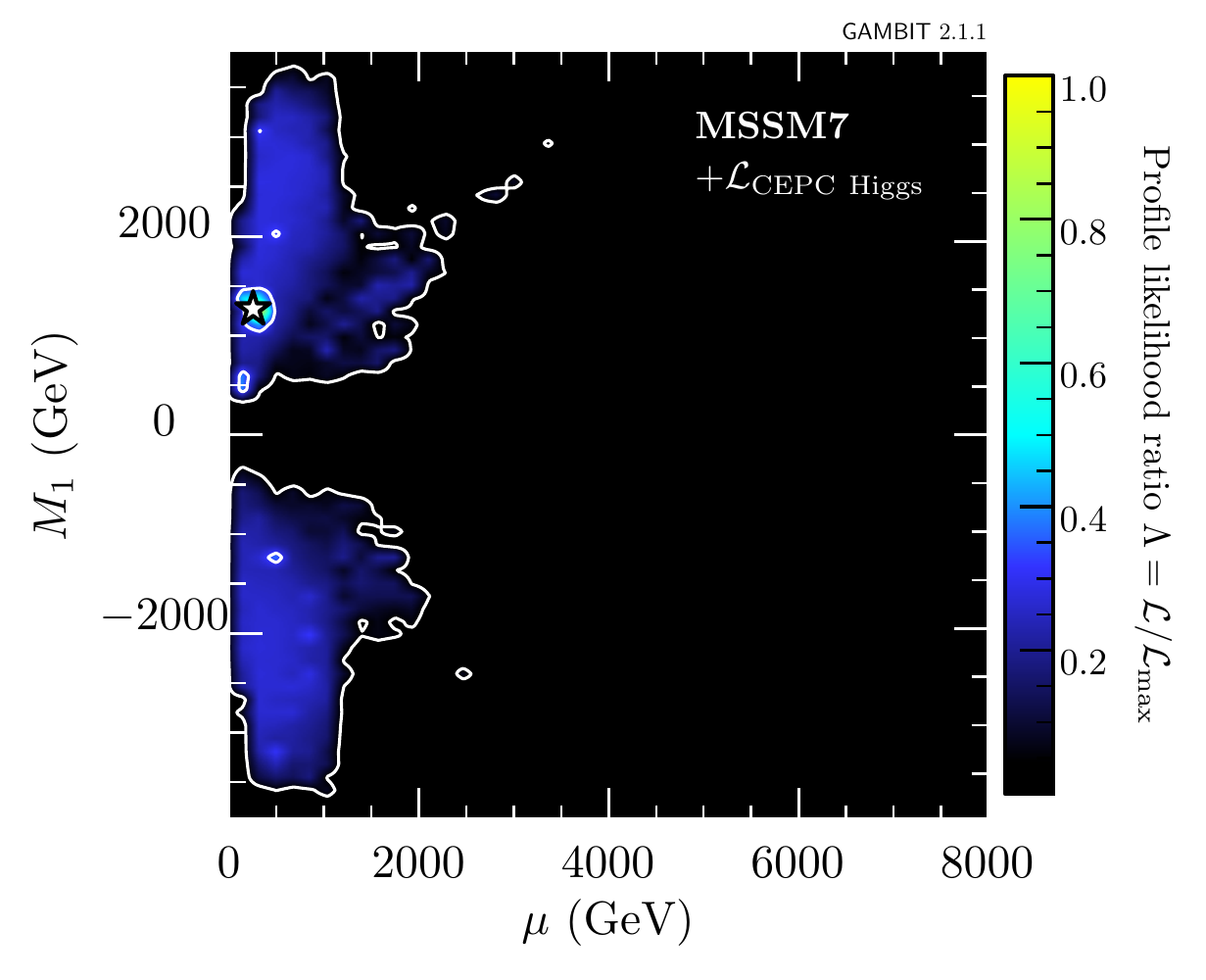}
  \\
  \includegraphics[width=0.32\textwidth]{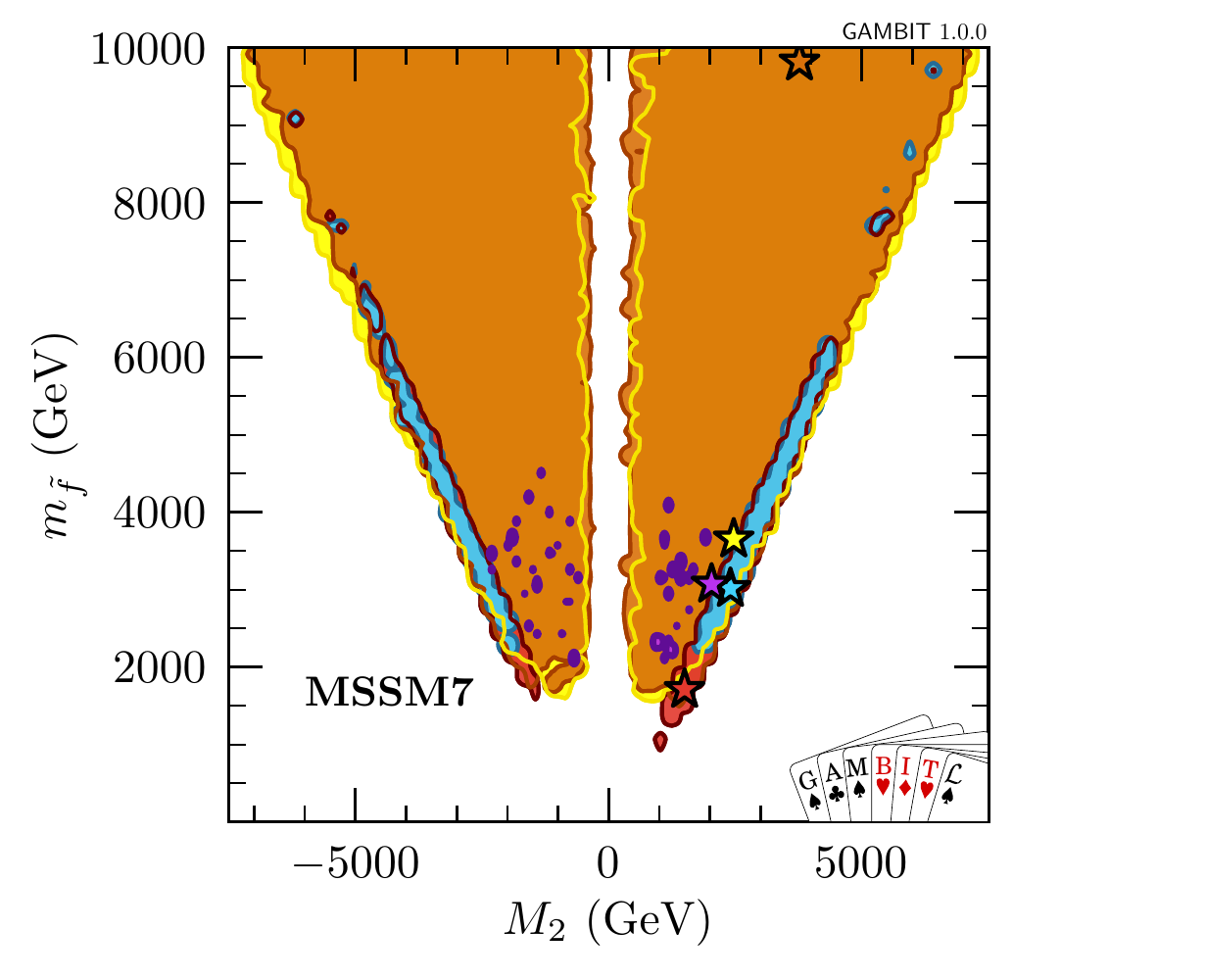}
  \includegraphics[width=0.32\textwidth]{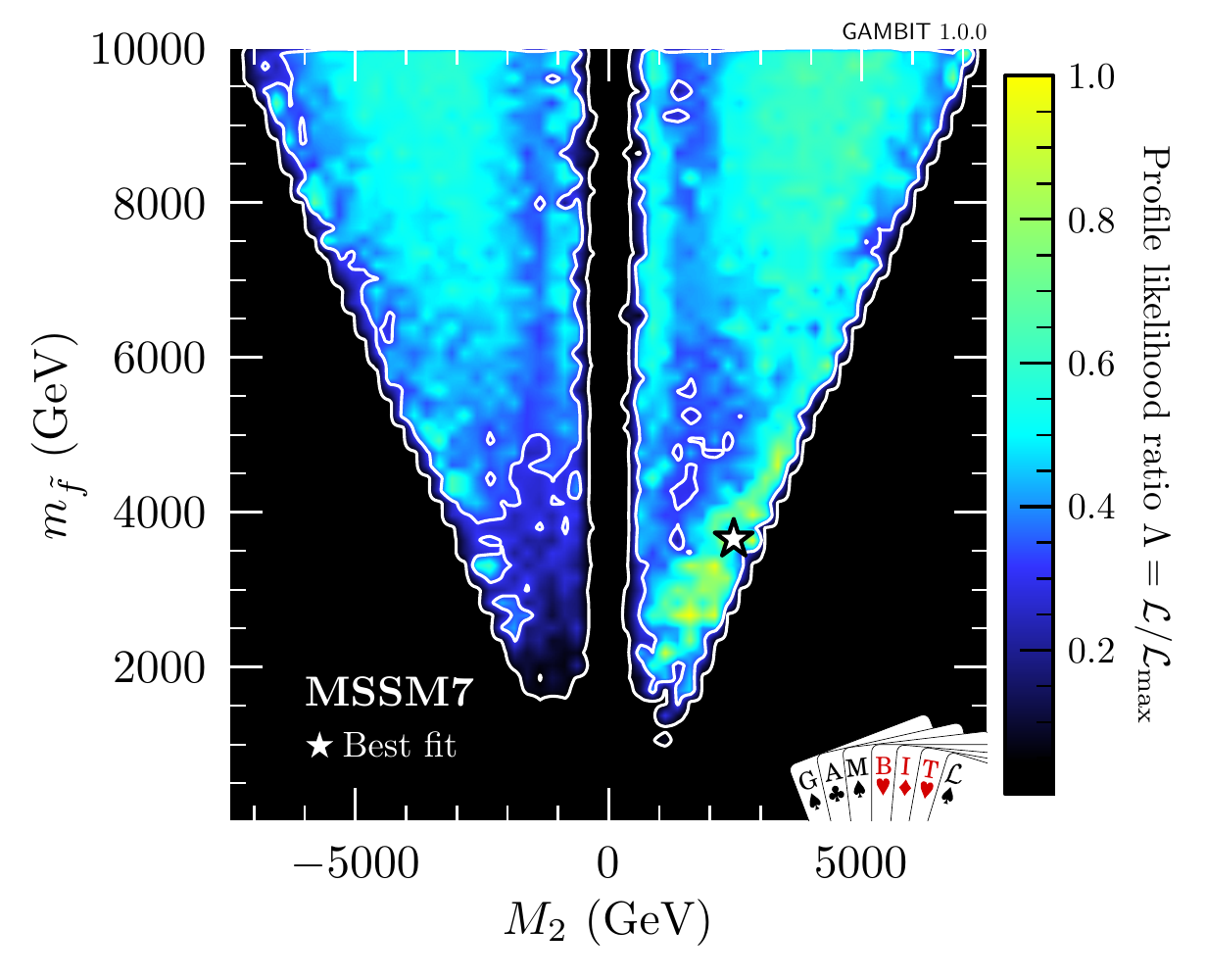}
  \includegraphics[width=0.32\textwidth]{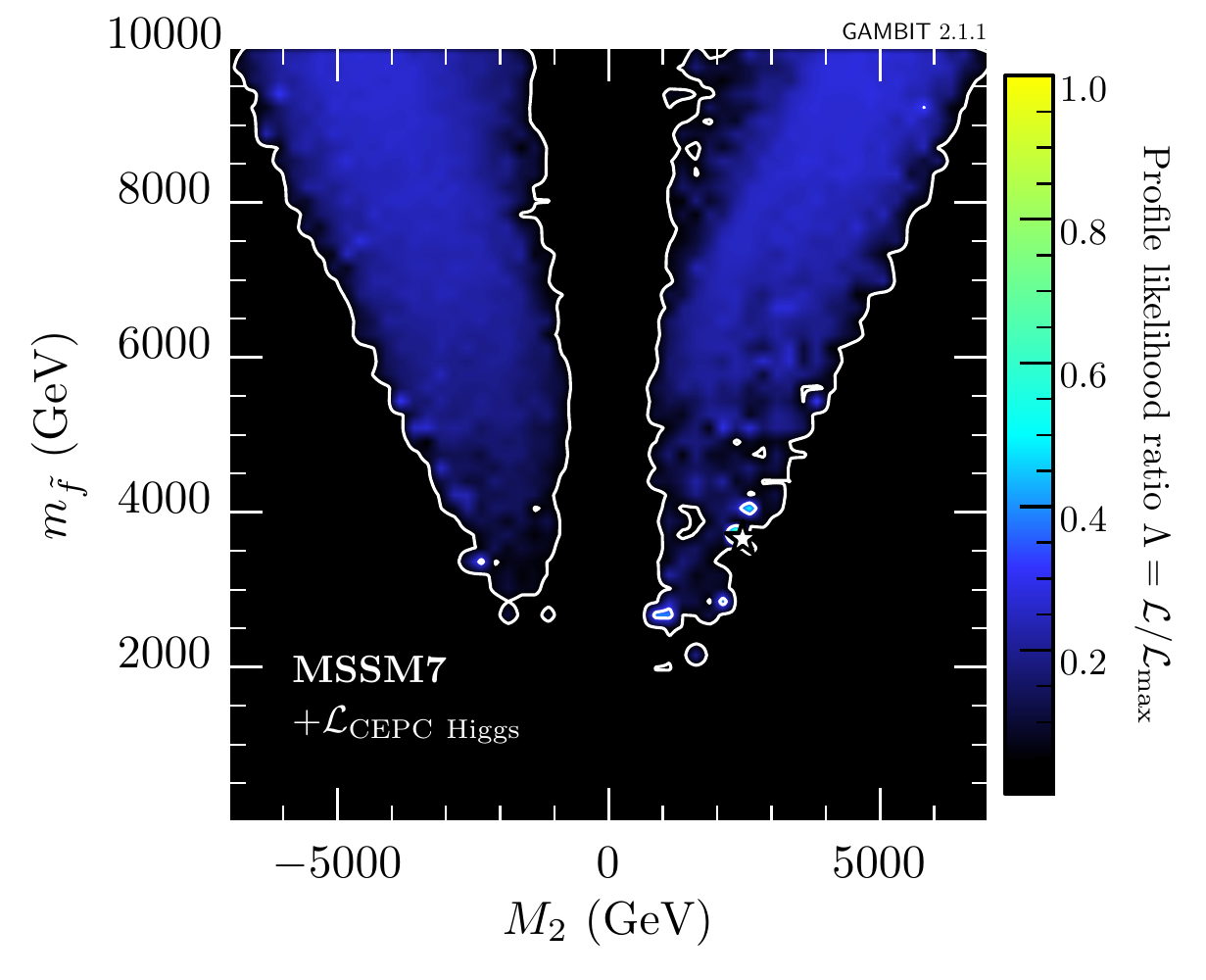}
  \\
  \includegraphics[width=0.32\textwidth]{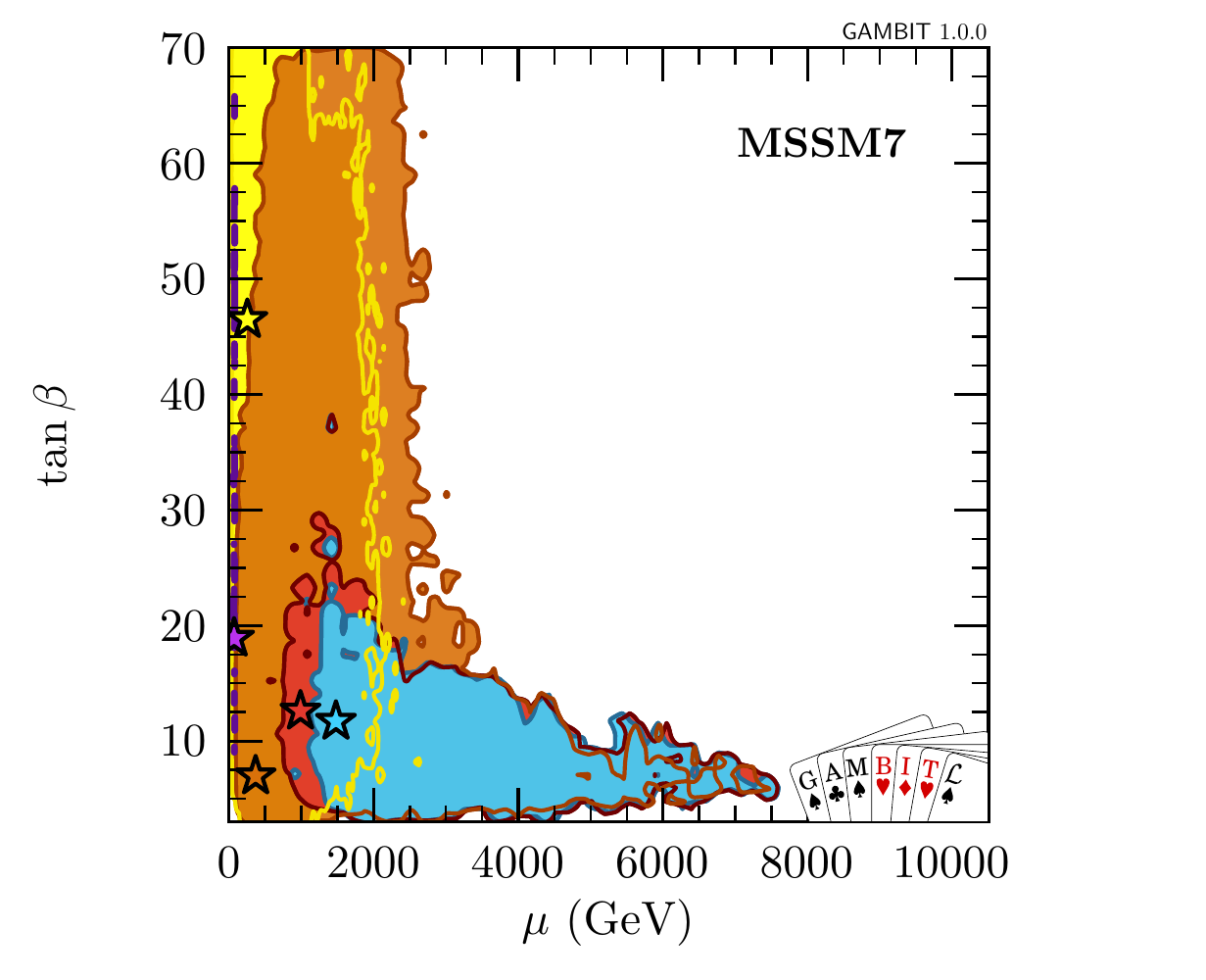}
  \includegraphics[width=0.32\textwidth]{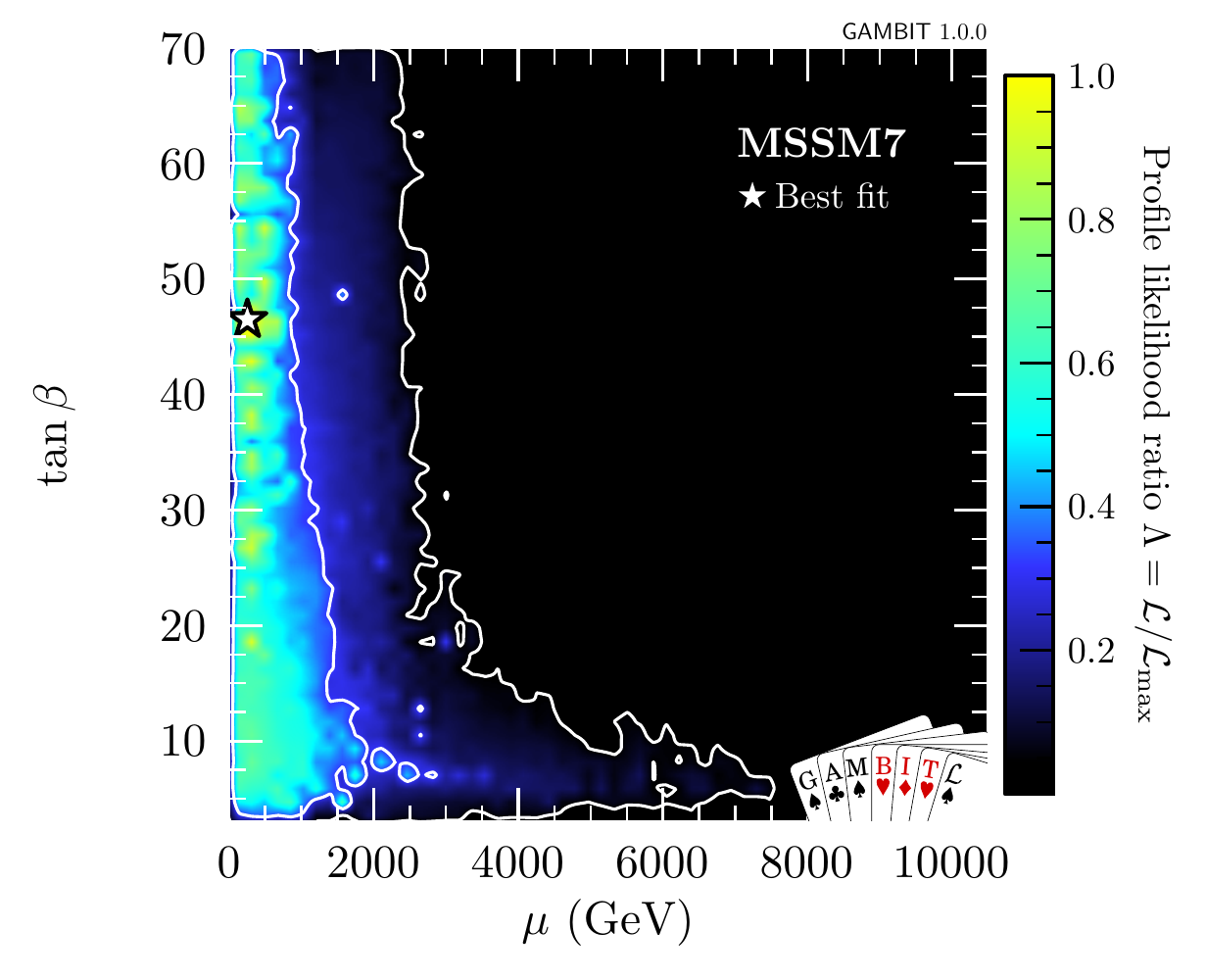}
  \includegraphics[width=0.32\textwidth]{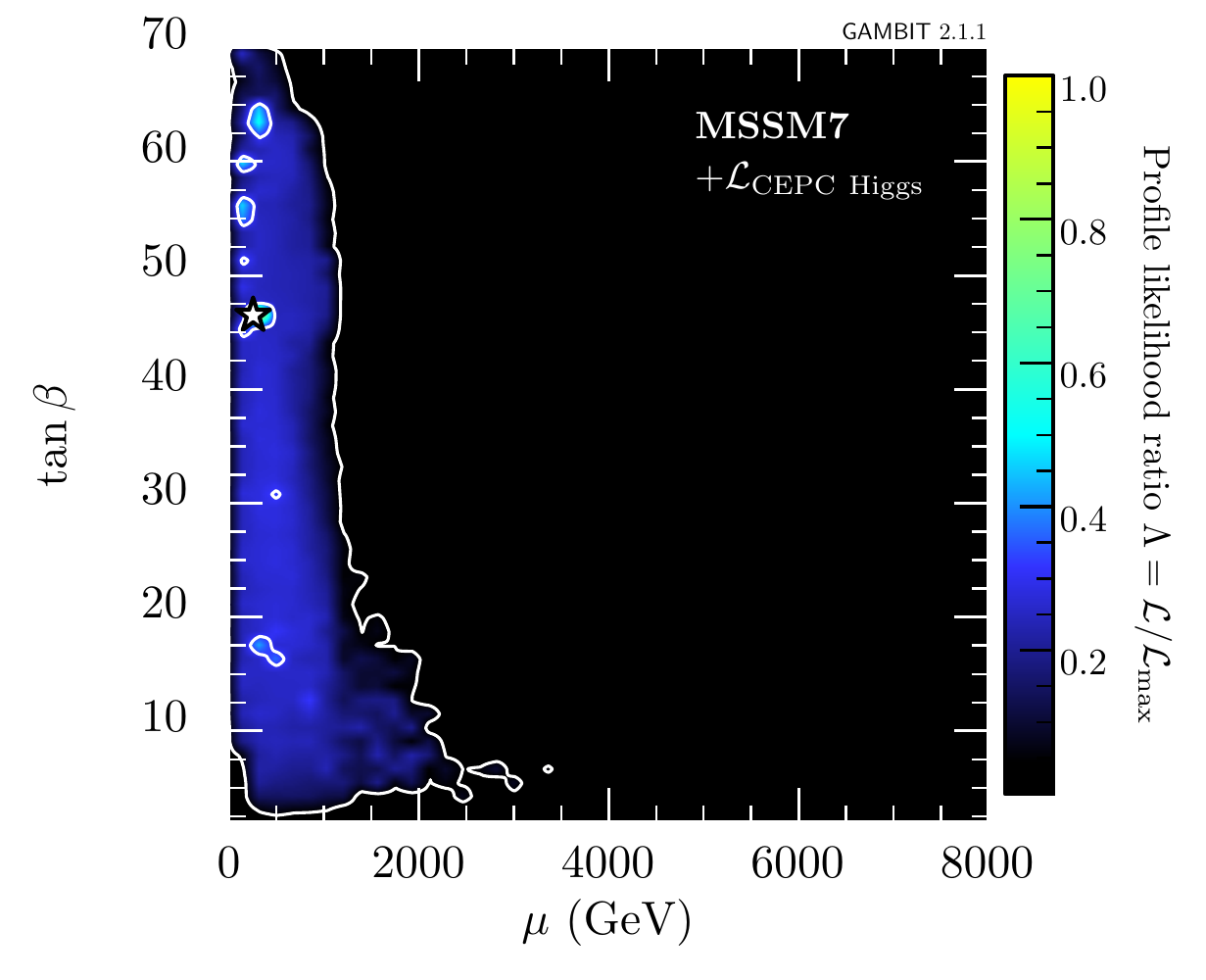}
  \\
  \includegraphics[width=0.32\textwidth]{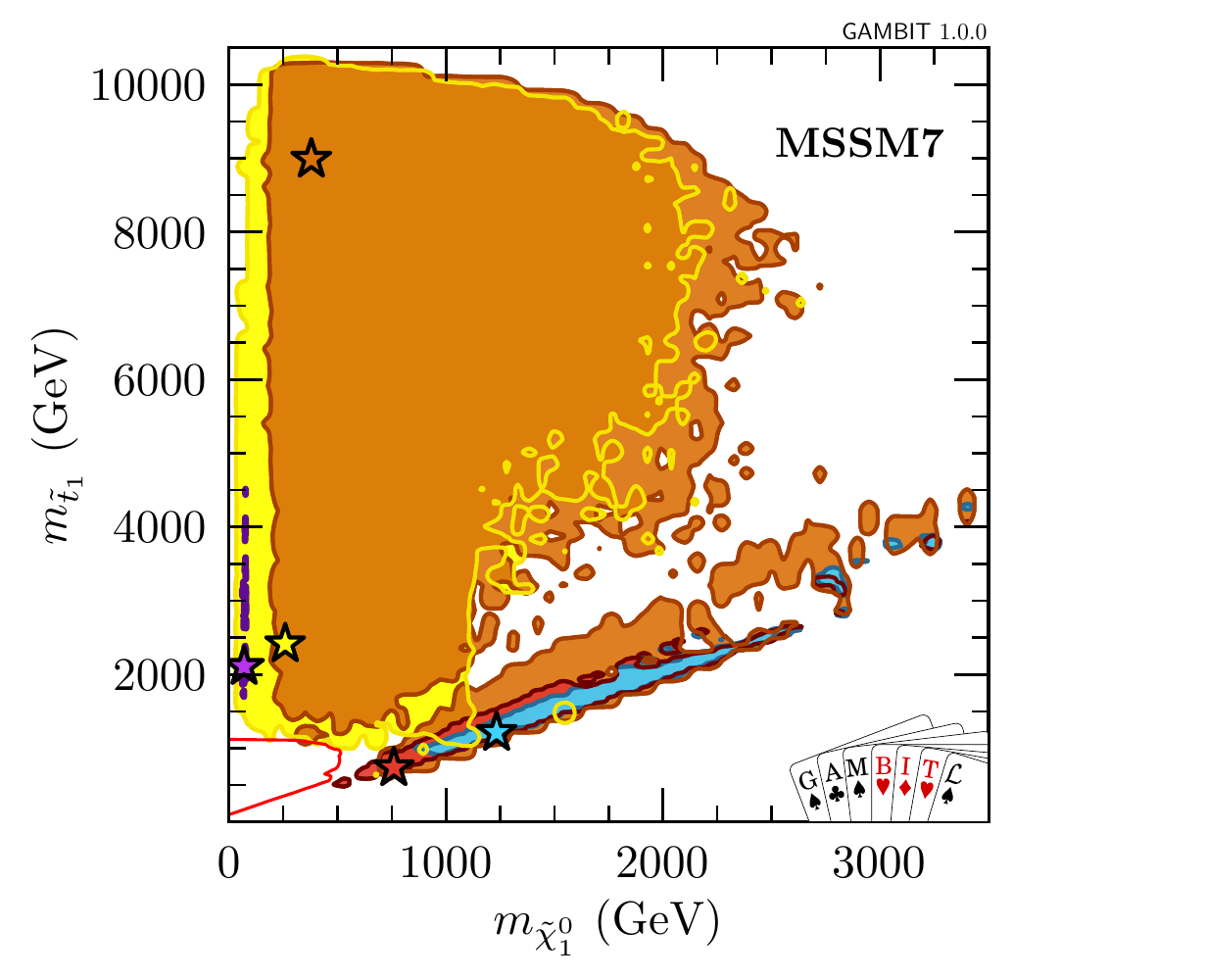}
  \includegraphics[width=0.32\textwidth]{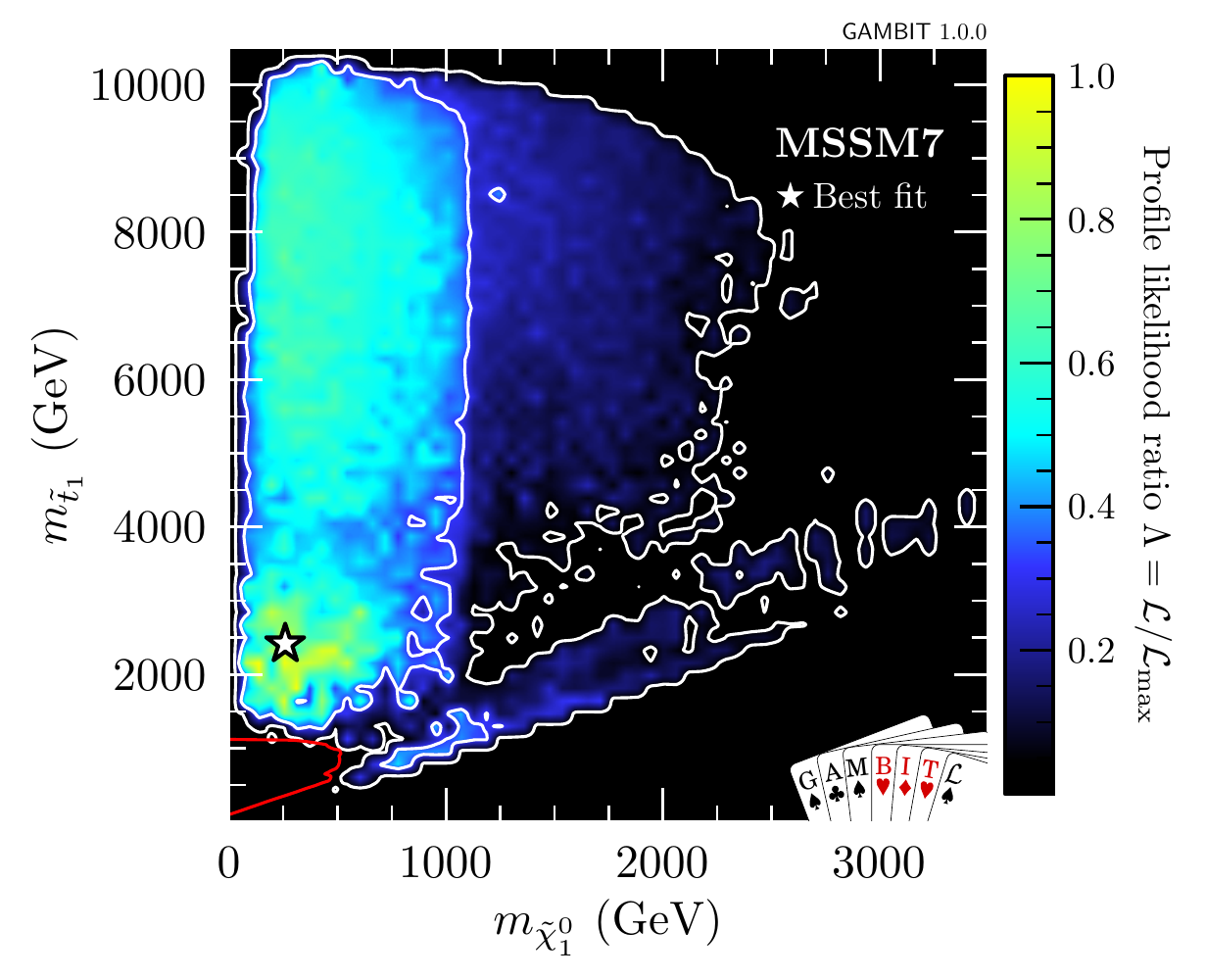}
  \includegraphics[width=0.32\textwidth]{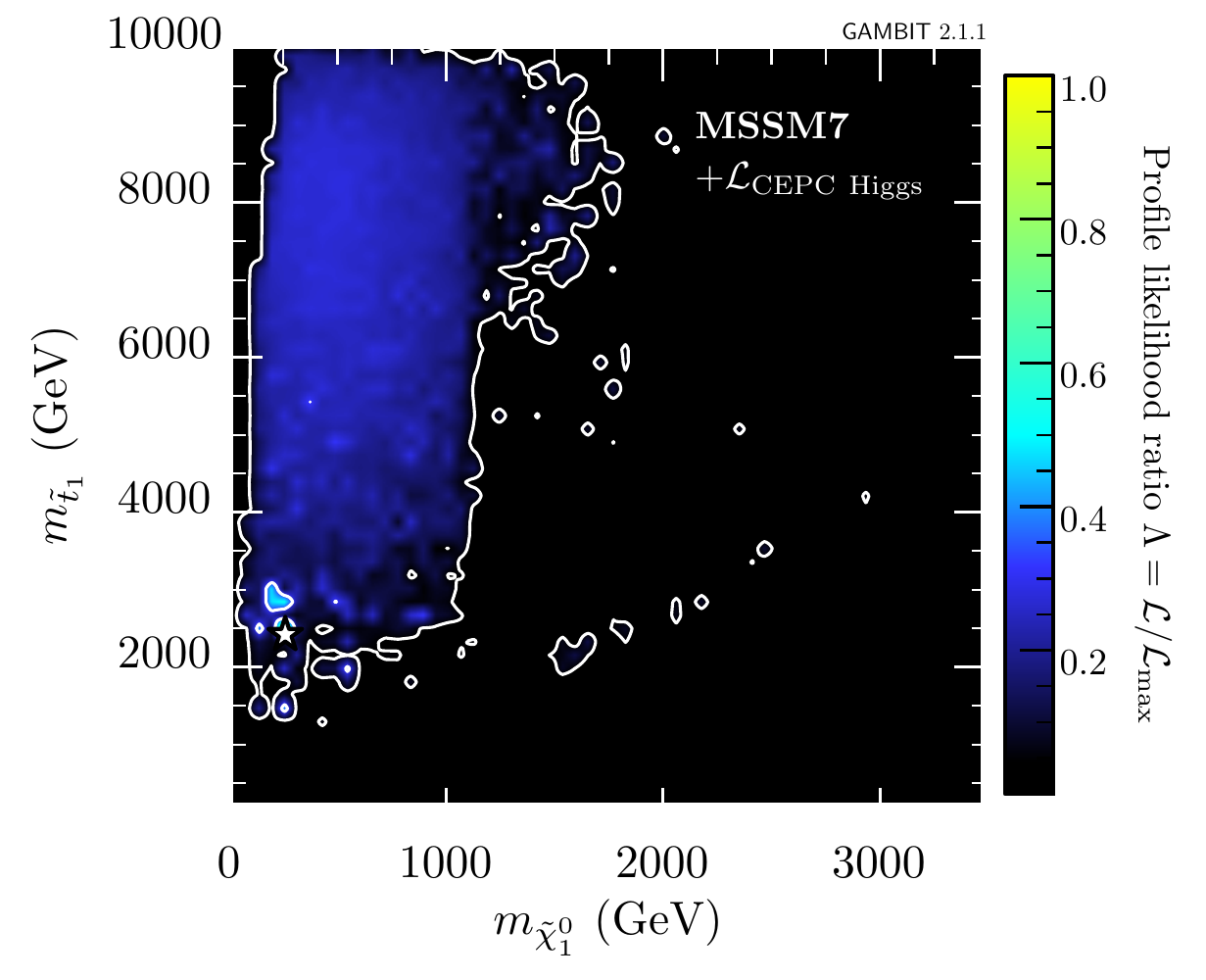}
  \\
  \includegraphics[height=4mm]{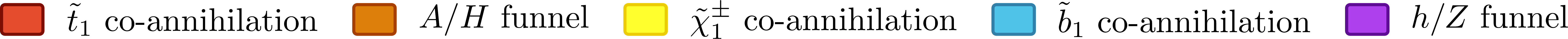}
  \caption{Profile likelihood ratio without the CEPC likelihood (middle panels, taken from~\cite{Athron:2017yua}) and 
with the CEPC likelihood (right panels) for the MSSM7.
Colour-coding in the left panels (taken from~\cite{Athron:2017yua}) shows the mechanisms 
active in models within the 95\% CL contour for avoiding thermal overproduction of neutralino dark matter. 
The overall best-fit point is indicated by a white star, while the best-fit points in each region are indicated 
by colored stars. The assumptions about the CEPC likelihood are the same as those in Figure~\ref{fig:2d_parameter_plots_cmssm}.}
  \label{fig:2d_parameter_plots_mssm7}
\end{figure}

The results of the MSSM7 scan are shown in Figure~\ref{fig:2d_parameter_plots_mssm7}, 
on the parameter planes of $\mu$--$M_1$, $M_2$--$m_{\tilde{f}}$ and $\mu$--$\tan\beta$, and on the 
mass plane of $\tilde{\chi}_1^0$--$\tilde{t}_1$. Here $\mu$ and $M_1$ are presented at the scale 
$M_{\rm SUSY} = \sqrt{m_{\tilde{t}_1}m_{\tilde{t}_2}}$. Two new regions 
\begin{itemize}
\item sbottom co-annihilation: $m_{\tilde{b}_1} \leq 1.2\,m_{\tilde\chi^0_1}$,
\item $h/Z$ funnel: $1.6\,m_{\tilde\chi^0_1} \leq m_\textrm{light} \leq 2.4\,m_{\tilde\chi^0_1}$,
\end{itemize}
appear, where `light' may be $h$ or $Z$. In the sbottom co-annihilation region, the lightest sfermion 
is $\tilde{t}_1$ and therefore it highly overlaps with the stop co-annihilation region. 
In contrast to the GUT-scale models, the overall best-fit point of the MSSM7 is located in the chargino co-annihilation 
region, and the corresponding best-fit likelihood is improved.

We see that the $1\sigma$ ranges are shrunk significantly, similar to the results of GUT-scale models,  
but the $2\sigma$ ranges of parameters are not visibly reduced by the CEPC constraints, except for the $\mu$ parameter. 
With free $m_{H_u}^2$ and $m_{H_d}^2$, the lightest chargino and neutralino can be higgsino-dominated for almost any 
value of $M_2$, $m_{\tilde{f}}$, $A_{d_3}$, $A_{u_3}$ and $\tan\beta$. Considering that we set the central values 
of Higgs measurements at CEPC to the values of the overall best-fit point, most of the chargino co-annihilation 
region survives. Thus, the ranges of those parameters cannot be restricted. On the other hand, the $\mu$ parameter 
has to be lighter than $M_1$ in the chargino co-annihilation region, so that the lightest chargino is not bino-dominated, 
shown in the top left panel of Figure~\ref{fig:2d_parameter_plots_mssm7}. When the other regions are excluded, 
the upper limit on $\mu$ drops accordingly.

As the $A/H$ funnel region and the chargino co-annihilation region overlap heavily in all the planes, part of the $A/H$ funnel regions escapes from the restriction of the CEPC Higgs measurements. Besides, there are a few samples satisfying the stop and sbottom co-annihilation conditions, which can be seen in the low panels of Figure~\ref{fig:2d_parameter_plots_mssm7}, but they fulfill the $A/H$ funnel selection at the same time. We also see one sample satisfying the $h/Z$ funnel and the chargino co-annihilation conditions simultaneously. 
Overall, Higgs measurements at the CEPC do not have much power to discriminate between different DM annihilation mechanisms in the MSSM7. 

\begin{figure}[t!]
  \centering
  \includegraphics[width=0.49\textwidth]{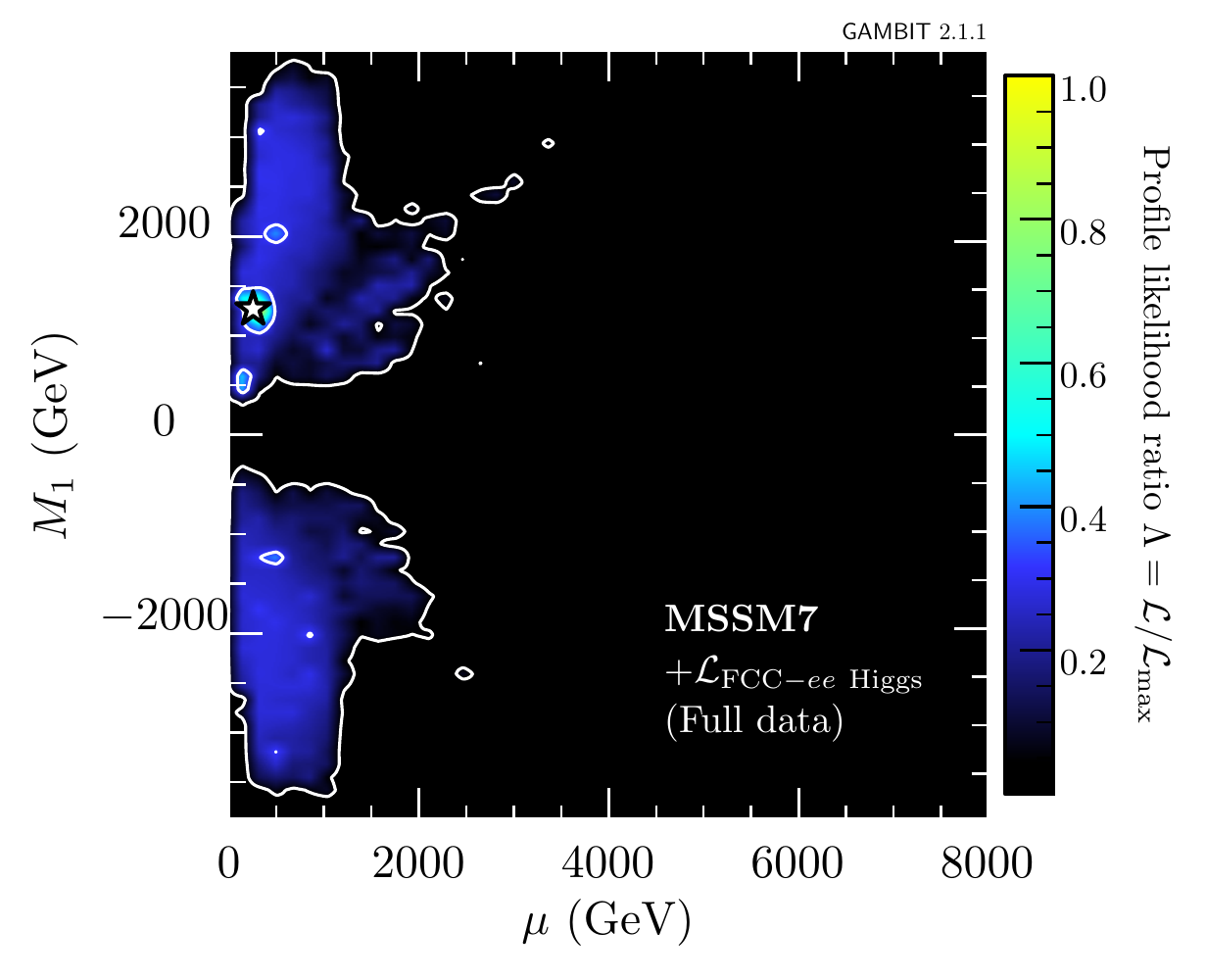}
  \includegraphics[width=0.49\textwidth]{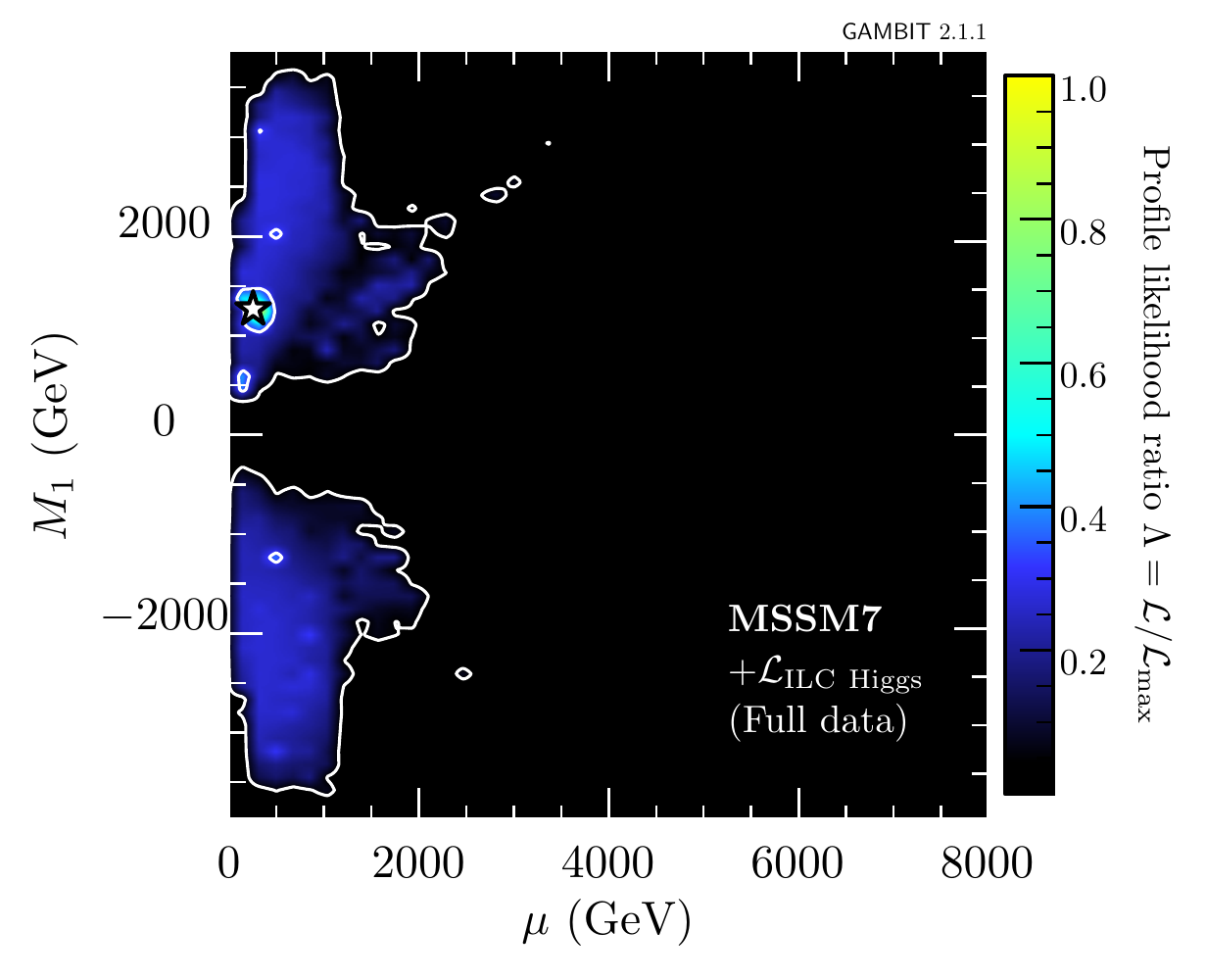}\\
  \caption{
  Profile likelihood ratio for the MSSM7, comparison between the potential reach of FCC-$ee$ and ILC. }
  \label{fig:compare_colliders}
\end{figure}

We present a brief comparison for the potential reach of the CEPC, FCC-$ee$ and ILC in Figure~\ref{fig:compare_colliders}, together with the top left of Figure~\ref{fig:2d_parameter_plots_mssm7}. The results include measurements at all proposed center-of-mass energies listed in Table~\ref{tab:mu_precision}. 
The CEPC and ILC result in slightly stronger constraints than the FCC-$ee$ and almost same contour regions, but for different reasons. The CEPC proposes a higher integrated luminosity at $\sqrt{s}=250\,\GeV$, while the ILC has more center of mass energy options. The difference of favored regions between these facilities are small due to the theoretical uncertainties which are larger than the anticipated precision in some signal channels. 

\begingroup
Before concluding, we consider the implications on our models of two outstanding experimental anomalies.  First, the surviving regions shown for all of these SUSY models cannot make a significant contribution to resolving the recently updated muon $g-2$ anomaly~\cite{Muong-2:2021ojo}, as the unified sfermion mass $m_{\tilde{f}}$ is pushed to a relatively large value by other constraints, such as LHC sparticle searches and $B$-physics constraints. An MSSM explanation of the muon $g-2$ anomaly at the $2\sigma$ level requires light EWinos and light sleptons (see, e.g., \cite{Athron:2021iuf, Wang:2021bcx, Endo:2021zal,Abdughani:2019wai}), which may be accessible at the future runs of the LHC \cite{Abdughani:2019wai,Cao:2022htd}. 
    
Second, the new CDF II measurement of the $W$-boson mass shows a $7\sigma$ deviation from the SM prediction~\cite{CDF:2022hxs}. This differs significantly from the central value used for $m_W$ in previous \gambit fits. Therefore, all samples in our favored regions disagree with the new measurement by at least $2\sigma$. In any case, it is difficult to explain the large deviation in the general MSSM ~\cite{Bagnaschi:2022qhb,Yang:2022gvz,Athron:2022isz}. 

Finally, let us remark on the issue of naturalness in our scenarios. We quantified the fine-tuning for the best fit points in the GUT-scale SUSY models through
\begin{equation}
    \Delta_\mu = \frac{\partial \ln M_Z^2}{\partial \ln \mu^2}
\end{equation}
Using \textsf{SuSpect}~\cite{Djouadi:2002ze}, we found about 4000 for the CMSSM, % 4056.0
700 for the NUHM1 and % 723.4  
900 for the NUHM2. % 876.8 
%In the $(m_0,m_{1/2})$ plane of CMSSM, the fine-tuning can be estimated from 
%\begin{align}
%\Delta_{\rm BG} &\approx ... {\rm ~When~} \tan\beta \approx 15 \\
%\Delta_{\rm BG} &\approx ... {\rm ~When~} %\tan\beta \approx  50.
%\end{align}
The fine-tuning in the favored regions of the CMSSM is generally larger than that in the NUHM1/2, because the condition $m_0^2=m_{H_u}^2=m_{H_d}^2$ leads to a strict constraint on $m_0$. Investigating the fine-tuning cost comprehensively is beyond the scope of this paper, but the fine-tuning of these benchmarks should be a reasonable indication of the typical level of tuning one expects in the surviving samples.
\endgroup

\section{Conclusions}
\label{sec:conclusion}

Utilizing the publicly available data for SUSY global fits from the GAMBIT community, we examined the potential impact of measurements at proposed Higgs factories on several constrained versions of the MSSM, namely the CMSSM, NUHM1, NUHM2 and MSSM7. 
We post-processed all the samples to calculate the likelihood for Higgs measurements from the CEPC as an example, 
and then compared profile likelihood ratios with and without this additional likelihood from the CEPC. The preferred 
regions in these models are significantly shrunk by the precise Higgs measurements at the CEPC. As a result, the possible dark matter annihilation mechanisms in the models and signs of $\mu$ parameter could be distinguished by CEPC measurements. Comparing results in different models, the constraints on model parameters are weaker in models with more input parameters, i.e., looser correlations between model parameters. The specific favored and excluded parameter regions highly depends on the assumed central values of the Higgs measurements, and are mildly dependent on assumptions about theoretical uncertainties.
Since the projected experimental uncertainties are even better than the current theoretical uncertainties, reaches of Higgs factories are dominated by the theoretical uncertainties.  Reducing these theory uncertainties is an important and significant challenge.  Under the assumption that the future theoretical uncertainties could be reduced to the same level of expected experimental uncertainties, we compared the impact of the CEPC, FCC-$ee$ and ILC on the MSSM7. We found that their reaches are quite similar, and slightly better with higher proposed luminosity or higher center-of-mass energy, as expected. 
In summary, with high-precision Higgs coupling measurements, future Higgs factories can significantly advance our understanding of the MSSM parameter space and mass spectrum, and could be complementary to dark matter searches and EW precision measurements.

\section*{Acknowledgements}
We thank Xuai Zhuang, Manqi Ruan, Zhijun Liang and Pat Scott for useful discussions. 
We acknowledge Beijing PARATERA Tech Company Limited for support of computing resources. 
This work was supported by the National Natural Science Foundation of China (NNSFC) under grant Nos.  12105248,
11821505, 12075300, 12147228 and 12150610460, by Peng-Huan-Wu Theoretical Physics Innovation Center (12047503), the Australian Research Council Future Fellowship grant FT160100274 and Discovery Project grants DP180102209 and DP210101636,  by the Key Research Program of the Chinese Academy of Sciences, Grant NO. XDPB15, and by a KIAS Individual Grant (PG084201) at Korea Institute for Advanced Study.

%\appendix
%\section{Appendix}
%\label{sec:app}

\addcontentsline{toc}{section}{References}
\bibliographystyle{JHEP}
\bibliography{refs}

\end{document}